\documentclass[12pt]{book}
\usepackage{fullpage}
\usepackage{amssymb}
\usepackage{epsf}
\usepackage{multicol}
\usepackage{graphicx}
\newcommand{\Case}[2]{{\textstyle \frac{#1}{#2}}}
\newcommand{\lP}{\ell_{\mathrm P}}
\begin{document}
\rightline{IMSc/2010/03/03}
\vskip 1.0cm
\centerline{\large{\bf Lectures on LQG/LQC}}
%\vskip .2in
%\centerline{\it In Preparation}
\vskip .3in
\centerline{Ghanashyam Date}
\vskip .1in
\centerline{The Institute of Mathematical Sciences, CIT Campus,
Chennai 600 113, India}
\vskip 2.0in
%\hrule
A School on Loop Quantum Gravity was held at the IMSc during Sept 8 --
18, 2009. In the first week a basic introduction to LQG was provided 
while in the second week the focus was on the two main application, to
cosmology (LQC) and to the black hole entropy. These notes are an
expanded written account of the lectures that I gave. These are
primarily meant for beginning researchers.
\vskip .3in
\tableofcontents
\newpage
%The course is planned along the following lines:
%
%\begin{enumerate}
%
\centerline{{\Large{\bf Preface}}}
\vskip 1.0cm

It has been felt for a while that our graduate students do not get an
opportunity to get an exposure to the non-perturbative, background
independent quantum theory of gravity at a pedagogic level. Although
there are several excellent reviews and lecture notes available, an
opportunity for complementing lectures by discussions is always an added
bonus for the students. With this in mind and taking into account of the
background preparation of the students, the {\em School on Loop Quantum
Gravity} was organized at IMSc, for a period of 10 days. The first 5
days were devoted to the basics of connection formulation and loop
quantization up to sketching steps involved in the quantization of the
Hamiltonian constraint. The next 5 days were devoted to the applications
to quantum cosmology and to the black hole entropy. In all 20 lectures
and 10 tutorials were planned, however some tutorials `became'
additional lectures. 

These notes are an expanded version of the topics that I covered. In
particular the material of chapter 2, sections 4.2.1, 4.2.2, 4.2.3 and
appendices 5.1, 5.3, 5.4 have been added. Email discussions on the
sections of chapter 4 and appendix 5.4 with Abhay Ashtekar, Martin
Bojowald and Madhavan Varadarajan have been very helpful and are
gratefully acknowledged. There could still be some differences in the
perceptions and formulations, what I have presented is my understanding
of the issues.

The other main lecturers at the school were: Prof Amit Ghosh, Saha
Institute of Nuclear Physics, Kolkata;  Dr Alok Laddha, Raman Research
Institute, Bengaluru; Parthasarathi Majumdar, SINP, Kolkata. In
addition, Prof Romesh Kaul, IMSc, Dr Kinjal Banerjee, IUCAA, Pune and
Ayan Chatterjee, SINP, Kolkata also gave a few lectures. Amit and Alok
discussed the connection formulation and loop quantization up to the
basic steps in the quantization of the Hamiltonian constraint. Partha,
Amit and Ayan discussed classical formulation of isolated horizons and
entropy associated with them. Romesh discussed the possibility of a
`vacuum structure for gravity' and Alok also briefly discussed the
Brown-Kuchar dust model.  It is a pleasure to acknowledge their
contributions.  At least some notes of the topics covered by these will
be available in not-too-distant a future. 

The funding for the school was provided by the Institute of Mathematical
Sciences under the XI$^{\mathrm{th}}$ Plan Project entitled {\em
Numerical Quantum Gravity and Cosmology}. It is envisaged that a more
specialized workshop at an advanced level will be held in the summer of
2010 with the possibility of a similar School being repeated one more
time.

\vfill
March 17, 2010 \hfill Ghanashyam Date
\vskip 0.3cm
\hrule
\chapter{General Remarks:} 
\section {Why a Quantum Theory of Gravity?} 

This is not a rhetorical question but it is intended to {\em identify
physical context} in which the classical theory of gravity, specifically
the Einstein Theory of Gravity also referred to as General Relativity,
is {\em inadequate} and calls for an extension. One has met with
inadequacies of classical theories many times and has seen how their
quantum versions have alleviated the inadequacies. For example, the
classical theory of charges and electromagnetic fields was quite
adequate until the hydrogen atom was found to have a central nucleus
with an electron going around it. Classical theory predicts that since
the electron is necessarily accelerated, it must radiate away its energy
and spiral into the proton. Indeed in about $10^{-9}$ seconds (!) the
classical trajectory of a (bound) electron would `end' in the proton. We
all know that this is physically wrong and the atoms are known to be
stable for billions of years. We also know that the `fault' lies not
with the `Coulomb law' (which does get modified) but with the classical
{\em framework} of using well defined trajectories to describe the
dynamical evolution for both the electron and the electromagnetic field.
Inadequacies of classical theories are also revealed in the black body
spectrum, specific heat of solids at low temperatures etc etc and the
appropriate quantum theory of matter and electromagnetism cures these
problems i.e.  gives results consistent with experimental observations.
The quantum nature of other interactions such as the strong and the weak
is also verified in nuclear and particle physics. What about the
gravitational interactions?

Gravitationally bound `atoms' can also be considered. If gravity is
described in the Newtonian manner, there is no gravitational radiation
from an accelerated motion and the inward spiralling problem will not
arise. But Einstein's theory of gravity is very different and
accelerated sources do radiate away energy and the stability issue
re-surfaces. Of course the `weakness' of gravitational interaction does
not threaten the existence of such gravitationally bound atoms if the
decay time is larger than the age of the universe, but in principle
possibility exists. In fact Einstein did suggest a need for a quantum
theory of gravity \cite{AbhayEinstein} almost immediately after GR was
constructed\footnote{At that time, the universe was supposed to be
eternal and hence, however small the gravitational instability,
existence of stable atoms would threaten GR unless gravitational
radiation is also terminated at certain stage.}. 

General relativity however uncovered two distinct contexts in which the
theory calls for an extension -- the context of (i) cosmological and
black hole singularities and (ii) entropy of black hole horizons.  Let
us take a little closer look at these contexts.

{\bf The cosmological context:} Under the assumption of {\em homogeneity
and isotropy}, the space-time metric is described in terms of a single
dynamical variable -- the scale factor. As long as the energy density is
positive and the pressure is not too negative (which is true for the
properties of known matter), in an expanding universe (an observational
fact), the scale factor vanishes at a finite time in the past. The
universe has `beginning', a finite age and the space-time curvature (or
gravity) is infinitely large. Thus, a homogeneous, isotropic universe
has {\em singular beginning}. If one relaxes isotropy but retains
homogeneity, one has several {\em types} of space-times. There are now a
maximum of {\em six} dynamical variables. The simplest of these, the
vacuum Bianchi I space-time, has three `scale factors' whose time
dependence is given by the (exact) Kasner solution. This is also
singular. As one evolves back in time, two of the scale factors vanish
while the third one diverges.  The most complex of these models, the
vacuum Bianchi IX space, is also singular and the backward evolution has
an {\em oscillatory} behaviour.  Like the Kasner solution, two scale
factors begin decreasing and the third one increasing. But after a
while, the three scale factors change their behaviour and a different
pair begins decreasing. This continues ad infinitum. If the {\em
non-diagonal} metric components are included, then the directions along
which contraction/expansion takes place are also `rotated'. If one
relaxes homogeneity as well, then a beautiful analysis done by
Belinskii-Khalatnikov-Lifschitz (BKL), shows that there exists a {\em
general solution} of the vacuum Einstein equations which can be
described as smaller and smaller portions of the spatial slice behaving
as a homogeneous, Bianchi IX solution. The BKL analysis in particular
shows that singular solutions found in the simpler situations are {\em
not} due to high degree of symmetry (homogeneity and isotropy), but even
without such symmetries, there exist general solutions which are
singular (diverging curvatures) and the nature of singularities can be
extremely complicated. 

During the sixties Geroch-Penrose-Hawking used another approach to
establish the {\em Singularity theorems} identifying conditions under
which singularities are inevitable consequence of classical GR. For
these theorems, singular space-times were defined as those {\em
in-extendible space-times which admit at least one causal (time-like or
null) geodesic which is incomplete.} Here incompleteness means that the
geodesic cannot be defined for all real values of an affine parameter.
There are three types of inputs in these theorems: (a) One restricts to
a class of space-times which are {\em causally well-behaved} eg are free
from closed causal curves. The space-times which are free of all causal
pathologies and are fully deterministic are the so-called {\em globally
hyperbolic} space-times. (b) The space-times are solutions of Einstein
equation with the matter stress tensor satisfying suitable {\em energy
condition(s)}. This incorporates that idea that gravity is attractive
(for positive mass/energy). These two types of conditions are general
requirements for a space-time model to be physically relevant. (c) the
third input is a condition that distinguishes specific physical context
such as an everywhere {\em expanding universe} or a gravitational
collapse which has proceeded far enough to develop {\em trapped
surfaces}. The presence of the last condition(s) shows that not every
solution satisfying the first two conditions is singular (e.g.  the
Minkowski space-time). Thus, singularity theorems do {\em not} imply
that gravitational interactions {\em always} produce singularities --
the (c) type condition is necessary. While inclusion of (c) will imply
singular space-times, it is {\em not} automatic that this condition is
{\em realized} in the physical world. In our physical world however
universe is expanding and it is widely believed that black holes also
exist and hence condition (c) {\em is } realized in nature. Thus,
physical contexts exist wherein classical GR is inadequate.

\underline {Remark 1:} The global hyperbolicity condition implies that
the space-time is {\em stably causal} i.e. a global time function exists
such that each hyper-surface of constant `time' is space-like.
Furthermore, a time function can be chosen such that the space-like
hypersurfaces are {\em Cauchy} surfaces. The topology of such
space-times is necessarily $\mathbb{R} \times \Sigma$.  A Hamiltonian
formulation makes sense only in such space-times. Thus, {\em not every
solution} of Einstein equation yields a physically acceptable space-time
(i.e. causally well behaved or globally hyperbolic). A Hamiltonian
evolution however constructs such space-times \footnote{One could {\em
analytically extend} such maximally Cauchy evolved solutions further.
These will have Cauchy Horizons, an example being Reissner-Nordstrom
space-time.}.

\underline {Remark 2:} In specific solutions, one encounters
singularities (regions of diverging curvatures) which are {\em
space-like} (positive mass Schwarzschild solution, homogeneous
cosmologies), {\em time-like} (negative mass Schwarzschild solution) or
even {\em null} (some of the Weyl class of solutions). Singularities
that {\em arise} in an evolution from {\em non-singular} initial
conditions are the ones which strongly display inadequacy of the theory.
Typically, these are the space-like singularities. Since the Hamiltonian
formulation is an initial value formulation, it can ``see'' only such
singularities. 

{\bf The Black hole context:} Black holes are objects whose `interiors'
are inaccessible to far away observers. More precisely, these are
space-time geometries that have a {\em horizon} which leave some regions
out of bounds for asymptotic observers. The special class of {\em
stationary} black holes are characterized by a few parameters -- mass
($M$), angular momentum ($J$) and electric charge (say) ($Q$).
Associated with their horizons are some characteristic parameters --
area of the horizon ($A$), surface gravity at the horizon ($\kappa$),
angular velocity at the horizon ($\Omega$) and electromagnetic potential
at the horizon ($\Phi$). In the seventies, a remarkable set of ``laws''
governing processes involving black holes were discovered. If in a
process a black hole is disturbed (by accreting mass, say) and returns
to a stationary state again, then the changes in the parameters obey:
\begin{equation} 
\delta M ~ = ~ \frac{\kappa}{8 \pi} \delta A + \Omega_H \delta J +
\Phi_H \delta Q ~ ~ , ~ ~ \delta A \ge 0\ .  
\end{equation}
These are very temptingly analogous to the laws of thermodynamics!
Especially after one also proves that $\kappa$ is constant over the
surface of the horizon. Bekenstein in fact suggested that area of the
horizon be identified with the entropy of a thermodynamic system. This
suggests that the surface gravity be identified with a temperature.
Hawking subsequently showed that when possibilities of quantum
instabilities are taken into account, a black hole can be thought of a
black body with temperature $T = \Case{\kappa \lP^2}{2 \pi}$ and hence
$S = \Case{1}{4} \Case{A}{\lP^2}, \lP^2 := G\hbar$. For all other
systems we know that thermodynamics is a manifestation of an underlying
statistical mechanics of a large number of {\em microscopic} degrees of
freedom. What are these micro-constituents of the black holes? Notice
that from far away, only the exterior of a horizon is accessible and so
also parameters such as mass and angular momentum. All detailed memory
of what collapsed to form the black hole is lost. So these
micro-constituents must be distinct from the  matter degrees of freedom.
They must represent ``atoms'' of {\em geometry}. But classical geometry
is continuous so how does a particulate nature arise? Perhaps, not just
the specific dynamics given by Einstein equation is inadequate but the
{\em framework of classical geometry} itself is inadequate. Note that
black hole horizons are {\em not} regions of high curvatures and
geodesic incompleteness occurs in their interiors. Thus, black hole
thermodynamics is a {\em qualitatively different situation.}

In summary, classical GR contains within its domain, physically
realizable physical context where the theory is inadequate. At least one
of its context involves {\em highly dynamical} geometries with high
curvatures, matter densities etc. Because of these features, it is hard
to imagine how any {\em perturbative approach} can be developed in these
contexts. Since gravity (or space-time geometry) is dynamical and a
perturbative approach is unlikely to be suitable, it is necessary to
have a quantum theory of gravity which of {\em does not use} any fixed
background space-time in its {\em construction}.

\section {An Essential Feature of Classical Gravity} 
Let us recall briefly that special relativity combines Newtonian notions
of space and time into a single entity, the {\em space-time} (Minkowski
space-time). The analysis of the geometry inferred in a rotating frame
indicates that the geometry is non-Euclidean. Principle of equivalence,
which identifies uniform acceleration with uniform gravity (in the
Newtonian sense), then implies that gravity affects the space-time
geometry and since matter affects gravity, it also affects geometry. In
the final formulation of the relativistic theory of gravity, the
space-time geometry, described by a metric tensor, is a dynamical
(changeable) entity with the Newtonian gravity being a manifestation of
the curvature. The law governing the matter-geometry interaction is
encoded in the Einstein equation. That all observers are on equal
footing to formulate the laws of physics implies that all quantities
(and equations) be tensor fields (and equations) with respect to {\em
general coordinate transformations.} Note that a general coordinate
transformation corresponds to a {\em change of chart} in the framework
of differentiable manifolds. 

Such coordinate transformations however have another interpretation in
terms of mapping of the manifold (or local regions thereof) into itself
-- the {\em active diffeomorphisms}. Under the action of such mappings,
the pull-backs and push-forwards, generate ``new'' tensor fields from
the old ones. That is, in a given neighbourhood, we will have the
original tensor field and the one obtained via pull-back/push-forward.
If $x \to y(x)$ represents the mapping in terms of local coordinates,
then the pulled-back (pushed-forward) quantities are related to the
original ones in precisely the same manner as general coordinate
transformation \footnote{Let $\phi: p \to \phi(p)$ be a smooth map of a
manifold onto itself. Given a function $f':M \to \mathbb{R}$, define
another function $f:M \to \mathbb{R}$ as: $f(p) := f'(\phi(p))$. This
function is the {\em pull-back} of $f'$ and is also denoted as $f :=
\phi^*(f')$. Likewise, given a vector field $X$ on $M$ define a new
vector field $X'$ as: $X'(f')|_{\phi(p)} := X(f)|_{p}$. The new vector
field is called the {\em push-forward} of $X$ and is also denoted as:
$X' = \phi_*(X)$. Now introduce local coordinates $x^i$ around $p$ and
$y^i$ around $\phi(p)$.  It is easy to see that the components of
tensors relative to these coordinates are related in exactly the same
manner as though $x \to y(x)$ is a change of chart.}.  If the dynamical
equations are covariant with respect to general coordinate
transformations (coordinate transform of a solution is also a solution),
they must also be covariant with respect to the active diffeomorphisms
i.e. a configuration and its transform under active diffeomorphism are
both solutions if any one of them is.  This has far reaching
implications. 

{\em The Einstein Hole Argument:} The active diffeos can be chosen such
that they map non-trivially only in some region (`sub-manifold') of the
manifold. Choose a region which is free of any matter. Assume that the
equations determining gravitational field and matter distribution are
also tensor equations (i.e. generally covariant). Consider a solution
which has certain curvature distribution inside our chosen `hole'. Make
a diffeo which is non-trivial only inside the hole and change the
curvature distribution. This will also be a solution by covariance.
Thus we get a situation that even though matter distribution is
unchanged, in a region where there is no matter, we can have two
`different' gravitational fields i.e. matter distribution does {\em not}
determine the gravitational field. But in the non-relativistic limit
Newtonian gravity is determined by matter. So {\em either} the equations
should {\em not} be covariant {\em or} {\em in the absence of any matter
available for `marking' points of a manifold, the `different'
distributions of curvature must be regarded as describing the `same'
gravitational field}. It is the latter possibility that remains once the
{\em covariance of the equations} is accepted. This in turn implies that
{\em it is the equivalence classes of solutions, with respect to the
space-time diffeomorphisms, that correspond to physical reality.} 

{\em Note}: The size of the hole in the hole argument is unimportant.
Also, the metric description itself does {\em not} play a role; one
could repeat the argument for any other field. All that is used is that
the fields are tensors under general coordinate transformations (chart
change), field equations are covariant and the fields are inhomogeneous
within a hole. Values of individual fields at any, manifold point are
irrelevant but values of fields at points {\em specified by other
fields} are invariant and thus physical. 

Since `points', not marked by any {\em dynamical entity}, have no
physical meaning, the only, physically meaningful, questions are of
relational nature.  That is, it is physically meaningless to ask what is
the curvature (or say electric field) ``here and now''. The meaningful
questions is what is the curvature where a certain field has a certain
value. If we had {\em any} particular field to be fixed (non-dynamical),
then with reference to that field we could ask the `here and now'
question. Such a field, constitutes a background. Note that the usual
non-gravitational theories or in the perturbative treatment of gravity,
the space-time geometry (metric) plays the role of a background. Since
in the general relativistic theory {\em all fields including the metric}
are fundamentally dynamical, such a theory is {\em necessarily
background free.} The twin features of the framework namely {\em all}
fields on a manifold being dynamical and the fundamental equation being
{\em generally covariant and deterministic}, implies covariance with
respect to active diffeomorphisms and physical characterization being in
terms of 4-diffeo equivalence classes of fields \footnote{Since we have
taken (differential) equations as specifying a presentation of a theory,
the manifold {\em cannot} be thought of as a background, but rather part
of the specification of the theory. If we take some transition
amplitudes (among topological spaces, sets, ...) as specifying a theory,
then the choice of a particular differential structure {\em will}
constitute a background since it is also a dynamical entity. We will
restrict to manifold category. This also explains why metric by itself
is not essential for background independence of gravity, a dynamical
tetrad with compatible spin connection would do just as well.}.

This is an essential feature of general relativity, much more
fundamental than the particular Einstein equations themselves. The
challenge is to construct a quantum theory which faithfully incorporates
this feature i.e. a quantum theory of gravity must be background
free (or at least recover background independence in the
classical limit).

This also poses challenges, because we have to construct observables
which are space-time diffeomorphism invariant. These alone could
characterise specific equivalence classes of space-times and this
problem is not understood even classically for spatially compact case
and in absence of matter! Note that curvature invariants, although
scalars, are local and {\em not} diffeo invariants.  Hence these cannot
be physical observables. Consequently identifying a physical state
corresponding to (say) Minkowski space-time is much harder.

For a more detailed discussion of these conceptual issues, see
\cite{Rovelli}.
\section {Towards the construction of a quantum theory}
Ultimately constructing a quantum theory of some phenomenon means
specifying a state space - (projective) Hilbert space, identifying
(self-adjoint) operators on it to correspond to physically observable
quantities, a notion of evolution or dynamics such that in a suitable
semi-classical approximation, the evolution of expectation values of a
class of observables tracks the corresponding classical evolution with
the quantum uncertainties less than the observational precision. Here
the classical evolution is the one specified by the classical
description of the phenomenon. Using such a framework, one can compute
matrix elements of suitable observables or transition amplitudes etc.
One familiar procedure is that of the {\em canonical quantization}.

In canonical quantization, typically we have a classical phase space
which is cotangent bundle, ${\cal T}^*Q$ of some configuration manifold
$Q$ and the Hilbert space is the space complex valued functions of $Q$
which are square integrable with respect to some suitable measure,
$d\mu$. When $Q \sim \mathbb{R}^N$, we have the familiar
$L_2(\mathbb{R}^N, d\mu_{\mathrm{Lebesgue}})$ which is unique thanks to
the Stone-von Neumann theorem.  This comfortable situation changes once
$Q$ becomes topologically non-trivial and/or becomes infinite
dimensional. The former can arise due to constraints while the latter
arises in a field theory. In relativistic field theory, the {\em
classical configuration space}, $Q$, is (say) the space of suitably
smooth tensor fields which is inadequate to describe the corresponding
quantum fields which can be arbitrarily non-smooth. Typically, $Q$ is
{\em extended} to a {\em quantum configuration space}, $\bar{Q}$, which
should admit a suitable measure. 

For a quantum theory of gravity, there are two additional features - (i)
we have a theory with first class constraints (i.e. a gauge theory) and
(ii) we would like to have background independence. 

In presence of constraints, the quantization procedure is a two step
process. In the first step one constructs a {\em kinematical Hilbert
space} on which are defined the constraint operators. The second step
aims to `solve' the constraints to get a quantum theory corresponding to
physical degrees of freedom. Again, {\em typically}, there are no
vectors in the kinematical Hilbert space which are annihilated by the
constraint operators and one is forced to consider {\em distributional
solutions}\footnote{Let $\Omega \subset H_{\mathrm{kin}}$ be a dense
subspace of the kinematical Hilbert space. Let $\Omega^*$ denote its
algebraic dual (space of linear functions on $\Omega$) so that $\Omega
\subset H_{\mathrm{kin}} \subset \Omega^*$. $\Omega$ is chosen so that
it contains the domains of the constraint operators as well as of other
operators of interest.  Distributional solutions of constraints are
those elements of $\Omega^*$ which evaluate to zero on all elements of
$\Omega$ of the form $\hat{C}|\psi\rangle, \forall\ \psi \in \Omega$.}.
The space of distributional solution however is {\em not} a Hilbert
space and another inner product needs to be defined on this space to
make it into the {\em physical Hilbert space}. The choice of this inner
product is limited by demanding that a suitable class of {\em Dirac
observables} -- operators which leave the space of solutions invariant
-- be self-adjoint.

There are many choices to be made along the way. The requirement of
background independence means that no non-dynamical fields should be
used in any step. This poses a severe challenge to constructing even the
kinematical Hilbert space. The connection formulation of gravity is of
great help as the quantum configuration space of a gauge theory,
$\overline{{\cal A}/{\cal G}}$ - space of generalized connections modulo
generalized gauge transformations - admits several measures and the
demand that the conjugate variables be represented by derivative
operators essentially singles out a unique measure - the
Ashtekar-Lewandowski measure, essentially constructed from the Haar
measure on compact groups. One has a natural choice of $\Omega$ and
(non-unique) definitions of constraint operators so that the kinematical
set-up is well founded. We will see the details of these steps.

We will begin with the Hamiltonian formulation of GR in terms of the
metric (ADM formulation).  Discover the redundant variables and make a
canonical transformation to a set of new variables (connection
formulation) which are amenable to background independent presentation.
These will lead us to the holonomy-flux variables and their Poisson
bracket algebra whose representation theory will give us a Hilbert
space.  This will complete the {\em first step} in the construction of
the quantum theory. Some of its novel features will be revealed through
the properties of the geometrical operators. Our study of the basic
formalism will conclude with the presentation of the constraint
operators on the kinematical Hilbert space. The dynamical aspects will
be studied through the simpler cases of homogeneous and isotropic
cosmology leading to the Big Bang singularity resolution. The other
application of quantum geometry, namely revealing the `atoms' of
geometry responsible for the black entropy will be discussed in the
second week along with the loop quantum cosmology. 
\chapter {Classical Hamiltonian Formulation}
We are familiar from usual spacial relativistic field theories (say a
scalar field) that solutions of the field equations can be viewed as an
evolution of fields, their spatial derivatives and their velocities from
one spatial slice to another one (``a Cauchy evolution''). In the
general relativistic case, one has to deal with space-times other than
the Minkowski space-time and eventually make the space-time itself to be
`dynamical'. However {\em not all space-times support this notion of
evolution.} 

To have a well defined, causal (no propagation faster than speed of
light in vacuum) and deterministic (given certain data at one instance,
the future data is {\em uniquely} determined), the space-time must be
free of {\em causal pathologies} such as (i) no closed causal (i.e.
time-like or null) curves (excludes by {\em chronology condition}); (ii)
no closed causal curves but causal curves which return arbitrarily close
to themselves (excluded by {\em strong causality}); (iii) strong
causality holds but when the space-time metric is made slightly `wider',
it is violated (excluded by {\em stable causality}).  All such
pathologies are absent in space-time which are {\em stably causal} i.e.
admit a differentiable function such that $\partial^{\mu} f$ is a
time-like vector field. This alone is still not sufficient to guarantee
the possibility of a {\em Cauchy Problem}. For this, one needs {\em
Globally Hyperbolic Space-times}. These are space-times which are which
admit a spatial hyper-surface such that events to the future (past) are
completely determined by data specified on it. Such space-time admit a
globally defined `time function' whose equal time surfaces are Cauchy
surfaces. It follows further that such space-times must have the
topology: ${\cal M} = \mathbb{R} \times \Sigma_3$. 

Thus, in order to have a well-posed causal theory of {\em matter
fields}, the space-times must be globally hyperbolic.

It is a non-trivial result of analysis of Einstein equation that
Einstein equation can be cast in a Hamiltonian form such that if initial
conditions are chosen to satisfy certain {\em constraints}, then
corresponding Hamiltonian evolution generates a solution (space-time) of
the Einstein equation\footnote{This is local existence and uniqueness
theorem for the Einstein equation. Since these are short time
evolutions, one cannot guarantee that largest possible space-time
constructed will be globally hyperbolic. However, if a globally
hyperbolic solution is to exist, one can perform a time + space
decomposition to put the equations in a Hamiltonian form. That Einstein
equation admit a well-posed initial value problem is a necessary
condition for globally hyperbolic solutions. That the equations are of
Hamiltonian form is an additional, non-trivial property. This follows
most directly via the Einstein-Hilbert action formulation.}. 

Although a Hamiltonian formulations can be specified ab initio by giving
a {\em phase space} which is a {\em symplectic manifold}, a {\em
Hamiltonian} function and specifying evolution by the Hamilton's
equations of motion, it is more common that a theory is specified in
terms an action which is a functional defined on a set of fields on a
space-time manifold. Typically this is expressed as an integral of a
Lagrangian density made up of finite order derivatives of a set of
tensor (spinor) fields. In such a manifestly space-time covariant
presentation, one needs to choose a ``time'' direction along which to
`evolve' data specified on a `equal time surface'. These data identify
the {\em configuration space variables and their velocities}. The
Hamiltonian formulations is then obtained from this {\em Lagrangian
formulation} by passing through a Legendre transform. This
identification of a time direction and a spatial slice on which data are
to be specified is referred to as a ``3 + 1 decomposition''.
\section{The 3 + 1 decomposition}
Let us assume that our would be space-time manifold is such as to admit
a smooth function $T: {\cal M} \to \mathbb{R}$ such that the $T =
constant$ level sets, generate a foliation. Different possible
$T$-functions will generate different foliations. For this to be
possible, we must have ${\cal M} \sim \mathbb{R} \times \Sigma_3$.

Now choose a vector field $t^{\mu}\partial_{\mu}$ which is {\em
transversal} to the foliation i.e. every integral curve of the vector
field intersects each of the leaves, transversally. Furthermore, locally
in the parameter of the curve, the leaves are intersected once and only
once. {\em Normalize} the vector field so that $t^{\mu} \partial_{\mu} T
= 1$. This ensures that values of the $T-$functions can be taken as a
``time'' parameter which we denote as $t$.

Fix a leaf $\Sigma_{t_0}$ and introduce coordinates, $x^a, a = 1, 2, 3$
on it. Carry these along the integral curves of the vector fields, to
the other leaves. This sets up a local coordinate system on ${\cal M}$
such that the normalized parametrization provides the coordinate $t$
while the integral curves themselves are labelled by the $\{x^a\}$. Note
that there is no metric so ${\cal M}$ is not yet a space-time. We have
only set up a coordinate system.

Choose tensors $g_{ab}, N^a, N$ on each of the leaves in a smooth manner
ad {\em define} a space-time metric via the line element:
\begin{equation}
ds^2 ~:=~ -N^2 dt^2 + \bar{g}_{ab}\left(dx^a + N^a dt\right)\left(dx^b +
N^b dt\right).
\end{equation}
Choosing $\bar{g}_{ab}$ to be {\em positive definite} and $N \ne 0$
ensures that the space-time metric $g_{\mu\nu}$ is {\em invertible}. Its
inverse is given by,
\begin{equation}
g^{tt} ~=~ -N^{-2} ~~,~~ g^{tb} ~=~ N^b N^{-2} ~~,~~ g^{ab} ~=~
\bar{g}^{ab} - N^{-2} N^a N^b ~~,~~ \bar{g}^{ac}\bar{g}_{cb} =
\delta^a_b ~ .
\end{equation} 
We now have a space-time.  The space-time metric is defined in terms of
10 independent functions and so there is no loss of generality. It is a
convenient parametrization for reasons given below, but alternative
parametrization are possible.

It follows that,
(i) The induced metric on the leaves is the Riemannian metric
$\bar{g}_{ab}$.

(ii) $n_{\mu} := \partial_{\mu} T$ is normal to the leaves, since for
any tangent vector $X^{\mu}\partial_{\mu}$, to $\Sigma_t$, $X^{\mu}
n_{\mu} = X^{\mu} \partial_{\mu} T = 0$. Thanks to the normalization of
$t^{\mu}\partial_{\mu}$, we have $n_{\mu} = (1, 0, 0, 0)$.

(iii) $n^{\mu} := g^{\mu\nu} n_{\nu} \Rightarrow n^{\mu} n_{\mu} =
g^{tt} = - N^{2} < 0$ and therefore the normal is {\em time-like} and
hence the leaves are {\em space-like}. The $N n^{\mu}$ is a unit
time-like vector.

(iv) The original transversal vector field can be decomposed as $t^{\mu}
= a n^{\mu} + \tilde{N}^{\mu}$ where $\tilde{N}^{\mu} n_{\mu} = 0$ and
hence $\tilde{N}^{\mu}$ is tangential to the leaves and $\tilde{N}^0 =
0$. This decomposition refers to $N^2$ as the {\em lapse} function and
$\tilde{N}^{\mu}$ as the {\em shift vector}. Next, $t^{\mu}n_{\mu} = 1
\Rightarrow a = - N^2$. The integral curve equation, $d_t x^{\mu} = -
N^2 n^{\mu} + \tilde{N}^{\mu}$ implies for $\mu = a$, $\tilde{N}^a = N^2
n^a = N^2 g^{at} = N^2 ( N^{-2} N^a ) = N^a$. This identifies the $N^a$
with the shift vector (which is spatial).

The particular parametrization of the space-time metric can be said to
be {\em adapted} to the pre-selected coordinate system. Since the
coordinate system is defined {\em without} any reference to any metric,
we can similarly parametrize other tensor fields, notably the {\em
co-tetrad}, $e_{\mu}^I$.

\underline{Note:} We chose an arbitrary foliation (through an arbitrary
choice of a ``Time'' function and then a transversal vector field to
enable us to choose coordinate on the manifold. The foliation provides
us with a normal $n_{\mu}$ and the transversal vector field can be
parametrized in terms of this normal, a lapse function and a shift
vector. Varying the lapse and shift varies the transversal vector field
{\em relative to the foliation}. If we also change the foliation, then
the normal changes and so must the shift vector. The changes induced by
lapse and shift correspond to making a space-time diffeomorphism and {\em
every infinitesimal space-time diffeomorphism can be generated by
infinitesimal changes in the lapse and shift.}

\section{Digression on tetrad formulation}
General relativity is formulated as theory consisting of tensorial
fields on a manifold and a second rank, symmetric, non-degenerate
(invertible) tensor field, $g_{\mu\nu}$ encoding gravitational
phenomena.  To do differential calculus on general tensor fields one
also needs to define a covariant derivative, $\nabla_{\mu}$ which
involves the introduction of an affine connection, $\Gamma^{\lambda}\,
_{\mu\nu}$, which is usually taken to symmetric and {\em metric
compatible} i.e.  $\nabla_{\lambda} g_{\mu\nu} = 0$. The non-abelian
gauge theories already introduce quantities which are not just tensors
with respect to general coordinate transformations but also transform
under the action of ``an internal'' group, eg a Higgs field $\Phi^a$, a
YM potential $A_{\mu}^a$, its corresponding field strength,
$F^a_{\mu\nu}$ etc. The index $a$ indicates a response to the (adjoint)
action of a group such as $SU(N)$.  Developing calculus for such
quantities, also needs a {\em gauge} covariant derivative and a
corresponding {\em gauge} connection eg $A_{\mu}^a$.

Consider now a quantity, $e^I_{\mu}(x)$ where $\mu$ responds to a
general coordinate transformation ($e^I_{\mu}$ transform as a covariant
rank 1 tensor) and the index $I$ responds to the {\em local} action of
the pseudo-orthogonal group, $SO(1,3)$ under the defining
representation. This quantity can also be thought of as a $4 \times 4$
matrix and we will take it to be an invertible matrix. This is referred
to as a {\em co-tetrad} while its inverse quantity, $e_I^{\mu}$ is
referred to as a {\em tetrad}: $e^I_{\mu} e_J^{\mu} = \delta^I_J,
e^I_{\mu} e_I^{\nu} = \delta^{\nu}_{\mu}$. It is possible to formulate
the theory of gravity in terms of a (co-)tetrad as follows.

{\bf(1)} Let $\Lambda^I_J \in SO(1,3)$ i.e. $\Lambda^I_K \Lambda^J_L
\eta^{KL} = \eta^{IJ}$ holds where $\eta^{IJ} = \mathrm{diag}(-1, 1, 1,
1)$. Then the co-tetrad transforms as:
\begin{equation}
(e')^{I}_{\mu}(x'(x)) ~ := ~ \Lambda^I_J(x) \frac{\partial
x'^{\nu}}{\partial x^{\mu}} e^J_{\nu}(x) \ .
\end{equation}
The $\Lambda-$ transformation is referred to as a {\em Local Lorentz
Transformation} (LLT) while $x \to x'(x)$ is the {\em General Coordinate
Transformation} (GCT). Clearly to have the derivatives of the co-tetrad
to transform covariantly under both sets of transformations, we need two
connections: an {\em affine connection} (not necessarily symmetric in
the lower indices) and a {\em Spin connection, $\omega_{\mu}\, ^{IJ}$}.
The spin connection is anti-symmetric in the $IJ$ indices and thus
transforms as the {\em adjoint representation} of the pseudo-orthogonal
group. The derivative covariant with respect to the LLT is denoted by
$D_{\mu}$, that with respect to GCT is denoted by the usual
$\nabla_{\mu}$ while the one with respect to both will be denoted by
${\cal D}_{\mu}$. The Lorentz indices will be raised/lowered using the
{\em Lorentz metric} $\eta^{IJ}, \eta_{IJ}, \mathrm{where} ~
\eta^{IK}\eta_{KJ} = \delta^I_J$.

Armed with the tetrad, the spin connection and the Lorentz metric, {\em
define} the following quantities:

$ 
\begin{array}{|l|ccl|}
\hline
& & & \nonumber \\
\mathrm{Torsion:} & T^I(e, \omega) & := & d e^I + \omega^I\, _J \wedge
e^J \\
& T^I\, _{\mu\nu} & = &\partial_{\mu} e^I_{\nu} + \omega_{\mu}\, ^I\,_J
e^J_{\nu} - ({\mu \leftrightarrow \nu}) \nonumber \\
& & & \nonumber \\
\hline
& & & \nonumber \\
\mathrm{Curvature:} & R^{IJ}(\omega) & := & d \omega^{IJ} + \omega^I\,
_K \wedge \omega^{K J} \\
& R^{IJ}~_{\mu\nu} & = & \partial_{\mu}\omega_{\nu}\, ^{IJ} +
\omega_{\mu}\, ^{I}\,_{K} \omega_{\nu}\, ^{KJ} - ({\mu \leftrightarrow
\nu}) \nonumber \\
& & & \nonumber \\
\hline
& & & \nonumber \\
\mathrm{Bianchi~Identity:} & (D R)^I\,_J & = & 0 ~:=~ dR^I\,_J +
\omega^I\,_K\wedge R^K\,_J + \omega_J\,^K\wedge R^I\,_K \\
& & & \nonumber \\
\mathrm{Cyclic~Identity:} & (D T)^I & = & R^I\,_J\wedge e^J ~~:=~~d T^I
+ \omega^I\,_J\wedge T^J \\
& & & \nonumber \\
\hline \hline
& & & \nonumber \\
\mathrm{Metric:} & g_{\mu\nu}(e, \eta) & := & e_{\mu}^I e_{\nu}^J
\eta_{IJ} \\
& & & \nonumber \\
\hline
& & & \nonumber \\
\mathrm{Christoffel\,Connection:} & \left\{\begin{array}{c}
\lambda\\{\mu \nu}\end{array}\right\}(g(e)) & := & \frac{1}{2}
g^{\lambda\alpha}\left( g_{\alpha\mu, \nu} + g_{\alpha\nu, \mu} -
g_{\mu\nu, \alpha}\right) \\
& & & \nonumber \\
\hline
& & & \nonumber \\
\mathrm{Affine~connection:} & \Gamma^{\lambda}\, _{\mu\nu} (e, \omega) &
:= & \left\{\begin{array}{c} \lambda\\{\mu \nu}\end{array}\right\}(g(e))
+ \frac{1}{2} g^{\lambda\alpha}\left(T_{\alpha\mu\nu} - T_{\mu\nu\alpha}
- T_{\nu\mu\alpha}\right) \\
\hspace{2.2cm}\mathrm{where,} & {\bf T_{\alpha\mu\nu}(e, \omega)} & := &
{\bf e_{I\alpha} T^I\, _{\mu\nu}(e, \omega) }\nonumber \\
& & & \nonumber \\
\hline \hline
& & & \nonumber \\
\mathrm{These~imply} &  & & \nonumber \\
& & & \nonumber \\
\mathrm{Matric~compatibility:} & \nabla_{\lambda} g_{\mu\nu}  & = & 0 \\
\mathrm{tetrad~compatibility:} & {\cal D}_{\mu} e^I_{\nu} & = & 0 ~~ :=
~~ \nabla_{\mu} e^I_{\nu} + \omega_{\mu}\,^I\,_J e^J_{\nu} -
\Gamma^{\lambda}\,_{\mu\nu} e^I_{\lambda} \\ 
& & & \nonumber \\
\hline
& & & \nonumber \\
\mathrm{Compatibility} ~~ \Rightarrow &
\hspace{1.0cm}R^{\alpha}\,_{\lambda\mu\nu}(\Gamma) & = & e^{\alpha}_I
e_{\lambda\, J} R^{IJ}\,_{\mu\nu}(\omega) \\
& & & \nonumber \\
\hline
\end{array}
$

No assumption about the torsion tensor is made.

{\bf(2)} It is possible to invert the torsion equation to `solve for'
the spin connection in terms of the tetrad, its derivatives and the
torsion tensor. All one needs to do is manipulate the combination
$T_{\lambda\mu\nu} + T_{\mu\nu\lambda} - T_{\nu\lambda\mu}$ and use the
invertibility of the (co-)tetrad. The result is:
\begin{eqnarray}
\omega_{\mu}\,^{IJ} & := & \hat{\omega}_{\mu}\,^{IJ}(e) +
K_{\mu}\,^{IJ}(e, T) \\
\hat{\omega}_{\mu}\,^{IJ} & := & \frac{1}{2}\left[ e^{\nu
I}\left(\partial_{\mu} e_{\nu}^J - \partial_{\nu} e_{\mu}^J\right)
- e^{\nu J}\left(\partial_{\mu} e_{\nu}^I - \partial_{\nu}
  e_{\mu}^I\right)
- e^{\nu I} e^{\lambda J} \left(\partial_{\nu}e^K_{\lambda} -
  \partial_{\lambda}e^K_{\nu} \right) e_{\mu K} \right] \\
K_{\mu}\,^{IJ} & := & - \frac{1}{2} e^{\nu I} e^{\lambda J} \left(
T_{\nu\lambda\mu} + T_{\lambda\mu\nu} - T_{\mu\nu\lambda} \right)
\end{eqnarray}
The $K$ is called the {\em con-torsion tensor} and $\hat{\omega}$ is the
{\em torsion-free spin connection} which is explicitly determined by the
tetrad. The Affine connection equation is the corresponding inversion of
the metric compatibility condition (covariant constancy of the metric)
to express the general affine connection in terms of the torsion free
Christoffel connection plus the torsion combinations.

Notice that a priori, we have two connections: the affine and the spin.
Both define corresponding and independent torsions ($T^I\,_{\mu\nu}$ and
the antisymmetric part of $\Gamma^{\lambda}\,_{\mu\nu}$). The
introduction of the metric as the `square' of the co-tetrad and the two
compatibility conditions together identify the two torsions. 

The last equation demonstrates that we can use the tetrad and the
co-tetrad to convert the Lorentz indices and the general tensor indices
into each other with the compatibility conditions ensuring the two
distinct curvatures also going into each other.

{\em Although we have referred to only 4 dimensions and Lorentz
signature metric, the definitions generalise to any dimensions and any
signature.}

{\bf(3)} Four dimensions have additional features available. One can
define {\em internal dual (Lorentz dual)} for anti-symmetric rank-2
Lorentz tensors apart from the usual {\em Hodge dual (space-time dual)}
for 2-forms.

Let, ${\cal E}^{IJKL}$ and ${\cal E}_{\mu\nu\alpha\beta}$ denote the
Levi-Civita symbols; These are completely antisymmetric in their indices
and we choose the conventions: ${\cal E}^{0123} = 1 = {\cal E}_{txyz}$.
The indices on these are raised and lowered by the Lorentz and the
space-time metric respectively. Using these we define:
\begin{eqnarray}
\tilde{X}^{IJ} & := & \frac{1}{2} {\cal E}^{IJ}\,_{KL} X^{KL}
\hspace{1.0cm} (\mathrm{Internal~Dual}) \\
(*X)_{\mu\nu} & := & \frac{1}{2} {\cal E}_{\mu\nu}\,^{\alpha\beta}
X_{\alpha\beta} \hspace{1.3cm} (\mathrm{Hodge~Dual}) 
\end{eqnarray}
{\bf(4)} From the tetrad and the spin connection, the following Local
Lorentz invariant four forms can be constructed whose integrals are
candidate terms for an action.
\begin{enumerate}
\item {\em Hilbert-Palatini:} 
${\cal L}_{HP}(e, \omega) ~:=~ \frac{1}{2}{\cal E}_{IJKL}
R^{IJ}(\omega)\wedge e^K\wedge e^L$. 

The variational equations following from this are equivalent to the
Einstein equations. The spin-connection equation implies that the
torsion vanishes and the tetrad equation implies the vanishing of the
Ricci tensor. The Hilbert-Palatini action is thus classically equivalent
to the Einstein-Hilbert action of the metric formulation. 

This term taken as an action with the tetrad and spin connection treated
as independent variables is sometimes referred to as the tetrad
formulation of gravity.
\item {\em Cosmological Constant:} ${\cal L}_{\Lambda}(e) ~ := ~
\frac{\Lambda}{4!}{\cal E}_{IJKL} e^I\wedge e^J\wedge e^K\wedge e^L$

This is the usual cosmological constant term, proportional to the volume
form.
\item {\em Euler Invariant:}
${\cal L}_{E}(\omega) ~ := ~ \frac{1}{2} {\cal E}_{IJKL} R^{IJ} \wedge
R^{KL}$. 

This 4-form is a {\em topological term} i.e. its variation under
arbitrary infinitesimal changes in the spin connection, is an exact form
and therefore the variation of its integral receives contributions only
from the {\em boundary values}. Furthermore, explicitly,
\begin{equation}
{\cal L}_{E}(\omega) ~ = ~ - d\left\{ \frac{1}{2}{\cal E}^I\,_{JMN}
\omega^{MN}\wedge\left(d \omega^J\,_I + \frac{2}{3}
\omega^J\,_K\wedge\omega^K\,_I\right) \right\}
\end{equation}
\item {\em Pontryagin Invariant:}
${\cal L}_{P}(\omega) ~ := ~ R^{IJ} \wedge R^{IJ}$. 

This 4-form is also a {\em topological term}.  Furthermore, explicitly,
\begin{equation}
{\cal L}_{P}(\omega) ~ = ~ - d\left\{ \omega^I\,_J\wedge\left(d
\omega^J\,_I + \frac{2}{3} \omega^J\,_K\wedge\omega^K\,_I\right)
\right\}
\end{equation} The terms enclosed within the braces is the {\em
Chern-Simmons} 3-form.
\item {\em Nieh-Yan Invariant:}
${\cal L}_{NY}(e, \omega) ~ := ~ T^I\wedge T_I - R^{IJ} \wedge e^I\wedge
e^J$. 

This 4-form is also a {\em topological term} which depends on both the
tetrad and the spin connection. It vanishes if torsion is zero (for zero
torsion, the second term vanishes by the cyclic identity.) Explicitly,
\begin{equation}
{\cal L}_{NY} ~ = ~ d\left\{ e_I\wedge T^I \right\}.
\end{equation} 
\end{enumerate}

Note that we have 5 different, Lorentz covariant 2-forms: $T^I,
\Sigma^{IJ} := e^I\wedge e^J, \tilde{\Sigma}^{IJ}, R^{IJ},
\tilde{R}^{IJ}$. From these, we can form the six Lorentz invariants:
$T^2, \Sigma^2 (= 0 = - \tilde{\Sigma}^2), \Sigma\wedge\tilde{\Sigma},
R^2 (= - \tilde{R}^2), R\wedge\tilde{R}, R\wedge\Sigma,
R\wedge\tilde{\Sigma}$.  If we are to get the Einstein equation (with a
cosmological constant), then the $T^2$ and $R\wedge\Sigma$ must be
combined into the Nieh-Yan combination,

{\bf(5)} We will note a parametrization of the tetrad, adapted to the
3+1 decomposition, which leads to the corresponding metric
decomposition. This can be derived from the identifications:
\begin{eqnarray}
e^I_t e^J_t \eta_{IJ} & := & - N^2 + \bar{g}_{ab} N^a N^b \nonumber \\
e^I_t e^J_a \eta_{IJ} & := & \bar{g}_{ab} N^b \\
e^I_a e^J_b \eta_{IJ} & := & \bar{g}_{ab} \nonumber
\end{eqnarray}
It follows,
\begin{center}
$
\begin{array}{|ccl|l|}
\hline
& & & \\
\mathrm{Co-tetrad:} & & & ~~\mathrm{Introduces}~n_I \\
& & & \\
e^I_t & := & N n^I + N^a V^I_a ~~&~~ n^I n^J \eta_{IJ} := -1~~,~~ n^I V^J_a
\eta_{IJ} = 0 \\
& & & \\
e^I_a & := & V^I_a ~~~(\mathrm{free}) & \\
& & & \\
\bar{g}_{ab} & := & V^I_a V^J_b \eta_{IJ} ~~&~~ \mathrm{is~invertible;} \\
& & & \\
\hline
& & & \\
\mathrm{Tetrad:} & & & ~~\mathrm{Defines}~~V_I^a \\
& & & \\
e_I^t & := & - N^{-1} n_I ~~&~~ n^I V^a_I := 0 \\
& & & \\
e_I^a & := & N^{-1} n_I N^a + V_I^a  & ~~ V^a_I V^I_b ~=~ \delta^a_b ~~,~~
V^a_I V^J_a ~=~ \delta_I^J + n_I n^J \\
& & & \\
\hline
\end{array}
$
\end{center}
In this parametrization, the 16 variables in the tetrad have been traded
with $V_a^I (12), N, N^a (4)$ and $n^I (4)$ variables with 4 conditions:
$n^2 = -1, n\cdot V_a = 0$. The conditions can be viewed as 4 conditions
on $n^I$ given freely chosen $V^I_a$ {\em or} one condition on $n^I$ and
3 conditions on $V^I_a$ given freely chosen spatial vector $n^i$.

Notice that we have the {\em normalized normal}, $Nn_{\mu}$ defined by
the foliation. From this we can construct an internal vector
$\tilde{n}^I := e_I^{\mu}(Nn_{\mu})$. In the parametrization, we have
also introduced an internal normalized time-like vector $n^I$,
determined by the freely chosen $V^I_a$. These two are related by the
parametrization of the tetrad as, $\tilde{n}^I = - n^I$.

We can view $n_{\mu}$ defined by the foliation and $n_I$ defined by a
choice of $V^I_a$ as two time-like normalized vectors in the $T^*({\cal
M})$. These are not identical in general and in particular $n_I$ is not
normal to the foliation. Demanding it to be so, puts a restriction on
the $V^I_a$: $n_I \propto n_{\mu} ~\Rightarrow~ n_i = 0, ~ n_0n^0 = -1$
and $n\cdot V_a = 0 ~\Rightarrow V^0_a = 0$. This implies that $V^I_a$
are confined to $T^*(\Sigma)$. This choice is the so-called {\em time
gauge}.

Finally, we reiterate that using the tetrad and co-tetrad we can freely
convert the Lorentz and the general coordinate indices into each other.
The normalized normal ($Nn_{\mu}$) can be used to define a {\em
projector}, $P^{\mu}_{\nu} := \delta^{\mu}_{\nu} + N^2 n^{\mu} n_{\nu}$
which projects space-time tensors onto {\em spatial tensors}. This would
lead to 3+1 decomposition (or parametrization) of all other tensorial
quantities.

\chapter {Symmetry Reduction} 
There are different uses of the term `symmetry reduction'.
Heuristically, if $S$ is a state space of a system, on which is
specified an action of a group, $G$, which preserves the defining
specification of the system (so that $G$ is its {\em symmetry group}),
then the space $S$ gets ``decomposed'' into orbits of $G$. The space of
orbits, $S/G$, is `smaller' than $S$ and could constitute a
simplification.  $S/G$, is thought of as a {\em symmetry reduction of
$S$ by $G$}. Alternatively, one could restrict to the subset of the so
called {\em invariant} states which may be thought of as a collection of
trivial orbits. In our context, we will be using the term in the latter
sense. The system could be classical or quantum mechanical.

For example, if $S$ is the quantum mechanical state space of a particle
with a rotationally invariant Hamiltonian, then the subspace of the
invariant states would be all the states with zero angular momentum. If
it is the phase space of a particle with a rotationally invariant
dynamics, then the only invariant `state' is the origin of the
configuration space with zero momentum. If however, $S$ denotes the
space of field configurations on a manifold, then the subset of
invariant configurations is non-trivial. If the quantum mechanical state
space of a system consists of {\em distributions} on a space of `test
functions', then invariant states could be defined as those
distributions whose support consists of invariant test functions.

If one obtains a reduction by restricting to invariant states (and
invariant observables) of a quantum system, one has followed the {\em
first quantize, then reduce} route and the reduced system can be thought
of as a {\em symmetric sector}. This is not always possible, since one
does not have adequate explicit control over the quantum system.
Alternatively, one can consider invariant subspace of a classical phase
space and construct a corresponding quantum theory. This is the {\em
first reduce, then quantize} route. In general, the relation between
these two approaches is unclear.

While the former approach is more desirable, in practice, it is the
latter approach which is followed commonly.  We will also follow this
approach.  However, we will follow the methods -- basic variables,
construction of quantum Hilbert space etc -- used in the full theory.
The viability of these simplified models are then thought to constitute
a test of the methods and premises of the full theory. The reduction of
the classical theory is carried out by requiring certain {\em
symmetries} to be exactly realized.

\section{Symmetry Reduced Models} 
We are already familiar with use of symmetries to simplify a problem.
For example, assuming spherical symmetry we choose coordinates and
metric components to simplify the Einstein equation and obtain the
Schwarzschild solution or using homogeneity and isotropy one obtains the
FRW solutions. Thus symmetry groups (isometries) allow us to classify
suitable ansatz for the basic variables of the theory. Note however that
we are not interested in solving Einstein equations, but rather in
obtaining a classical action with fewer degrees of freedom and
constructing a corresponding quantum theory. In the context of spherical
symmetry for example, this corresponds to restricting to only
spherically symmetric form of 3-metrics: $ds^2 = \Lambda^2(t,r)dr^2 +
R^2(t,r)( d\theta^2 + \mathrm{sin}^2\theta d\phi^2)$ and reducing the
Einstein-Hilbert action to get an action in terms of the two field
degrees of freedoms -- $\Lambda(r), R(r)$. Such reductions of degrees of
freedom is termed {\em mini-superspace} model if the degrees of freedom
is {\em finite} and a {\em midi-superspace} model, if the degrees of
freedom is still infinite i.e. a lower dimensional field theory. The
former occur in {\em homogeneous cosmologies} while examples of the
latter include spherical symmetry, certain inhomogeneous cosmological
models such as the Gowdy models, Einstein-Rosen waves etc. Needless to
say  that the midi-superspace models are still very complicated. We will
concentrate on the mini-superspace models and specifically on
(spatially) homogeneous cosmologies. We begin by defining spatially
homogeneous space-times which are not necessarily solutions of Einstein
equation.

\subsection{Spatially homogeneous models} 
A four dimensional space-time is said to be spatially homogeneous if (a)
it can be foliated by a 1-parameter family of space-like hypersurfaces,
$\Sigma_t$ and (b) possessing a (Lie) group of isometries such that for
each $t$ and any two points $p, q \in \Sigma_t$ there exist an isometry
of the space-time metric which maps $p$ to $q$.  The isometry group $G$
is then said to act {\em transitively} on each of the $\Sigma_t$. If the
group element connecting $p, q$ is unique, the group action is said to
be {\em simply transitive} (otherwise multiply transitive). Spatially
homogeneous space-times are further divided into two types. 

A spatially homogeneous space-time is said to be of a {\bf Bianchi type}
if the group of isometries contains a subgroup (possibly itself), $G^*$,
which acts simply transitively on $\Sigma_t$ otherwise it is said to be
of the {\bf Kantowski-Sachs type} (interior of Schwarzschild solution).
It turns out that except for the special case of $\Sigma \sim S^2 \times
\mathbb{R}$ and $G = SO(3) \times \mathbb{R}$, in all other cases one
has a Bianchi type space-time. 

Transitive action implies that there must be at least three independent
Killing vectors at each point of $\Sigma_t$ since $\Sigma_t$ is three
dimensional. But there could be additional Killing vectors which vanish
at a point. These Killing vectors generate the {\em isotropy} (or
stability) subgroup, $H$ of $G$. Since $H$ will induce a transformation
on the tangent spaces to the spatial slices, it must be a subgroup of
$SO(3)$ and  thus dimension of $G$ can be at most 6 and at least 3 since
the dimension of $G^*$ is always 3.  All 3 dimensional Lie groups have
been classified by Bianchi into 9 types. The classification goes along
the following lines\cite{LandauLifshitz}. 

A Lie algebra (or connected component of a Lie group) is characterised
by structure constants $C^I_{~JK}$ with respect to a basis $X_I$,
satisfying the antisymmetry and Jacobi identity namely,
\[ 
\left[X_J, X_K\right] ~ = ~ C^I_{~JK} X_I~~;~~ C^I_{~JK} = C^I_{~KJ}
~~;~~ \sum_{(IJK)} C^N_{~IL}C^L_{~JK} = 0 ~~,~~ I, J, K = 1, 2, 3\ .
\]
Using the availability of the Levi-Civita symbols, ${\cal E}_{IJK}, \
{\cal E}^{IJK}, ~ {\cal E}_{123} = 1 = {\cal E}^{123}$, we can write the
structure constants as, 
\begin{eqnarray}
C^I_{~JK} & = & {\cal E}_{JKL}C^{LI} ~~,~~C^{IJ} := M^{IJ} + {\cal
E}^{IJK}A_K
\end{eqnarray}
Thus, the 9 structure constants are traded for 6 $M^{IJ}$ (symmetric in
$IJ$) and the 3 $A_K$. This has used only antisymmetry. The Jacobi
identity implies, $M^{IJ}A_J = 0$. 

Noting that the structure constants are subject to linear
transformations induced by linear transformations, $X_I \to
S_I^{~J}X_J$, on the basis of the Lie algebra, the symmetric $M^{IJ}$
can be diagonalized by orthogonal transformations and the non-zero
eigenvalues can be further scaled to $\pm 1$: $M^{IJ} = n^I\delta^{IJ}$.
The condition $M^{IJ}A_J = 0$ implies that {\em either} $A_I = 0$ ({\bf
Class A}) {\em or} $A_I \neq 0$ ({\bf class B}) in which case $M^{IJ}$
has a zero eigenvalue and we may take the non-zero eigenvector $A_I$ to
be along the ``1st'' axis, i.e. $A_I = a\delta_{I, 1}$ and $n^1 = 0$.
This leads to,
\[
\left[X_J, X_K\right] ~ = ~ n^I {\cal E}_{IJK} X_I + X_J A_K - X_K A_J
~.
\]

In the class A, there are precisely 6 possibilities organized by the
{\em rank of the matrix} -- 0, 1, 2, 3 and {\em signature} for ranks 2,
3 viz $(++, +-)$ and $(+++, ++-)$. The eigenvalues of $M^{IJ}$ can be
taken to be $n^I = \pm 1, 0$.

In the class B, the rank of $M^{IJ}$ cannot be 3 and the possibilities
are restricted to the ranks 0, 1, 2 and signatures $(++, +-)$ for rank
2. If the rank of $M$ is 0, all three eigenvalues are zero and scaling
$X_1$, we can arrange $a = 1$. For rank 1, taking $n_3$ to be the
non-zero eigenvalue, scaling $X_1, X_3$ ensures $a = 1$.  For rank 2
however, ($n_2 = \pm 1, n_3 = \pm 1$), no scaling can preserve $n_2,
n_3$ and set $a = 1$ (of course $a = 1$ is possible). 

Here is a table of the classification of Riemannian, homogeneous
3-geometries\cite{LandauLifshitz}:
\begin{center}
\begin{tabular}{|c|c|c|c|c|l|}
\hline 
Type & a & $n_1$ & $n_2$ & $n_3$ & Remarks \\
\hline 
{\bf Class A} & ~\hspace{0.75cm}~ & ~\hspace{0.75cm}~&~\hspace{0.75cm}
~&~\hspace{0.75cm}~ & \\
\hline 
I & 0 & 0 & 0 & 0 & Euclidean space \\
\hline 
 & & & & & (Leads to the Kasner space-time)\\
\hline 
II & 0 & 1 & 0 & 0 & 	 \\
\hline 
VII$_0$ & 0 & 1 & 1 & 0 & \\
\hline 
VI$_0$ & 0 & 1 & -1 & 0 & \\
\hline 
IX & 0 & 1 & 1 & 1 & $S^3$ is a special case (with isotropy)\\
\hline 
 & & & & & (Central to BKL Scenario)\\
\hline 
VIII & 0 & 1 & 1 & -1 &  \\
\hline 
{\bf Class B} & ~\hspace{1.0cm}~ & ~\hspace{1.0cm}~&~\hspace{1.0cm}
~&~\hspace{1.0cm}~ & \\
\hline 
V & 1 & 0 & 0 & 0 & $H^3$ a special case (with isotropy)\\
\hline 
IV & 1 & 0 & 0 & 1 &  \\
\hline 
VII$_a$ & a & 0 & 1 & 1 & \\
\hline 
III & 1 & 0 & 1 & -1 & sub-case of type VI$_a$ \\
\hline 
VI$_a$ & a & 0 & 1 & -1 & \\
\hline 
\end{tabular}
\end{center}

Of interests to us are the so called {\em class A} models which are
characterised by the structure constants satisfying $C^I\,_{I J} = 2 A_J
= 0$ \footnote{The remaining, Class B models are thought not have a
Hamiltonian formulation and hence are not amenable to analysis by
canonical methods \cite{AshtekarSamuel}.}. 

When $H = SO(3)$, one has homogeneity {\em and} isotropy i.e. FRW
space-times. We know that these come in three varieties depending on the
constant spatial curvature. The spatially flat case is of type Bianchi I
while positively curved case is of type Bianchi IX. (The negatively
curved case is in class B, type V). 

The metrics of the general Bianchi type space-times have at the most 6
degrees of freedom thus constituting mini-superspaces. The spatial
metrics {\em can be put} in the form: $ds^2 = g_{IJ}(t) e^I_i e^J_j dx^i
dx^j$, where $e^I_i dx^i$ are the so called Maurer-Cartan forms on the
group manifold $G^*$, satisfying $d e^I = - \Case{1}{2} C^I _{JK} e^J
\wedge e^K$. When one further restricts to {\em diagonal} $g_{IJ}$ one
gets the so-called {\em diagonal Bianchi models}.

{\em Remark:} One should notice that restricting to a subclass of
metrics amounts to introducing {\em background structures} from the
perspective of the full theory. In the present case, these structures
are the symmetry group and the coordinates adapted to the group action
(which allowed the metric to be put in the specific form). This is
unavoidable and constitutes a specification of the reduced model.  From
the perspective of a reduced model, these structures are {\em
non-dynamical, analogous to the manifold structure for the full theory}
and therefore do not automatically violate background independence.
Instead, the background independence now means that quantization
procedure should not depend the metric $g_{IJ}$ which is a dynamical
variable.

Our basic variables however are not the 3-metric and the extrinsic
curvatures.  They are the $SU(2)$ connection and the densitized triad.
In the metric variables, the natural notion of symmetry is isometry
while in the connection formulation it is the {\em group of
automorphisms} of the $SU(2)$ bundle. Thus, the cosmological models will
now be understood to be characterised by groups of automorphisms of the
$SU(2)$ bundle which acts on the base manifold $\Sigma$ transitively.
The task is to characterise the connection and triad variables which are
{\em invariant} under the group action (just as isometries mean
invariant metrics). This requires more mathematical machinery and we
will only state the conclusions\footnote{A few essential details from
Forgacs and Manton are summarised in the appendix.}.

For the Bianchi models, the invariant connections and densitized triad
are of the form:
\begin{equation}
A_a^i(t, x) ~ := ~ \Phi^i_I(t) {\omega}^I_a(x)~ ~ , ~ ~ E^a_i(t,x) ~ :=
~ \sqrt{g_0}(x) p_i^I(t) X^a_I(x).
\end{equation} 
In the above equation, $a$ refers to spatial coordinate index, $i$
refers to the adjoint representation of $SU(2)$ and $I$ refers to the
adjoint index of the Lie algebra of the symmetry (sub) group $G^*$ (and
hence takes 3 values). The $\omega^I_a dx^a$ are the Maurer-Cartan
1-forms (left-invariant 1-forms) on $\Sigma_t$ identified with the group
manifold while $X_I^a\Case{\partial}{\partial x^a}$ are the
corresponding invariant vector fields dual to the 1-forms, i.e.
$\omega^I(X_J) = \omega^I_a X_J^a = \delta^I_J$. The $g_0(x)$ is the
determinant of the invariant metric on the symmetry group and provides
the necessary density weight. It is regarded as a fiducial quantity and
will drop out later. All the coordinate dependence resides in these
forms, vector fields and the fiducial metric while the coefficients
containing the $t$ dependence are the basic dynamical
variables\footnote{Similar decomposition is made for all quantities with
spatial and the Lorentz indices. The {\em contravariant} spatial index
is expressed using the invariant vector fields and the covariant one
using the invariant 1-forms. In particular, $K_a^i := K^i_I \omega_a^I
~,~ \Gamma^i_a := \gamma^i_I \omega_a^I$.}.

If we have isotropy in addition, then the degrees of freedom are further
reduced: $\Phi^i_I := c\Lambda^i_I, \ p_i^I := p \Lambda^I_i $ and there
is only one degree of freedom left.  Here the $\Lambda$'s are a set of
orthonormal vectors satisfying, $\Lambda_I^i \Lambda_J^i = \delta_{IJ},
\Lambda_I^i \Lambda^j_J \Lambda^k_K \epsilon_{ijk} = \epsilon_{IJK}$.
The phase space variables $c, p$ are gauge invariant.

The intermediate case of {\em diagonal models} arises from a {\em
choice} $\Phi^i_I := c_I \Lambda^i_I, p_i^I := p^I \Lambda_i^I$ (no sum
over I).  The residual (SU(2)) gauge transformations act on the
$\Lambda$'s and leaving the $c_I, p^I$ as the {\em gauge invariant}
phase space variables thereby solving the Gauss constraint at the
outset\footnote{There is still a discrete invariance remaining and
involves changing the sign of two of the triad and connection
components.}. Thus there are only 3 degrees of freedom
\cite{MartinHomogeneous}.

Having identified relevant degrees of freedom parameterising quantities
invariant under symmetry transformation, the next task is to obtain the
symplectic structure (basic Poisson brackets) and simplify the
expressions for the constraints.

{\em Symplectic form:} In the full theory, this is given by $(8\pi G
\gamma)^{-1}\int_{\Sigma} d^3x \dot{A}^i_a(t,x)E_i^a(t,x)$. Direct
substitution gives,
\begin{equation}
\frac{1}{\kappa\gamma}\int_{\Sigma} d^3x \dot{A}^i_a(t,x)E_i^a(t,x) ~ =
~ \frac{1}{\kappa\gamma}\dot{\Phi}^i_I p_i^I\ \left\{\int_{\Sigma} d^3x
\sqrt{g_0}\right\} \ ,
~ \Rightarrow ~ \{\Phi^i_I, p^J_j\} = \frac{\kappa \gamma}{V_0}
\delta^i_j\delta^J_I\ .  \end{equation}
The quantity in the braces is the {\em fiducial} volume, $V_0$, of
$\Sigma_t$. For spatially flat, isotropic case, the slice is non-compact
and the fiducial volume is infinite. This problem is addressed by
restricting to a finite cell whose fiducial volume is finite. One has to
ensure that the final results do {\em not} depend on the fiducial
cell\footnote{This is discussed in more details in section
\ref{CellIndependence}}.  The dependence on the fiducial volume is
gotten rid off by redefining the basic variables as $\Phi \to \Phi
V_0^{-1/3}, \ p \to p V_0^{-2/3}$.  If we have isotropy, the symplectic
form would become $\Case{3}{\kappa\gamma}V_0 \dot{c}p$ which leads to
(after rescaling) to the Poisson bracket, $\{c, p\} =
\Case{\kappa\gamma}{3}$. With this rescaling understood, we will now
effectively put $V_0 = 1$. 

{\em Curvature:} The curvature corresponding to the invariant connection
above, is obtained as:
\begin{eqnarray}
F^i & := & d A^i + \frac{1}{2} \epsilon^i\,_{jk} A^j \wedge A^k  ~~:=~~
\frac{1}{2}F^i_{JK}\omega^J\wedge\omega^K \\
%
%\Arrowvert & & \nonumber \\
%
\therefore ~~ F^i_{JK} & = & -\Phi^i_I C^I\,_{JK} + \epsilon^i\,_{jk}
\Phi^j_J\Phi^k_K
\end{eqnarray}

{\em Gauss Constraint:} The full theory expression is:
\begin{eqnarray}
G(\Lambda) & := & \int_{\Sigma}
\Lambda^i\left\{\frac{1}{\kappa\gamma}\left(\partial_a E^a_i +
\epsilon_{ij}\,^k A^j_a E^a_k\right)\right\} \nonumber \\
& = & \frac{\Lambda^i}{\kappa\gamma}\left[ p^I_i
\underbrace{\int_{\Sigma} \partial_a(\sqrt{g_0} X^a_I)} ~ + ~
\epsilon_{ij}\,^k p^I_k \Phi^i_J\underbrace{\int_{\Sigma} \sqrt{g_0}
X^a_I \omega^J_a}\right] \nonumber \\
& & \hspace{1.7cm} - V_0 C^J\,_{IJ} \hspace{4.0cm} V_0 \\
\label{GaussOne}
\therefore ~~~ G_i & = & (\kappa\gamma)^{-1}\left\{-p^I_iC^J_{~IJ} +
\epsilon_{ij}^{~k}\Phi^j_Ip^I_k\right\} \ .
\end{eqnarray}

Notice that for the class A models, the first term is zero and for the
{\em diagonal} models the second term vanishes as well (since $\epsilon$
is antisymmetric in $j, k$ while the $\Lambda$ factors are symmetric in
$j, k$). There are no continuous gauge invariances left.  Note that the
first term in eqn (\ref{GaussOne}), is a surface term which {\em could}
vanish if $\Sigma$ has no boundaries. But this would not be true for say
spatially flat models which will have the $\Sigma$ as a cell on which
the invariant vector fields need not vanish. The integrand however is
proportional to $C^J_{~IJ}$ and these vanish for the class A models.

{\em Diffeo Constraint:}
\begin{eqnarray}
C_{\mathrm{diff}}(\vec{N}) & := & \frac{1}{\kappa\gamma}\int_{\Sigma}
N^a(x) E^b_i(x) F^i_{ab}(x) - \int_{\Sigma} N^a(x) A^i_a(x) G_i(x)
\nonumber \\
N^a(t,x) & := & N^I(t)X^a_I(x) \\
\therefore~~N^I C_I & = &
\frac{N^I}{\kappa\gamma}\left[\left(C^K\,_{JK}\Phi^i_I +
C^K\,_{IJ}\Phi^i_K\right)p^J_i\right] 
\end{eqnarray}
This constraint again vanishes for {\em diagonal}, class A models.

{\em Hamiltonian Constraint:} The full theory Hamiltonian constraint is
given by,
\begin{equation}
C_{\mathrm{Ham}}(N) ~ := ~ \frac{1}{2\kappa}\int_{\Sigma} N
\frac{E^a_iE^b_j}{\sqrt|\mathrm{det}q|}\left[\epsilon^{ij}_{\,k}
F^k_{ab} - 2(1 + \gamma^2) K^i_{[a} K^j_{b]}\right] 
\end{equation}
To carry out the integration, we need to note the expressions:
\begin{equation}
\sqrt{g_0} ~ = ~ \frac{1}{3!}\epsilon_{IJK}\epsilon^{abc}
\omega_a^I\omega_b^J\omega_c^K ~~,~~ \frac{1}{\sqrt{g_0}} ~ =~
\epsilon_{abc}\epsilon^{IJK}X^a_I X^b_J X^c_K .
\end{equation}
This leads to,
\begin{equation}
\mathrm{det} q ~ = ~ \mathrm{det}(E^a_i) ~ := ~ \frac{1}{3!}
\epsilon_{abc}\epsilon^{ijk} (g_0)^{3/2} X^a_I X^b_J X^c_K p^I_i p^J_j
p^K_k ~~=~~
\frac{1}{3!} g_0 \epsilon^{ijk}\epsilon_{IJK} p^I_i p^J_j p^K_k
\end{equation} 
Now the integration can be carried out immediately to give,
\begin{eqnarray}
H_{\mathrm{grav}} & = &
\frac{N}{2\kappa}\left[\frac{p^I_ip^J_j}{\sqrt{\frac{1}{6}|
\epsilon^{ijk}\epsilon_{IJK} p^I_i p^J_j p^K_k|}}
\left\{\epsilon^{ij}\,k F^k\,_{IJ} - 2 (1 + \gamma^2) K^{[i}_I
K^{j]}_J\right\} \right]
\end{eqnarray}
In the above, $K^i_I = \gamma^{-1}( \Phi^i_I - \Gamma^i_I)$. 
These expressions are valid for general Bianchi models. 

At this stage, we could in principle attempt to carry out the usual
Schrodinger quantization with $\Phi^i_I$ being multiplicative operators
and $P^I_i$ being the derivative operators. Both transform covariantly
under the action of SU(2).  

However, we can also imagine `specializing the holonomy-flux variables'
of the full theory, for these symmetric fields. It is natural to choose
edges along the symmetry directions i.e. along integral curves of the
$X^a_I$ vector fields. It follows that due to homogeneity, the path
ordered exponentials, holonomies, become just the {\em ordinary
exponentials}, $h_I(\Phi) := h_{e_I}(\Phi) := {\cal
P}\mathrm{exp}\{\int_{e_I}\Phi^i_I\tau_i\omega^I_a dx^a\} =
\mathrm{exp}\{\Phi_I^i(t)\tau_i\int_{e_I}\omega_a^Idx^a$\}. There is no
sum over $I$ in these expressions. These can be further expressed using
the identity $e^{i\theta\hat{n}\cdot\vec{\sigma}} = \mathrm{cos}(\theta)
+ i \hat{n}\cdot\vec{\sigma}\mathrm{sin}\theta$. The holonomy is then
given in terms of $\theta \sim \sqrt{\Phi^i_I\Phi^i_I}$ which is gauge
invariant and two angular, gauge variant components corresponding to the
direction $\hat{n} \sim $ unit vector in the direction of $\Phi^i_I$. A
simplification occurs if we further restrict to the diagonal models:
$\Phi^i_I := c_I\Lambda^i_I$ which makes the $\hat{n} = \vec{\Lambda}_I$
and now the matrix elements of these holonomies can be obtained from the
elementary functions, $e^{\mu_{(I)}c_I/2}$. These have been termed as
the {\em point holonomies}. The fluxes through surfaces perpendicular to
the symmetry directions, likewise simplify to $E_{S_{JK}}(f) = p^I
\Lambda_i^I f^i\int_{S_{JK}} \sqrt{g_0(x)} \epsilon_{abc} X^a_IdS^{bc}
\propto p^I$.  Unlike the flux operators in the full theory, these
fluxes Poisson commute among themselves.  Thus, in the diagonal models,
we can extract gauge invariant phase space coordinates, with the
holonomies and fluxes having the usual Poisson algebra. In quantum
theory, a useful triad representation can then be set-up.

{\em Point holonomies and commuting flux variables are new features}
which arise in the (diagonal) mini-superspace reduction. These are also
responsible for the relative ease of analysis possible for these models.
This will be discussed more below. 

What about inhomogeneous models? There have fewer efforts regarding
these. Among the inhomogeneous models, the reduction for the Gowdy model
on 3-torus can be seen in \cite{GowdyClassical}, while spherical
symmetric model can be seen in \cite{Spherical}. Martin's lattice model
is briefly discussed in the appendix.

\chapter {Singularity Resolution in Quantum Theory}
The most detailed analysis of the singularity resolution is available
for the homogeneous and isotropic geometry coupled to a massless scalar
and this is the case that we discuss below. Prior to 2005, the
singularity resolution  was understood as the deterministic nature of
fundamental equation (the Hamiltonian constraint) and in terms of an
effective picture deduced either from the WKB approximation of by taking
expectation values of the Hamiltonian. In this sense, resolution of
singularities was seen for (i) FRW coupled to a scalar field with
arbitrary positive semidefinite potential and (ii) diagonalised Bianchi
class A (anisotropic) models. These resolutions were seen as an
implication of the {\em inverse triad quantum corrections} which were
present in the matter sector (and in the curvature for non-flat models).
Post 2005, it was realized, at least for the FRW case, that {\em the
holonomy corrections by themselves could also resolve singularities}.
This required restriction to massless scalar and treating it as a clock
variable, thereby paving the way for construction of physical states,
Dirac observables and physical expectation values. Although restricted
to special matter, it allows completion of the quantization program to
the physical level and throws light on {\em how} a quantum singularity
resolution may be viewed. For this reason, we this case is discussed in
detail. Subsequently, Madhavan also showed another quantization for the
same case, also completed to physical level, wherein holonomy
corrections are absent and singularity resolution is achieved by inverse
triad corrections only. There are also some issues which have been
better understood in the past few years. These are briefly summarised
and discussed in sections \ref{CellIndependence} and \ref{LatticeView} . 
\section{FRW, Classical Theory} \label{ClassicalFRW}
{\em Classical model:} Using coordinates adapted to the spatially
homogeneous slicing of the space-time, the metric and the extrinsic
curvature are given by, 
\begin{equation} \label{FRWMetric}
ds^2 := - dt^2 + a^2(t)\left\{dr^2 + r^2d\Omega^2\right\} ~~:= ~~ -dt^2
+ a^2(t) ds^2_{\mathrm{comoving}}\ .  
\end{equation}
Starting from the usual Einstein-Hilbert action and scalar matter for
definiteness, one can get to the Hamiltonian as,
\begin{eqnarray}
S & := & \int dt \int_{\mathrm{cell}} dx^3 \sqrt{|det
g_{\mu\nu}|}\left\{ \frac{R(g)}{16 \pi G} + \frac{1}{2}\dot{\phi}^2 -
V(\phi) \right\} \nonumber\\
& = & V_0\int dt \left\{\frac{3}{8 \pi G}(- a\dot{a}^2) + \frac{1}{2}a^3
\dot{\phi}^2 - V(\phi) a^3 \right\} \\
p_a & = & - \frac{3 V_0}{4 \pi G} a \dot{a} ~ ~ , ~ ~ p_{\phi} ~ = ~ V_0
a^3 \dot{\phi} ~ ~,~ ~ V_0 ~ := ~ \int_{\mathrm{cell}}
d^3x\sqrt{g_{\mathrm{comoving}}} ~ ; \nonumber\\
H (a, p_a, \phi, p_{\phi}) & = & H_{\mathrm{grav}} + H_{\mathrm{matter}}
\nonumber \\
& = & \left[- \frac{2 \pi G}{3} \frac{p_a^2}{V_0 a} \right] +
\left[\frac{1}{2} \frac{p_{\phi}^2}{a^3 V_0} + a^3 V_0 V(\phi)\right] \\
& = & \left(\frac{3 V_0 a^3}{8 \pi G}\right)\left[ -
\frac{\dot{a}^2}{a^2} + \left(\frac{8 \pi G}{3}\right)
\left(\frac{H_{\mathrm{matter}}}{ V_0 a^3}\right)\right]
\end{eqnarray}
Thus, $H = 0 \leftrightarrow $ Friedmann Equation. For the spatially
flat model, one has to choose a fiducial cell whose fiducial volume is
denoted by $V_0$.

In the connection formulation, instead of the metric one uses the
densitized triad i.e. instead of the scale factor $a$ one has
$\tilde{p}, |\tilde{p}| := a^2$ while the connection variable is
related to the extrinsic curvature as: $\tilde{c} := \gamma \dot{a}$
(the spin connections is absent for the flat model). Their Poisson
bracket is given by $\{\tilde{c}, \tilde{p}\} = (8\pi G \gamma)/(3
V_0)$. The arbitrary fiducial volume can be absorbed away by defining $c
:= V_0^{1/3} \tilde{c}, ~ p := V_0^{2/3}\tilde{p}$. Here, $\gamma$ is
the Barbero-Immirzi parameter which is dimensionless and is determined
from the Black hole entropy computations to be approximately $0.23$
\cite{BHEntropy}. From now on we put $8\pi G := \kappa$. The classical
Hamiltonian is then given by,
\begin{equation}\label{ClassHam}
H ~ = ~ \left[- \frac{3}{\kappa}\left( \gamma^{-2}c^2 \sqrt{|p|}\right)
\right] + \left[\frac{1}{2}|p|^{-3/2} p_{\phi}^2 + |p|^{3/2}
V(\phi)\right] \ .
\end{equation}
For future comparison, we now take the potential for the scalar field,
$V(\phi)$ to be zero as well.

One can obtain the Hamilton's equations of motion and solve them easily.
On the constrained surface ($H = 0$), eliminating $c$ in favour of $p$
and $ p_{\phi}$, one has,
\begin{eqnarray}
c ~ = ~ \pm \gamma \sqrt{\frac{\kappa}{6}} \frac{|p_{\phi}|}{|p|} ~ & ,
& ~ \dot{p} ~ = ~ \pm 2 \sqrt{\frac{\kappa}{6}} |p_{\phi}||p|^{-1/2} \ .
\nonumber \\
\dot{\phi} ~ = ~ p_{\phi} |p|^{-3/2}~ & , & ~ \dot{p_{\phi}} ~ = ~ 0\ ,
\\
\frac{d p}{d \phi} ~ = ~ \pm \sqrt{\frac{2\kappa}{3}} |p| ~ &
\Rightarrow & ~  
{\bf p(\phi) ~ = ~ p_* e^{\pm \sqrt{\frac{2\kappa}{3}}( \phi - \phi_*)}}
\label{ClassRelationalSoln}
\end{eqnarray}
Since $\phi$ is a monotonic function of the synchronous time $t$, it can
be taken as a new ``time'' variable. The solution is determined by
$p(\phi)$ which is (i) independent of the constant $p_{\phi}$ and (ii)
passes through $p = 0$ as $\phi \to \pm \infty$ (expanding/contracting
solutions).  It is immediate that, along these curves, $p(\phi)$, the
energy density ($p^{-6}p_{\phi}^2/2$) and the extrinsic curvature
diverge as $p \to 0$.  Furthermore, the divergence of the density
implies that $\phi(t)$ is {\em incomplete} i.e. $t$ ranges over a
semi-infinite interval as $\phi$ ranges over the full real
line\footnote{For the FRW metric, integral curves of $\partial_t$ are
time-like geodesics and hence incompleteness with respect to $t$ is
synonymous with geodesic incompleteness.}.  

Thus a singularity is signalled by {\em every}  solution $p(\phi)$
passing through $p = 0$ in {\em finite} synchronous time (or
equivalently by the density diverging somewhere along any solution).  A
natural way to ensure that {\em all} solutions are non-singular is to
ensure that either of the two terms in the Hamiltonian constraint is
{\em bounded}. Question is: {\em If and how does a quantum theory
replace the Big Bang singularity by something non-singular?}

There are at least two ways to explore this question. One can imagine
computing corrections to the Hamiltonian constraint such that individual
terms in the effective constraint are bounded. This approach presupposes
the classical framework and thus will have a {\em domain of validity of
these corrections}. Alternatively and more satisfactorily, one should be
able to define suitable Dirac observables whose expectation values will
generate the analogue of $p(\phi)$ curves along which physical
quantities such as energy density, remain bounded. Both are discussed
below.

\section{FRW, Quantum Theory}

{\bf Schrodinger Quantization:} In the standard Schrodinger
quantization, one can introduce wave functions of $p, \phi$ and quantize
the Hamiltonian operator by $c \to i\hbar \kappa\gamma/3 \partial_p ~,~
p_{\phi} \to -i \hbar \partial_{\phi}$, in equation (\ref{ClassHam}).
With a choice of operator ordering, $\hat{H}\Psi(p, \phi) = 0$ leads to
the Wheeler-De Witt partial differential equation which has singular
coefficients. We will return to this later.

{\bf Loop Quantization:} The background independent quantization of Loop
Quantum Gravity however suggest a different quantization of the
isotropic model. One should look for a Hilbert space on which only
exponentials of $c$ (holonomies of the connection) are well defined
operators and not $\hat{c}$. Such a Hilbert space is obtained as the
representation space of the C* algebra of holonomies. In the present
context this algebra is the algebra of {\em almost periodic functions of
$c$}, finite linear combinations of functions of the form $e^{i\lambda_j
c}, \lambda_j \in \mathbb{R}$.  Inner product (analogue of the
Ashtekar-Lewandowski measure) on the space of the almost periodic
functions is given by: 
\begin{equation}
(\Psi, \Phi) ~ := ~ \lim_{T \to \infty} ~ \frac{1}{2T} ~ \int_{-T}^{T}
dc ~ \Psi^{*}(c) \Phi(c) \ .
\end{equation}
The single exponentials form an orthonormal set. Let us denote it as,
$\langle c|\mu\rangle := $exp$\{ \Case{i}{2} \mu c \}, \mu \in
\mathbb{R}$. The holonomy-flux representation can now be made explicit
as:
\begin{eqnarray} \label{HolonomyFluxRepren}
\hat{p}|\mu\rangle & = &
\frac{1}{6}\gamma\lP^2\mu|\mu\rangle~~,~~\langle\mu|\mu'\rangle ~=~
\delta_{\mu, \mu'}~~,~~\mu \in \mathbb{R} \nonumber \\
\widehat{h_{\nu}}|\mu\rangle & := & \widehat{e^{\Case{i}{2}\nu
c}}|\mu\rangle ~ = ~ |\mu + \nu\rangle 
\end{eqnarray} 
Notice that that the triad operator has every real number as a {\em
proper eigenvalue} (i.e. has a corresponding {\em normalizable}
eigenvector, the spectrum is {\em discrete}). This implies that the
holonomy operator, is {\em not weakly continuous} in the label $\nu$
i.e. arbitrary matrix elements of $\hat{h}_{\nu}$ are not continuous
functions of $\nu$. Therefore one {\em cannot} define a $\hat{c}$
operator. Note that the {\em volume operator}, is given by $\hat{V} :=
|p|^{3/2}$.

{\bf Inverse Triad Operator:} The fact that spectrum of the triad
operator is {\em discrete}, has a major implication: {\em inverses of
positive powers of triad operators do not exist} \footnote{The domain of
the inverse power operator(s) will have to exclude the subspace
corresponding to the zero eigenvalue of the triad operator. But this
makes the domain {\em non-dense} and its adjoint cannot be defined
\cite{ABL}.}.  These have to be defined by using alternative classical
expressions and promoting them to quantum operators. This can be done
with at least one parameter worth of freedom, eg.
\begin{equation}
|p|^{-1} ~ = ~ \left[ \frac{3}{\kappa \gamma l}\{c, |p|^l\}\right]^{1/(1
-l)} ~,~ l \in (0, 1)\ .  
\end{equation}
Only positive powers of $|p|$ appear now. However, this still cannot be
used for quantization since there is no $\hat{c}$ operator. One must use
holonomies: $h_j(c) ~ := ~ e^{\mu_0 c \Lambda^i\tau_i}\ ,$ where
$\tau_i$ are anti-hermitian generators of $SU(2)$ in the $j^{th}$
representation satisfying $\mathrm{Tr}_j (\tau_i \tau_j) = - \Case{1}{3}
j (j + 1) (2j + 1) \delta_{ij}$, $\Lambda^i$ is a unit vector specifying
a direction in the Lie algebra of $SU(2)$ and $\mu_0$ is the coordinate
length of the edge used in defining the holonomy. It is a fraction of
$V_0^{1/3}$. Using the holonomies,
\begin{eqnarray} 
|p|^{-1} & = & (8 \pi G \mu_0\gamma l)^{\frac{1}{l - 1}} \left[
\frac{3}{j(j + 1)(2j + 1)} \mathrm{Tr}_j \Lambda\cdot\tau \
h_j\left\{h_j^{-1}, |p|^l\right\}\right]^{\frac{1}{1 - l}} \ ,
\end{eqnarray}
which can be promoted to an operator. Two parameters, $\mu_0 \in
\mathbb{R}$ and $j \in \mathbb{N}/2$, have crept in and we have a three
parameter family of inverse triad operators. The definitions are:
\begin{eqnarray} 
\widehat{|p|^{-1}_{(jl)}} |\mu\rangle & = &
\left(\frac{2j\mu_0}{6}\gamma\lP^2\right)^{-1} (F_{l}(q))^{\frac{1}{1
-l}} |\mu\rangle ~ ~ , ~ ~ q := \frac{\mu}{2\mu_0j} ~ := ~
\frac{p}{2jp_0}~ ~, \label{InvTriad}\\
F_l(q) & := & \frac{3}{2l}\left[ ~ ~ \frac{1}{l + 2}\left\{ (q + 1)^{l +
2} - |q - 1|^{l + 2}\right\} \right. \nonumber \\ & & \hspace{0.7cm}
\left. - \frac{1}{l + 1} q \left\{ (q + 1)^{l + 1} - \mathrm{sgn}(q -1)
|q - 1|^{l + 1}\right\} ~ ~ \right] \nonumber \\
F_l( q \gg 1 ) & \approx & \left[q^{-1}\right]^{1 - l} \ ,
\label{Ffunction} \\
F_l( q \approx 0 ) & \approx & \left[\frac{3q}{l + 1}\right] \ .
\nonumber 
\end{eqnarray}
All these operators obviously commute with $\hat{p}$ and their
eigenvalues are bounded above. This implies that the matter densities
(and also intrinsic curvatures for more general homogeneous models),
remain bounded over the classically singular region. Most of the
phenomenological novelties are consequences of this particular feature
predominantly anchored in the matter sector. In the effective
Hamiltonian computations, this modification will imply the second term
in the Hamiltonian constraint (\ref{ClassHam}) is rendered bounded
implying singularity avoidance.

We have also introduced two scales: $p_0 := \Case{1}{6}\mu_0\lP^2$ and
$2jp_0 := \Case{1}{6}\mu_0 (2j) \lP^2$. The regime $|p| \ll p_0$ is
termed the {\em deep quantum regime}, $p \gg 2jp_0$ is termed the {\em
classical regime} and $p_0 \lesssim |p| \lesssim 2jp_0$ is termed the
{\em semi-classical regime}.  The modifications due to the inverse triad
defined above are strong in the semi-classical and the deep quantum
regimes. For $j = 1/2$ the semi-classical regime is absent.  Note that
such scales are not available for the Schrodinger quantization.

{\bf The Gravitational Constraint:} Since $\hat{c}$ operator does not
exist, the gravitational Hamiltonian (the first bracket in
eq.(\ref{ClassHam})), has to be expressed in an equivalent form using
holonomies. For this, let us go back to the full theory Hamiltonian:
\begin{equation} \label{IsoGrav}
\frac{6}{\gamma^2} c^2\sqrt{p} ~ = ~ \gamma^{-2}\int_{\mathrm{cell}}d^3x
\frac{\epsilon_{ijk} E^{ai} E^{bj} F^k_{ab}}{\sqrt{|\mathrm{det}E|}}
\end{equation}
Now use the two identities:
\begin{eqnarray}
\frac{\epsilon_{ijk} E^{ai} E^{bj}}{\sqrt{|\mathrm{det}E|}} & = & \sum_k
\frac{4 \ \mathrm{sgn} \ p}{\kappa \gamma \mu_0 V_0^{1/3}}~
\epsilon^{abc} ~ \omega^k_c \ \mathrm{Tr}\left( h^{(\mu_0)}_k \left\{
\left(h^{(\mu_0)}_k\right)^{-1}, V\right\} \tau_i \right) \\
F^k_{ab} & = & -2 \lim_{A_{\square} \to 0} \mathrm{Tr}
\left(\frac{h^{(\mu_0)}_{\square_{ij}} - 1}{\mu_0^2 V_0^{2/3}}\right)
\tau^{k} ~\omega_a^{i} ~ \omega_b^j \\
h^{(\mu_0)}_{\square_{ij}} & := & h^{(\mu_0)}_i h^{(\mu_0)}_j
\left(h^{(\mu_0)}_i\right)^{-1} \left(h^{(\mu_0)}_j\right)^{-1}
\end{eqnarray}
In the above, the fiducial cell is thought to have been sub-divided into
smaller cells of side $\mu_0\ell, \ell := V_0^{1/3}$. The area of a
plaquette is $A_{\square} = \ell^2\mu_0^2$. The plaquette is to be
shrunk such that its area goes to zero. The superscript $(\mu_0)$ on the
holonomies is to remind that the length of the edge is $\mu_0$. The
1-forms $\omega^i_a$ are the fiducial 1-forms whose square gives the
fiducial metric and the $\epsilon^{abc}$ is the (fiducial) metric
dependent Levi-Civita density.

In quantum geometry however there is a gap in the spectrum of area
operator and thus it is {\em not} appropriate to take the area to zero,
but at the most to the smallest possible eigenvalue. Independently, if
we force the limit, it will imply $\mu_0 \to 0$ which in turn amounts to
defining $\hat{c}$ operator which does {\em not} exist on the Hilbert
space.

Substituting these in the (\ref{IsoGrav}) and carrying out the
integration over the cell leads to (suppressing the $\mu_0$ superscript
on the holonomies),
\begin{eqnarray} \label{RegularisedHam}
H_{\mathrm{grav}} & = & - \frac{4}{8 \pi G \gamma^3 \mu_0^3} \sum_{ijk}
\epsilon^{ijk}\mathrm{Tr}\left(h_i h_j h_i^{-1}h_j^{-1}h_k\{h_k^{-1},
V\}\right)
\end{eqnarray}
In the above, we have used $j = 1/2$ representation for the holonomies
and $V$ denotes the volume function. In the limit $\mu_0 \to 0$ one gets
back the classical expression \footnote{The expression for the
Hamiltonian constraint follows exactly from the full theory procedure.
Starting from the equation (\ref{IsoGrav}), the integral will be
replaced by a sum over smaller cells of a triangulation. The size
parameter of the cells will drop out thanks to density weight 1 of the
Hamiltonian. Due to homogeneity, contribution from each cell will be the
same and hence the total sum will be number of cells of the
triangulation times the contribution of one cell. There are exactly
$\mu_0^{-3}$ cubical (say) cells with side of length $\mu_0 V^{1/3}_0$
and this produces the factor of $\mu_0^{-3}$ in equation
(\ref{RegularisedHam}). The $V_0$ of course disappears as in the full
theory \label{SubCells}}. 

If we promote this expression to a quantum operator (modulo ordering
ambiguities) on the LQC Hilbert space constructed above, then we {\em
cannot} take the limit $\mu_0 \to 0$ because it would imply that
$\hat{c}$ exist which we have shown to be impossible. Thus, at the
quantum level we should {\em not} take the limit $\mu_0 \to 0$. The best
we can do is to take reduce $\mu_0$ such that the area reaches its
smallest possible (and non-zero due to the gap) eigenvalue $\Delta :=
(2\sqrt{3} \pi \gamma) \lP^2$. But which area do we consider, the
fiducial or the physical? These are related by a factor of $|p|$. It
seems appropriate to choose the physical area, which implies that we
must take $\mu_0$ to be a function $\bar{\mu}$ of $p$ given by,
$\bar{\mu}(p) := \sqrt{\Delta/|p|}$. Note that this is {\em one}
prescription to interpret the limitation on shrinking of the plaquette.
There are others which will be mentioned later. In the following we will
continue to use the $\mu_0$ notation and replace it by $\bar{\mu}(p)$
when needed.

While promoting this expression to operators, there is a choice of
factor ordering involved and many are possible.  We will present two
choices of ordering: the {\em non-symmetric} one which keeps the
holonomies on the left as used in the existing choice for the full
theory, and the particular {\em symmetric} one used in \cite{APSOne}. 
\begin{eqnarray}
\hat{H}^{\mathrm{non-sym}}_{\mathrm{grav}} & = & \frac{24 i }{\gamma^3
\mu_0^3 \lP^2} \mathrm{sin}^2 \mu_0c \left( \mathrm{sin}\frac{\mu_0c}{2}
\hat{V} \mathrm{cos}\frac{\mu_0c}{2} - \mathrm{cos}\frac{\mu_0c}{2}
\hat{V} \mathrm{sin}\frac{\mu_0c}{2} \right) \\
\hat{H}^{\mathrm{sym}}_{\mathrm{grav}} & = & \frac{24 i
(\mathrm{sgn}(p))}{\gamma^3 \mu_0^3 \lP^2} \mathrm{sin} \mu_0c \left(
\mathrm{sin}\frac{\mu_0c}{2} \hat{V} \mathrm{cos}\frac{\mu_0c}{2} -
\mathrm{cos}\frac{\mu_0c}{2} \hat{V} \mathrm{sin}\frac{\mu_0c}{2}
\right) \mathrm{sin} \mu_0c \end{eqnarray}
At the quantum level, $\mu_0$ cannot be taken to zero since $\hat{c}$
operator does not exist. The action of the Hamiltonian operators on
$|\mu\rangle$ is obtained as,
\begin{eqnarray}
\hat{H}^{\mathrm{non-sym}}_{\mathrm{grav}}|\mu\rangle & = &
\frac{3}{\mu_0^3\gamma^3\lP^2}\left(V_{\mu + \mu_0} - V_{\mu -
\mu_0}\right) \left( |\mu + 4 \mu_0\rangle - 2 | \mu\rangle + |\mu -
4\mu_0\rangle\right) \\
\hat{H}^{\mathrm{sym}}_{\mathrm{grav}}|\mu\rangle & = &
\frac{3}{\mu_0^3\gamma^3\lP^2}\left[ \left|V_{\mu + 3\mu_0} - V_{\mu +
\mu_0}\right||\mu + 4 \mu_0\rangle + \left|V_{\mu - \mu_0} - V_{\mu -
3\mu_0}\right||\mu - 4 \mu_0\rangle \right. \nonumber \\ & & \left.
\hspace{1.5cm} - \left\{\left|V_{\mu + 3\mu_0} - V_{\mu + \mu_0}\right|
+ \left|V_{\mu - \mu_0} - V_{\mu - 3\mu_0}\right|\right\}|\mu\rangle
\right]
\end{eqnarray}
where $V_{\mu} := (\Case{1}{6}\gamma \lP^2 |\mu|)^{3/2}$ denotes the
eigenvalue of $\hat{V}$. 

We also have the Hilbert space for the matter degrees which for us is a
single scalar, $\phi$ and the full kinematical Hilbert space is the
tensor product of the $L_2(\mathbb{R}_{\mathrm{Bohr}},
d\mu_{\mathrm{Bohr}}) \otimes {\cal H}_{\mathrm{matter}}$.

{\bf Wheeler-DeWitt Difference Equation:} Let $|\Psi\rangle :=
\sum_{\mu} \Psi(\mu, \phi) |\mu\rangle$, where the sum is over a
countable subset of $\mathbb{R}$, the coefficients $\Psi(\mu, \phi)$ are
valued in the matter Hilbert space and the argument $\phi$ is a reminder
of that. The Hamiltonian constraint is imposed on these $|\Psi\rangle$
which leads to the {\em Wheeler-DeWitt equation} for the coefficients.
Thanks to the presence of the trigonometric operators, this equation is
a {\em difference equation}.  In the Schrodinger quantization, this
would be a differential equation.

For the non-symmetric operator we get,
$$
A(\mu + 4 \mu_0) \Psi(\mu + 4\mu_0, \phi) - 2 A(\mu) \Psi(\mu, \phi) +
A(\mu - 4 \mu_0) \Psi(\mu - 4\mu_0, \phi) 
$$
\begin{equation} 
~ = ~ -\frac{2 \kappa}{3}\mu_0^3 \gamma^3\lP^2 H_{matter}(\mu)\Psi(\mu,
\phi)
\end{equation} 
where, $A(\mu) := V_{\mu + \mu_0} - V_{\mu - \mu_0}$ and vanishes for
$\mu = 0$.

For the symmetric operator one gets,
$$
f_+(\mu) \Psi(\mu + 4\mu_0, \phi) + f_0(\mu) \Psi(\mu, \phi) + f_-(\mu)
\Psi(\mu - 4\mu_0, \phi) 
$$
\begin{equation} 
~ = ~ -\frac{2 \kappa}{3}\mu_0^3 \gamma^3\lP^2 H_{matter}(\mu)\Psi(\mu,
\phi) \hspace{3.0cm} \mbox{where,}
\end{equation} 
$$ f_+(\mu) ~ := ~ \left| V_{\mu + 3\mu_0} - V_{\mu + \mu_0} \right| ~ ,
~ f_-(\mu) ~ := ~ f_+(\mu - 4 \mu_0) ~, ~ f_0 ~ := ~  - f_+(\mu) -
f_-(\mu) \ .  $$
Notice that $f_+(-2\mu_0) = 0 = f_-(2\mu_0)$, but $f_0(\mu)$ is never
zero. The absolute values have entered due to the sgn($p$) factor. 

The difference equations relate $\Psi(\mu)$'s only for $\mu$'s in a
``{\em lattice}'', ${\cal L}_{\hat{\mu}} := \{\mu = \hat{\mu} + 4\mu_0
n, n \in \mathbb{Z}\}$ and the coefficients labelled by different
lattices are completely independent. The $\hat{\mu} \in [0, 4\mu_0)$,
label different {\em superselected} sectors. 

The equations are effectively second order difference equations and the
$\Psi(\mu, \phi)$ are determined by specifying $\Psi$ for two
consecutive values of $\mu$ eg for $ \mu = \hat{\mu} + 4 \mu_0 N$ and $
\mu = \hat{\mu} + 4 \mu_0 (N + 1)$.  Since the highest (lowest) order
coefficients vanishes for some $\mu$, then the corresponding component
$\Psi(\mu, \phi)$ is undetermined by the equation. Potentially this
could introduce an arbitrariness in extending the $\Psi$ specified by
data in the classical regime (eg $\mu \gg 2j $) to the negative $\mu$.
Potentially, maintaining determinism of the quantum wave function, is
one of the restrictive criteria for choosing the ordering. 

For the non-symmetric case, the highest (lowest) $A$ coefficients vanish
for their argument equal to zero thus leaving the corresponding $\Psi$
component undetermined. However, this undetermined component is
decoupled from the others. Thus apart from admitting the trivial
solution $\Psi(\mu, \phi) := \Phi(\phi)\delta_{\mu, 0}, ~ \forall \mu$,
all other non-trivial solutions are completely determined by giving two
consecutive components: $\Psi(\hat{\mu}, \phi), \Psi(\hat{\mu} + 4\mu_0,
\phi)$.

For the symmetric case, due to these properties of the $f_{\pm,0}(\mu)$,
it looks as if the difference equation is {\em non-deterministic} if
$\mu = 2\mu_0 + 4\mu_0 n, n \in \mathbb{Z}$. This is because for $\mu =
-2\mu_0$, $\Psi(2\mu_0, \phi)$ is undetermined by the lower order
$\Psi$'s and this coefficient enters in the determination of
$\Psi(2\mu_0, \phi)$.  However, the symmetric operator also commutes
with the parity operator: $(\Pi\Psi)(\mu, \phi) := \Psi(-\mu, \phi)$.
Consequently, $\Psi(2\mu_0, \phi)$ is determined by $\Psi(-2\mu_0,
\phi)$.  Thus, we can restrict to $\mu = 2\mu_0 + 4k\mu_0, k \ge 0$
where the equation {\em is} deterministic.  

In both cases then, the space of solutions of the constraint equation,
is completely determined by giving appropriate data for large $|\mu|$
i.e. in the classical regime. Such a deterministic nature of the
constraint equation has been taken as a necessary condition for
non-singularity at the quantum level \footnote{For contrast, if one just
symmetrizes the non-symmetric operator (without the sgn factor), one
gets a difference equation which {\em is non-deterministic}. Note that
this issue arises only in {\em one} superselection sector so may not
really be an issue. However, requiring deterministic equation in {\em
all} sectors could be invoked as a criterion to discriminate between
different factor ordering.}. 

{\bf Effective Hamiltonian:} By introducing an interpolating, slowly
varying smooth function, $\Psi(p (\mu) := \Case{1}{6}\gamma\lP^2\mu)$,
and keeping only the first non-vanishing terms, one deduces the
Wheeler-De Witt differential equation (with a modified matter
Hamiltonian) from the above difference equation. Making a WKB
approximation, one infers an effective Hamiltonian which matches with
the classical Hamiltonian for large volume ($\mu \gg \mu_0$) and small
extrinsic curvature (derivative of the WKB phase is small). There are
terms of $o(\hbar^0)$ which contain arbitrary powers of the first
derivative of the phase which can all be summed up. The resulting
effective Hamiltonian now contains modifications of the classical
gravitational Hamiltonian, apart from the modifications in the matter
Hamiltonian due to the inverse powers of the triad. The largest possible
domain of validity of effective Hamiltonian so deduced must have $|p|
\gtrsim p_0$ \cite{SemiClass,EffHam}.

An effective Hamiltonian can alternatively obtained by computing
expectation values of the Hamiltonian operator in semi-classical states
peaked in classical regimes \cite{Willis}. The leading order effective
Hamiltonian that one obtains is (spatially flat case):
\begin{eqnarray}
H^{\mathrm{non-sym}}_{\mathrm{eff}} & = & - \frac{1}{16 \pi
G}\left(\frac{6}{\mu_0^3 \gamma^3 \lP^2}\right) \left[ B_+(p)
\mathrm{sin}^2(\mu_0c) + \left( A(p) - \frac{1}{2}B_+(p) \right) \right]
+ H_{\mathrm{matter}} \ ; \nonumber\\
B_+(p) & := & A(p + 4 p_0) + A(p - 4 p_0) ~,~ A(p) ~ := ~ (|p +
p_0|^{3/2} - |p - p_0|^{3/2}) \ , \\
p & := & \frac{1}{6}\gamma \lP^2 \mu ~ ~ , ~ ~ p_0 ~ := ~
\frac{1}{6}\gamma \lP^2 \mu_0 \ . \nonumber
\end{eqnarray}
For the symmetric operator, the effective Hamiltonian is the same as
above except that $B_+(p) \to f_+(p) + f_-(p)$ and $2 A(p) \to f_+(p) +
f_-(p)$. 

The second bracket in the square bracket, is the quantum geometry
potential which is negative and higher order in $\lP$ but is important
in the small volume regime and plays a role in the genericness of bounce
deduced from the effective Hamiltonian \cite{GenBounce}. This term is
absent in the effective Hamiltonian deduced from the symmetric
constraint.  The matter Hamiltonian will typically have the eigenvalues
of powers of inverse triad operator which depend on the ambiguity
parameters $j, l$.

We already see that the quantum modifications are such that both the
matter (due to inverse volume corrections) and the gravitational part
(due to holonomy corrections) in the effective Hamiltonian, are rendered
bounded and effective dynamics must be non-singular.

For large values of the triad, $p \gg p_0$, $B_+(p) \sim 6 p_0 \sqrt{p}
-~o(p^{-3/2})$ while $A(p) \sim 3 p_0 \sqrt{p} - o(p^{-3/2})$. In this
regime, the effective Hamiltonians deduced from both symmetric and
non-symmetric ordering are the same\footnote{The effective Hamiltonian
then reduces to $-\Case{3}{\kappa}\gamma^{-2}\sqrt{p} \left[\mu_0^{-2}
\mathrm{sin}^2(\mu_0 c)\right]$. This is also the Hamiltonian in eq.
(\ref{RegularisedHam}) for non-zero $\mu_0$.}. The classical Hamiltonian
is obtained for $\mu_0 \to 0$. From this, one can obtain the equations
of motion and by computing the left hand side of the Friedmann equation,
infer the effective energy density. For $p \gg p_0$ one obtains
\footnote{For $p$ in the semi-classical regime, one should include the
contribution of the quantum geometry potential present in the
non-symmetric ordering, especially for examining the bounce possibility
\cite{EffHam}.},
\begin{equation} \label{EffDensity}
\frac{3}{8\pi G}\left(\frac{\dot{a}^2}{a^2}\right) ~ := ~
\rho_{\mathrm{eff}} ~ = ~
\left(\frac{H_{\mathrm{matter}}}{p^{3/2}}\right)\left\{1 - \frac{8 \pi G
\mu_0^2\gamma^2}{3} p
\left(\frac{H_{\mathrm{matter}}}{p^{3/2}}\right)\right\} ~ ~,~ ~ p :=
a^2/4\ .
\end{equation}

The effective density is quadratic in the classical density, $\rho_{cl}
:= H_\mathrm{matter} p^{-3/2}$. This modification is due to the quantum
correction in the gravitational Hamiltonian (due to the sin$^2$
feature). This is over and above the corrections hidden in the matter
Hamiltonian (due to the ``inverse volume'' modifications). As noted
before, we have two scales: $p_0$ controlled by $\mu_0$ in the
gravitational part and $2p_0 j$ in the matter part. For large $j$ it is
possible that we can have $p_0 \ll p \ll 2p_0j $ in which case the above
expressions will hold with $j$ dependent corrections in the matter
Hamiltonian. In this semi-classical regime, the corrections from sin$^2$
term are smaller in comparison to those from inverse volume. If $p \gg
2p_0j$ then the matter Hamiltonian is also the classical expression. For
$j = 1/2$, there is only the $p \gg p_0$ regime and $\rho_{cl}$ is
genuinely the classical density.

To summarize: 

(1) The connection formulation, in the homogeneous and isotropic
context, uses variables ($c, p \in \mathbb{R}$) in terms of which the
classical singularity ($p = 0$) is in the {\em interior} of the phase
space. By contrast, in the ADM variables ($a \ge 0, K$), in the same
context, the classical singularity ($a = 0$) is on the {\em boundary}.
This requires a boundary condition on the quantum wave functions to be
specified in the deep quantum region where the classical framework is
suspect. When the singularity is in the interior, only a continuation of
the quantum wave function is required, given its specification in the
semi-classical region. 

(2) The connection variables also strongly motivate the very different
{\em loop quantization}. Its immediate implications are two types of
corrections - the holonomy corrections and the inverse triad
corrections. {\em Either } of these is sufficient to indicate a bounce
in the effective Hamiltonian picture. The same use of holonomies make
the Wheeler-DeWitt equation, a {\em difference equation}.

(3) The analysis at the level of effective Hamiltonian already indicates
(i) replacement of big bang by big bounce; (ii) natural prediction of an
inflationary (accelerated) phase; (iii) singularity resolution for more
general homogeneous models with curvature.

(4) There are at least three distinct ambiguity parameters: $\mu_0$
related to the fiducial length of the loop used in writing the
holonomies; $j$ entering in the choice of $SU(2)$ representation which
is chosen to be $1/2$ in the gravitational sector and some large value
in the matter sector; $l$ entering in writing the inverse powers in
terms of Poisson brackets. The first one was thought to be determined by
the area gap from the full theory. The $j = 1/2$ in the gravitational
Hamiltonian seems needed to avoid high order difference equation and
larger $j$ values are hinted to be problematic in the study of a three
dimensional model \cite{LargeJ}.  Given this, the choice of a high value
of $j$ in the matter Hamiltonian seems unnatural\footnote{For an
alternative view on using large values of $j$, see reference
\cite{MartinLattice}.}. Nevertheless the higher values of $j$ in
the matter sector allow for a larger semi-classical regime. The $l$ does
not play as significant a role.

(5) LQC being a constrained theory, it would be more appropriate if
singularity resolution is formulated and demonstrated in terms of
physical expectation values of physical (Dirac) operators i.e. in terms
of ``gauge invariant quantities''. This can be done at present with
self-adjoint constraint i.e. a symmetric ordering and for free, massless
scalar matter.

{\bf Physical quantities and Singularity Resolution:}
When the Hamiltonian is a constraint, at the classical level itself, the
notion of dynamics in terms of the `time translations' generated by the
Hamiltonian is devoid of any {\em physical} meaning.  Furthermore, at
the quantum level when one attempts to impose the constraint as
$\hat{H}|\Psi\rangle = 0$, typically one finds that there are no
solutions in the Hilbert space on which $\hat{H}$ is defined - the
solutions are generically distributional. One then has to consider the
space of all distributional solutions, define a new physical inner
product to turn it into a Hilbert space (the physical Hilbert space),
define operators on the space of solutions (which must thus act
invariantly) which are self-adjoint (physical operators) and compute
expectation values, uncertainties etc of these operators to make
physical predictions.  Clearly, the space of solutions depends on the
quantization of the constraint and there is an arbitrariness in the
choice of physical inner product. This is usually chosen so that a
complete set of Dirac observables (as deduced from the classical theory)
are self-adjoint. This is greatly simplified if the constraint has a
{\em separable} form with respect to some degree of freedom\footnote{A
general abstract procedure using group averaging is also available.  }.
For LQC (and also for the Wheeler-De Witt quantum cosmology), such a
simplification is available for a free, massless scalar matter:
$H_{\mathrm{matter}}(\phi, p_{\phi}) := \Case{1}{2} p_{\phi}^2
|p|^{-3/2}$. Let us sketch the steps schematically, focusing on the
spatially flat model for simplicity \cite{APSOne,APSTwo}. 
\begin{enumerate}
\item {\em Fundamental constraint equation:} 

The classical constraint equations is:
\begin{equation} \label{ClassHamConstraint}
 - \frac{6}{\gamma^2} c^2 \sqrt{|p|} + 8\pi G \ p_{\phi}^2 \ |p|^{-3/2}
   ~ = ~ 0 ~ = ~ C_{\mathrm{grav}} + C_{\mathrm{matter}}\ ;
\end{equation}
The corresponding quantum equation for the wave function, $\Psi(p,
\phi)$ is:
\begin{equation}
8\pi G \hat{p}_{\phi}^2 \Psi(p, \phi) ~ = ~ [\tilde{B}(p)]^{-1}
\hat{C}_{\mathrm{grav}} \Psi(p, \phi)~ ~ , ~ ~ [\tilde{B}(p)] \mbox{ is
eigenvalue of } \widehat{|p|^{-3/2}} \ ;
\end{equation}
Putting $\hat{p}_{\phi} = - i \hbar \partial_{\phi}$, $p :=
\Case{\gamma\lP^2}{6}\mu$  and $\tilde{B}(p) :=
(\Case{\gamma\lP^2}{6})^{-3/2} B(\mu)$, the equation can be written in a
separated form as \footnote{Our primary goal here is to compare the
classical geometry (Wheeler-DeWitt quantization) and quantum geometry
(loop quantization). Consequently, the gravitational constraint is
quantized in two different ways but {\em for simplicity}, the matter
sector is quantized in the usual Schrodinger way. In both quantizations,
$p_{\phi} = -i\hbar\partial_{\phi}$ and there is no $\phi$ dependence in
the matter Hamiltonian, so the two quantum Hamiltonian will have
identical expressions. However, using $\phi$ as labelling an `emergent
time' would be questionable.},
\begin{equation} 
\frac{\partial^2 \Psi(\mu, \phi)}{\partial \phi^2} ~ = ~
[B(\mu)]^{-1}\left[8\pi G \left(\frac{\gamma}{6}\right)^{3/2}\lP^{-1}
\hat{C}_{\mathrm{grav}}\right]\Psi(\mu, \phi) ~ := ~ -
\hat{\Theta}(\mu)\Psi(\mu, \phi).
\end{equation}
The $\hat{\Theta}$ operator for different quantizations is different.
For Schrodinger quantization (Wheeler-De Witt), with a particular factor
ordering suggested by the continuum limit of the difference equation,
the operator $\hat{\Theta}(\mu)$ is given by,
\begin{equation}
\hat{\Theta}_{\mathrm{Sch}}(\mu)\Psi(\mu, \phi) ~ = ~ - \frac{16\pi
G}{3}|\mu|^{3/2} \partial_{\mu} \left(\sqrt{\mu}\
\partial_{\mu}\Psi(\mu, \phi)\right)
\end{equation}
while for LQC, with symmetric ordering, it is given by,
\begin{eqnarray}
\hat{\Theta}_{\mathrm{LQC}}(\mu)\Psi(\mu, \phi) & = & - [B(\mu)]^{-1}
\left\{ C^+(\mu) \Psi(\mu + 4 \mu_0, \phi) + C^0(\mu) \Psi(\mu, \phi) +
\right. \nonumber \\ 
& & \left. \hspace{5.0cm} C^-(\mu) \Psi(\mu - 4 \mu_0, \phi) \right\} \
, \nonumber \\ 
C^+(\mu) & := & \frac{\pi G}{9 \mu_0^3} \left| ~ |\mu + 3\mu_0|^{3/2} -
|\mu + \mu_0|^{3/2} \right| \ , \\
C^-(\mu) & := & C^+(\mu - 4 \mu_0) ~ ~ , ~ ~ C^0(\mu) ~ := ~ - C^+(\mu)
- C^-(\mu) \ . \nonumber
\end{eqnarray}
Note that in the Schrodinger quantization, the $B_{\mathrm{Sch}}(\mu) =
|\mu|^{-3/2}$ diverges at $\mu = 0$ while in LQC,
$B_{\mathrm{LQC}}(\mu)$ vanishes for all allowed choices of ambiguity
parameters. In both cases, $B(\mu) \sim |\mu|^{-3/2}$ as $|\mu| \to
\infty$.

\item {\em Inner product and General solution:} 

The operator $\hat{\Theta}$ turns out to be a self-adjoint, positive
definite operator on the space of functions $\Psi(\mu, \phi)$ for each
fixed $\phi$ with an inner product scaled by $B(\mu)$. That is, for the
Schrodinger quantization, it is an operator on $L^2(\mathbb{R},
B_{\mathrm{Sch}}(\mu) d\mu)$ while for LQC it is an operator on
$L^2(\mathbb{R}_{\mathrm{Bohr}},
B_{\mathrm{Bohr}}(\mu)d\mu_{\mathrm{Bohr}}).$ Because of this, the
operator has a complete set of eigenvectors\footnote{For the Schrodinger
quantization, the explicit eigenfunctions are: $e_k(\mu) :=
\Case{|\mu|^{1/4}}{4\pi} e^{ik\ell n|\mu|}$ and the eigenvalues are:
$\omega^2(k) := \Case{2\kappa}{3}(k^2 + 1/16)$
\cite{APSTwo}.\label{WDWEigenfunctions}}:
\begin{equation} \label{EigenEq}
\hat{\Theta}e_k(\mu) = \omega^2(k) e_k(\mu)~~,~~ \langle
e_k|e_{k'}\rangle = \delta(k, k')~~,~~ k,\, k' \in \mathbb{R}. 
\end{equation}
Consequently, the general solution of the fundamental constraint
equation can be expressed as
\begin{equation} \label{GeneralSolution}
\Psi(\mu, \phi) ~ = ~ \int dk ~ \tilde{\Psi}_+(k) e_k(\mu)
e^{i\omega(k)\phi} + \tilde{\Psi}_-(k) \bar{e}_k(\mu)  e^{-i
\omega(k)\phi} \ .
\end{equation}
The orthonormality relations among the $e_k(\mu)$ are in the
corresponding Hilbert spaces. Different quantizations differ in the form
of the eigenfunctions, possibly the spectrum itself and of course
$\omega(k)$. In general, these solutions are {\em not} normalizable in
$L^2(\mathbb{R}_{\mathrm{Bohr}}\times\mathbb{R},
B_{\mathrm{Bohr}}(\mu)d\mu_{\mathrm{Bohr}}\times d\mu)$, i.e. these are
distributional.

\underline{Remarks:} 

(1) The $\hat{\Theta}$ operator acts in each of the superselected sector
(thanks to the difference equation structure) and these are {\em
separable}. Hence Dirac-$\delta$ appears in general when the label takes
{\em continuous} values.

(2) The group averaging can be seen as follows. Given any $f(\mu, \phi)$
one defines its {\em group average}, 
\[ \Psi_f(\mu, \phi) ~ := ~ \int_{-\infty}^{\infty}d\lambda \
e^{i\lambda \hat{C}_{\mathrm{tot}} }\ f(\mu, \phi) ~~;~~ 
\hat{C}_{\mathrm{tot}} ~ := ~ \frac{\partial^2}{\partial \phi^2} +
\hat{\Theta} ~~, 
\]
The eigenfunctions of the constraint operator are of a product form
thanks to separability,
\[ \frac{\partial^2}{\partial \phi^2}g_{\sigma}(\phi) ~ = ~ - {\sigma}^2
g_{\sigma}(\phi) ~~,~~ \hat{\Theta}e_{k}(\mu) = \omega^2(k) e_{k}(\mu)
~~,~~ g_{\sigma}(\phi) = e^{i {\sigma}\phi}, ~~ k, {\sigma} ~ \in ~
\mathbb{R} 
\]
Expanding the general function $f(\mu, \phi)$ in the eigenbasis of the
constraint operator,
\[
f(\mu, \phi) ~ := ~ \int dk \int d\sigma  \ A_{k, {\sigma}}\
g_{\sigma}(\phi)e_{k}(\mu) ~~,~~ 
\]
implies,
\begin{eqnarray}
\Psi_f(\mu, \phi) & = & \int dk \int d\sigma \ A_{k, {\sigma}}
g_{\sigma}(\phi)e_{k}(\mu)\delta({\sigma}^2 - \omega^2(k)) \nonumber \\ 
& = & \int_{-\infty}^{\infty} \frac{dk}{2|\omega(k)|} \left[
\left\{A_{k, \omega(k)} e^{i\omega(k) \phi} + A_{k, - \omega(k)}
e^{-i\omega(k) \phi} \right\} e_{\omega}(\mu) \right]
\end{eqnarray}
In the second equation above, we have carried out the integration over
$\sigma$ using $\delta(\sigma^2 - \omega^2(k)) = \Case{1}{2|\omega(k)|}
(\delta(\sigma - \omega(k)) + \delta(\sigma + \omega(k))$. Clearly the
group average of a general function reproduces precisely the general
solution given in equation (\ref{GeneralSolution}).

\item {\em Choice of Dirac observables:}

Since the classical kinematical phase space is 4 dimensional and we have
a single first class constraint, the phase space of physical states
(reduced phase space) is two dimensional and we need two functions to
coordinatize this space. We should thus look for two (classical) Dirac
observables: functions on the kinematical phase space whose Poisson
bracket with the Hamiltonian constraint vanishes on the constraint
surface. Specific values of these functions serve to label the physics
states. Thus physical observables are values of the Dirac observables.
Classically, the Dirac observables can be obtained as follows.

Our constraint is: $p_{\phi}^2/2 - \Case{3}{\kappa}\gamma^{-2} c^2 p^2
\simeq 0$. A Dirac observable is a function $f(\phi, p_{\phi}, c, p)$
whose Poisson bracket with the constraint vanishes on the constraint
surface.  We can describe the constraint surface by solving for $c$ as:
$c(\phi, p_{\phi}, p) := \pm \sqrt{\kappa/6} \gamma p_{\phi} p^{-1}$
(say).  Consequently it should suffice to consider the Dirac observables
to be a function of $(\phi, p_{\phi}, p)$ (we need only two independent
Dirac observables). Then the Dirac observables are defined by the
differential equation,
\[
\frac{\partial f}{\partial\phi} \pm \sqrt{2\kappa/3}\frac{\partial
f}{\partial\ell n p} = 0 ~~\Rightarrow~~ f = f(\zeta,
p_{\phi})~~,~~\zeta := \phi \mp \sqrt{\frac{3}{2\kappa}}\ell n p ~.
\]
Evidently, $f = p_{\phi}$ is a Dirac observable. For the second one, we
can choose any function of $\zeta$. A particularly convenient choice is:
$f(\zeta(\phi, p)) := \# \mathrm{exp}\{\mp \sqrt{2\kappa/3}\zeta\} = p \
\mathrm{exp}\{\mp \sqrt{2\kappa/3}(\phi - \phi_0)\}$. These Dirac
observables taking a particular value, say $p_*$, define curves in the
$(p, \phi)$ plane, $p(\phi) := p_*\mathrm{exp}\{\pm
\sqrt{2\kappa/3}(\phi - \phi_0)\}$ which are the classical solutions in
(\ref{ClassRelationalSoln}).

In the quantum theory, the notion of Dirac observable is that it is an
operator which maps solutions of the constraint to (other) solutions. We
already have the general solution in eq (\ref{GeneralSolution}) which is
obtained via unitary evolution (in $\phi$) from an initial
$\Psi_{\pm}(\phi_0, \mu)$. Hence a Dirac observable is constructed by
defining an operator on $\Psi_{\pm}(\phi_0, \mu)$ to generate a new
`initial state', and evolving the new state by the unitary operator,
thereby constructing a new solution of the constraint. This procedure is
followed for the two Dirac operators corresponding to $f_1(\phi,
p_{\phi}, p) := p_{\phi}$ and $f_2(\phi, p_{\phi}, p) := p \
\mathrm{exp}\{\mp \sqrt{2\kappa/3}(\phi - \phi_0)\}$. Notice that {\em
at} $\phi = \phi_0$, these become the functions: $p_{\phi}$ and $p$
respectively and they act on the wavefunctions as the differential
operator $-i\hbar\partial_{\phi}$ and as the multiplicative operator
$p$, respectively. Explicit construction is as follows.

Consider initial data, $\Psi_{\pm}(\mu, \phi_0)$, with the corresponding
solution is denoted by,
\[ \Psi(\mu, \phi) :=  e^{ i\sqrt{\hat{\Theta}}(\phi -
\phi_0)}\Psi_+(\mu, \phi_0) + e^{ - i\sqrt{\hat{\Theta}}(\phi -
\phi_0)}\Psi_-(\mu, \phi_0) \ .
\]
Generate {\em new initial data} via the actions of $\hat{p}_{\phi},
\widehat{|\mu|_{\phi_0}}$ as,
\begin{equation}
\widehat{|\mu|_{\phi_0}}\Psi_{\pm}(\mu, \phi_0) ~ := ~
|\mu|\Psi_{\pm}(\mu, \phi_0) ~ ~ , ~ ~ \hat{p}_{\phi} \Psi_{\pm}(\mu,
\phi_0) ~ := ~ \hbar \sqrt{\hat{\Theta}}\Psi_{\pm}(\mu, \phi_0)\ .
\end{equation}
Evolve these respectively, by $e^{\pm i\sqrt{\hat{\Theta}}(\phi -
\phi_0)}$. By construction, these are solutions of the constraints,
being of the form of (\ref{GeneralSolution}). Explicitly,
\begin{eqnarray}
\hat{p}_{\phi}\Psi(\mu, \phi) & := & e^{i\sqrt{\hat{\Theta}}(\phi -
\phi_0)} (\hbar\sqrt{\hat{\Theta}}) \Psi_+(\mu, \phi_0) + e^{-
i\sqrt{\hat{\Theta}}(\phi - \phi_0)} (- \hbar\sqrt{\hat{\Theta}})
\Psi_-(\mu, \phi_0) \nonumber \\
& = & -i \hbar\partial_{\phi}\Psi(\mu, \phi) \ , \\
\widehat{|\mu|_{\phi_0}}\Psi(\mu, \phi) & := &
e^{i\sqrt{\hat{\Theta}}(\phi - \phi_0)} |\mu| \Psi_+(\mu, \phi_0) + e^{-
i\sqrt{\hat{\Theta}}(\phi - \phi_0)} |\mu| \Psi_-(\mu, \phi_0)
\end{eqnarray}
Expectation values and uncertainties of these operators are used to
track the quantum `evolution'.

\item {\em Physical inner product:}

It follows that the Dirac operators defined on the space of solutions
are self-adjoint if we define a {\em physical inner product} on the
space of solutions as:
\begin{equation}
\langle\Psi|\Psi'\rangle_{\mathrm{phys}} ~ := ~ ``\int_{\phi = \phi_0}
d\mu B(\mu)" ~ ~ \bar{\Psi}(\mu, \phi) \Psi'(\mu, \phi) \ .
\end{equation}
Thus the eigenvalues of the inverse volume operator crucially enter the
definition of the physical inner product. For Schrodinger quantization,
the integral is really an integral while for LQC it is actually a sum
over $\mu$ taking values in a lattice. The inner product is independent
of the choice of $\phi_0$.

A complete set of physical operators and physical inner product has now
been specified and physical questions can be phrased in terms of
(physical) expectation values of functions of these operators.

\item {\em Semi-classical states:}

To discuss semi-classical regime, typically one {\em defines}
semi-classical states: physical states such that a chosen set of
self-adjoint operators have specified expectation values with
uncertainties bounded by specified tolerances. A natural choice of
operators for us are the two Dirac operators defined above. 

%It is easy
%to construct semi-classical states with respect to these operators
%\footnote{{\bf Caution:} Strictly speaking, the states given below are
%not in the domain of the relevant operators, so the equations have to be
%understood with care \cite{Madhavan}.}. 

To be definite, let us consider the Wheeler-De Witt quantization. An
arbitrary wavefunction at some given $\phi_0$ is expressed as an
integral over $k$ of the eigenfunctions $e_k(\mu)$ multiplied by some
function of $k$ and by a phase factor $e^{i\omega(k)\phi_0}$. The inner
product involves integral over $\mu$.  We can arrange to have a peak at
some particular value $|\mu_*|$ by introducing a factor of $-
\sqrt{3/2\kappa}|\mu_*|$ so that for large $k$, the $\mu-$integral will
be dominated by $\mu \sim \mu_*$. A large value of $k$ can be picked-up
by choosing the function of $k$ to be a suitable Gaussian. This
particular large value can be related to a desired $p_{\phi}^*$. Here
are the expressions: 
\begin{eqnarray}
\Psi_{\mathrm{semi}}(\mu, \phi_0) & := & \int dk e^{- \frac{(k -
k^*)^2}{2 \sigma^2}} e_k(\mu) e^{i \omega (\phi_0 - \phi^*)} \\ k^* & =
& -\sqrt{3/2\kappa} \hbar^{-1} p^*_{\phi} ~ ~ ,  ~ ~ \phi^* = \phi_0
-\sqrt{3/2\kappa} \ell n|\mu^*| \ .
\end{eqnarray}
The Gaussian allows the integrand to be approximated by 
\[
e_{k^*}(\mu) e^{i\omega(k^*)(\phi_0 - \phi_*)} \sim e^{i\ell n|\mu| k^*
+ i(\omega(k^*)\sqrt{3/2\kappa})\ell n|\mu_*|} \sim e^{ik^*\ell
n|\mu/\mu_*|}
\] 
where, in the last equality we used: $k^*$ is large and $\omega(k^*)
\simeq -\sqrt{2\kappa/3}\ k^*$ (see the footnote
\ref{WDWEigenfunctions}). The integral in the inner product will now
pick-up contribution from near $\mu \simeq \mu_*$. With these
observations, it is easy to verify that the `initial' semiclassical wave
function given above gives $\langle \hat{p}_{\phi}\rangle = p_{\phi}^*$
and $\langle \widehat{|\mu|_{\phi_0}}\rangle  = \mu_*$. The initial
semiclassical wavefunction evolves into $\Psi_{\mathrm{semi}}(\mu,
\phi)$ which is same as the initial wave function with $\phi_0 \to \phi$
\footnote{In the above heuristic reasoning for the form of the
semiclassical state, we have glossed over some technical issues such as
whether the states exhibited are in the domain of the $\hat{\Theta}$
operator which requires carefully stipulating conditions on the function
of $k$ (which has been taken to be a Gaussian). These are too technical
to go into here and are not expected to affect the corresponding
discussion for LQC. A discussion of these issues may be seen in
\cite{Madhavan}.}.

For LQC, the $e_k(\mu)$ functions are different \cite{APSTwo} and the
physical expectation values are to be evaluated using the physical inner
product defined in the LQC context.

\item {\em Evolution of physical quantities:}

We have now the physical wave function, evolved from
$\Psi_{\mathrm{semin}}$. Since it retains the form of the initial
wavefunction, the $k$ integral can be approximated as before and thus
will lead to same expectation value for $\hat{p}_{\phi}$ for all $\phi$.
For the expectation value of $\langle
\widehat{|\mu|_{\phi_0}}\rangle(\phi)$, the $\mu$ integral will be
saturated by the new phase, $\phi - \phi_0 + \sqrt{3/2\kappa}\ \ell
n|\mu/\mu_*| \simeq 0$. And this we recognize as precisely the solution
(\ref{ClassRelationalSoln}). 

Thus, in the WdW quantization, the classical relational evolution, $p =
p(\phi)$ is reproduced by the expectation values of the Dirac
observables. 

{\em Exercise: compute/estimate the uncertainties of the Dirac
observables in the semiclassical state given above.}.

\item {\em Resolution of Big Bang Singularity:}

A classical solution is obtained as a curve in $(\mu, \phi)$ plane,
different curves being labelled by the points ($\mu^*, \phi^*$) in the
plane. The curves are independent of the constant value of
$p^{*}_{\phi}$ These curves are already given in
(\ref{ClassRelationalSoln}).

Quantum mechanically, we first select a semi-classical solution,
$\Psi_{\mathrm{semi}}(p_{\phi}^*, \mu^*: \phi)$ in which the expectation
values of the Dirac operators, at $\phi = \phi_0$, are $p_{\phi}^*$ and
$\mu^*$ respectively. These values serve as labels for the
semi-classical solution. The former one continues to be $p_{\phi}^*$ for
all $\phi$ whereas $\langle\widehat{|\mu|_{\phi_0}}\rangle(\phi) =:
|\mu|_{p_{\phi}^*, \mu^*}(\phi)$, determines a curve in the $(\mu,
\phi)$ plane. (We determined this curve above, using stationarity of the
phase for the Schrodinger quantization). In general one expects this
curve to be different from the classical curve in the region of small
$\mu$ (small volume). This is what happens for the loop quantized
theory.

The result of the computations is that Schrodinger quantization, the
curve $ |\mu|_{p_{\phi}^*, \mu^*}(\phi)$, does approach the $\mu = 0$
axis asymptotically. However for LQC, the curve {\em bounces away} from
the $\mu = 0$ axis. In this sense -- and now inferred in terms of
physical quantities -- the Big Bang singularity is resolved in LQC.  It
also turns out that for large enough values of $p^*_{\phi}$, the quantum
trajectories constructed by the above procedure are well approximated by
the trajectories by the effective Hamiltonian. All these statements are
for semi-classical solutions which are peaked at large $\mu_*$ at late
times.
\end{enumerate}

Two further features are noteworthy as they corroborate the suggestions
from the effective Hamiltonian analysis. 

First one is revealed by computing expectation value of the matter
density operator, $\rho_{\mathrm{matter}} :=
\Case{1}{2}\widehat{(p_{\phi}^*)^2 |p|^{-3}}$, at the bounce value of
$|p|$. It turns out that this value is sensitive to the value of
$p_{\phi}^*$ and can be made arbitrarily {\em small} by choosing
$p_{\phi}^*$ to be {\em large}. Physically this is unsatisfactory as
quantum effects are {\em not} expected to be significant for matter
density very small compared to the Planck density. This is traced to the
quantization of the gravitational Hamiltonian, in particular to the step
which introduces the ambiguity parameter $\mu_0$. A novel solution
proposed in the ``improved quantization'', removes this undesirable
feature. 

The second one refers to the role of quantum modifications in the
gravitational Hamiltonian compared to those in the matter Hamiltonian
(the inverse volume modification or $B(\mu)$). The former is much more
significant than the latter. So much so, that even if one uses the
$B(\mu)$ from the Schrodinger quantization (i.e. switch-off the inverse
volume modifications), one still obtains the bounce.  So bounce is seen
as the consequence of $\hat{\Theta}$ being different and as far as
qualitative singularity resolution is concerned, the inverse volume
modifications are {\em un-important}. As the effective picture (for
symmetric constraint) showed, the bounce occurs in the classical region
(for $j = 1/2$) where the inverse volume corrections can be neglected.
For an exact model which seeks to understand why the bounces are seen,
please see \cite{MartinExact}.

{\bf Improved Quantization:}
The undesirable features of the bounce coming from the classical region,
can be seen readily using the effective Hamiltonian, as remarked
earlier. To see the effects of modifications from the gravitational
Hamiltonian, choose $j = 1/2$ and consider the Friedmann equation
derived from the effective Hamiltonian leading to the effective energy
density (\ref{EffDensity}), with matter Hamiltonian given by
$H_{\mathrm{matter}} = \Case{1}{2}p^2_{\phi} |p|^{-3/2}$. The positivity
of the effective density implies that $p \ge p_*$ with $p_*$ determined
by vanishing of the effective energy density: $\rho_* := \rho_{cl}(p_*)
= (\Case{8\pi G \mu_0^2 \gamma^2}{3} p_*)^{-1}$. This leads to $|p_*| =
\sqrt{\Case{4 \pi G \mu_0^2 \gamma^2}{3}} |p_{\phi}|$ and $\rho_* =
\sqrt{2} (\frac{8\pi G \mu_0^2 \gamma^2}{3})^{-3/2} |p_{\phi}|^{-1}$.
One sees that for large $|p_{\phi}|$, the bounce scale $|p_*|$ can be
large and the maximum density -- density at bounce -- could be small.
Thus, {\em within the model}, there exist a possibility of seeing
quantum effects (bounce) even when neither the energy density nor the
bounce scale are comparable to the corresponding Planck quantities and
this is an undesirable feature of the model. This feature is independent
of factor ordering as long as the bounce occurs in the classical regime. 

One may notice that {\em if} we replace $\mu_0 \to \bar{\mu}(p) :=
\sqrt{\Delta/|p|}$ where $\Delta$ is a constant, then the effective
density vanishes when $\rho_{cl}$ equals the critical value
$\rho_{\mathrm{crit}} := (\Case{8\pi G \Delta \gamma^2}{3})^{-1}$, which
is independent of matter Hamiltonian. The bounce scale $p_*$ is
determined by $\rho_* = \rho_{\mathrm{crit}}$ which gives $|p_*| =
(\Case{p_{\phi}^2}{2\rho_{\mathrm{crit}}})^{1/3}$. Now although the
bounce scale can again be large depending upon $p_{\phi}$, the density
at bounce is always the universal value determined by $\Delta$. This is
a rather nice feature in that quantum geometry effects are revealed when
matter density (which couples to gravity) reaches a universal, critical
value regardless of the dynamical variables describing matter. For a
suitable choice of $\Delta$ one can ensure that a bounce always happens
when {\em the energy density} becomes comparable to the Planck density.
In this manner, one can retain the good feature (bounce) even for $j =
1/2$ thus ``effectively fixing'' an ambiguity parameter and also trade
another ambiguity parameter $\mu_0$ for $\Delta$. This is precisely what
is achieved by the ``improved quantization'' of the gravitational
Hamiltonian \cite{APSThree}.

The place where the quantization procedure is modified is when one
expresses the curvature in terms of the holonomies along a loop around a
``plaquette''. One shrinks the plaquette in the limiting procedure.  One
now makes an important departure: the plaquette should be shrunk only
till the physical area (as distinct from a fiducial one) reaches its
minimum possible value which is given by the area gap in the known
spectrum of area operator in quantum geometry: $\Delta = 2\sqrt{3}\pi
\gamma G\hbar$. Since the plaquette is a square of fiducial length
$\mu_0$, its physical area is $\mu_0^2|p|$ and this should set be to
$\Delta$. Since $|p|$ is a dynamical variable, $\mu_0$ cannot be a
constant and is to be thought of a function on the phase space,
$\bar{\mu}(p) := \sqrt{\Delta/|p|}$. Thus we need to define an operator
corresponding to the classical expression: $h_f :=
\mathrm{exp}(i\Case{1}{2}f(p)c)$, we have taken a general function
$f(p)$. This is little non-trivial since there is no $\hat{c}$ operator
and $c, p$ are conjugate variables.

Observe that the usual holonomy operator effects a shift in the argument
of eigenstates of the triad operator and formally the operator looks
like exp$(\nu \Case{d}{d\mu})$ i.e. it effects the action of a {\em
finite} diffeomorphism generated by a vector field on the wavefunction. 
We will take this as a guiding principle. 

Let $\Phi_f$ denote a diffeomorphism effecting a {\em unit parameter
shift} along the integral curve of the vector field
$f(\mu)\Case{d}{d\mu}$ and $\Phi_f^*$, the corresponding pull-back map.
We define $\widehat{h_f}\Psi(\mu) := [\Phi^*_f(\Psi)] :=
\Psi(\Phi_f(\mu))$. As argued above, for a constant function, this
reduces to the usual action (\ref{HolonomyFluxRepren}). It can be
checked directly that this action is also {\em unitary} in the
kinematical Hilbert space: $(\Phi, \Psi) = \sum_{\mu} \Phi^*(\mu)
\Psi(\mu)$ where the sum is over a countable set (this follows from
$|\Psi\rangle := \sum_{\mu \in \mathrm{countable~subset} \subset
\mathbb{R}} \Psi(\mu)|\mu\rangle$)\footnote{For a comparison in
Schrodinger quantization, see remarks in \cite{APSTwo}.}.

To compute a unit parameter shift due to the diffeomorphism generated by
$f(\mu)$, solve the equation
\begin{equation} \label{AffineParameter}
\int_{\mu}^{\mu'} \frac{dx}{f(x)} ~=~ \int_v^{v + 1} dv = 1
\end{equation}
This will give $\mu' := \Phi_f(\mu)$. 

For the specific choice of $f(p) := \bar{\mu}(p) := \sqrt{\Delta}
|p|^{-1/2},\  \Delta := \gamma\sqrt{3}\lP^2/4,\  p(\mu) =
\gamma\lP^2\mu/6$, one gets,
\begin{eqnarray}
\sqrt{\frac{\Delta}{|p|}} & = & \sqrt{\frac{3\sqrt{3}}{2}}|\mu|^{-1/2}
~=: ~ f(\mu) \hspace{2.0cm} \Rightarrow \nonumber \\
\mathrm{sgn}(\mu')|\mu'|^{3/2} & = & \mathrm{sgn}(\mu)|\mu|^{3/2} +
K^{-1} ~~\hspace{2.0cm} K := \frac{2}{3}\sqrt{\frac{2}{3\sqrt{3}}} \\
\left[\widehat{e^{f(\mu)\frac{d}{d\mu}}}\Psi\right](\mu) & := &
\Psi(\mu'). \nonumber
\end{eqnarray}
It is evident from the above that if we define $v := K
\mathrm{sgn}(\mu)|\mu|^{3/2}$, then the middle eqn reads: $v' = v + 1$.
This suggests that we use $|v\rangle$ as a basis instead of
$|\mu\rangle$. Apart from the constant $K$, and the sgn, $v$ is related
to the eigenvalue of the volume operator $|p|^{3/2}$. Note that $v$ as a
function of $\mu$ is one-to-one and on-to. Using the $h_{\bar{\mu}}$
operator and a basis labelled by volume eigenvalues, the Hamiltonian
constraint is defined and difference equation is obtained as before. The
relevant expressions are:
\begin{eqnarray}
v & := & K \mathrm{sgn}(\mu) |\mu|^{3/2}~ ~; \label{VDefn}\\
\hat{V}|v\rangle & = &
\left(\frac{\gamma}{6}\right)^{3/2}\frac{\lP^3}{K}\ |v|~ |v\rangle~,~ \\
\widehat{e^{ik\Case{\bar{\mu}}{2}c}}\Psi(v) & := & \Psi(v + k)~, \\
\left.\widehat{|p|^{-1/2}}\right|_{j = 1/2, l = 3/4}\Psi(v) & = &
\frac{3}{2}\left(\frac{\gamma\lP^2}{6}\right)^{-1/2}
K^{1/3}|v|^{1/3}\left| |v + 1|^{1/3} - |v - 1|^{1/3}\right| \Psi(v) \\
B(v) & = & \left(\frac{3}{2}\right)^{3/2} K |v| \left| |v + 1|^{1/3} -
|v - 1|^{1/3}\right|^3 \label{ImprovedInverse}\\
\hat{\Theta}_{\mathrm{Improved}}\Psi(v, \phi) & = & - [B(v)]^{-1}
\left\{ C^+(v) \Psi(v + 4, \phi) + C^0(v) \Psi(v, \phi) + \right.
\nonumber \\ 
& & \left. \hspace{5.0cm} C^-(v) \Psi(v - 4, \phi) \right\} \ , \\ 
C^+(v) & := & \frac{3 \pi K G}{8} |v + 2| \left| ~ |v + 1| - |v + 3|
\right| \ , \\
C^-(v) & := & C^+(v - 4) ~ ~ , ~ ~ C^0(v) ~ := ~ - C^+(v) - C^-(v) \ . 
\end{eqnarray}

Thus the main changes in the quantization of the Hamiltonian constraint
are: (1) replace $\mu_0 \to \bar{\mu} := \sqrt{\Delta/|p|}$ in the
holonomies; (2) {\em choose} symmetric ordering for the gravitational
constraint; and (3) {\em choose} $j = 1/2$ in both gravitational
Hamiltonian and the matter Hamiltonian (in the definition of inverse
powers of triad operator). The ``improvement'' refers to the first
point. This model is singularity free at the level of the fundamental
constraint equation (even though the leading coefficients of the
difference equation do vanish, because the parity symmetry again saves
the day); the densities continue to be bounded above -- and now with a
bound independent of matter parameters; the effective picture continues
to be singularity free and with undesirable features removed and the
classical Big Bang being replaced by a quantum bounce is established in
terms of {\em physical} quantities.

{\em There is yet another spin on the story of singularity resolution!}
\subsection {Madhavan Quantization \cite{Madhavan}:}
The improved quantization scheme works primarily through the holonomy
corrections, so much so that even if the inverse volume corrections in
the matter are turned-off by hand, the singularity resolution continues.
Madhavan works within the same kinematical Hilbert space of LQC but
treats the Hamiltonian constraint differently, exploiting its specific,
simple form for the massless scalar matter. In his quantization of the
Hamiltonian constraint, it is the inverse volume corrections that are
responsible for singularity resolution (also in terms of physical
quantities) and holonomy corrections are by-passed completely.

He observes that the classical Hamiltonian constraint
(\ref{ClassHamConstraint}), is quadratic in $c$ and is of the form of
difference of two squares. It can therefore be written as a product,
\[ 
C_{\mathrm{tot}} = - C_+C_- ~~,~~ C_{\pm} := - \sqrt{\frac{6}{\gamma}} c
|p|^{1/4} \pm \sqrt{\kappa} \frac{p_{\phi}}{|p|^{3/4}}
\]
The $C_{\pm}$ are linear in $c$. Since to define physical Hilbert space,
a general procedure is to average over the group generated by the
constraint, and this involves exponentiation of the constraint, one can
directly define these operators which involve factors of the same form
as the $h_f$ operators of the improved quantization! 

The key differences in Madhavan quantization are: (1) The Hamiltonian
constraint is regulated differently from the analogue of LQG and as a
consequence, {\em there are no holonomy corrections} (sin$^2(c)$).
However, the inverse triad corrections can be incorporated in the
definition of $\hat{C}_{\pm}$ through the $\hat{p}_{\phi}$ term. One
again uses the volume eigenvalues basis for the inverse triad
definition; (2) the physical states are constructed directly by group
averaging using the well-defined unitary operators, $e^{i\alpha
\hat{C}_{\pm}}$, consequently {\em there is no difference/differential
equation to be solved}; (3) One of the Dirac observables, $p_{\phi}$ is
the same but another one is somewhat different. Nevertheless, {\em
classical solutions} can be derived from their expectation values; (4)
The issue of independence from the fiducial cell (discussed in the next
subsection) is also addressed differently.

The results are: (i) without inverse triad modifications, the classical
(singular) solution is recovered; (ii) with inverse triad modifications,
there {\em is} extension of the solution past the classical singularity
with the energy density remaining bounded all through, making the
extension non-singular. {\em There is no bounce, but a regular
extension!}

Although Madhavan's procedure of bypassing the holonomy corrections
completely, is tied to the particular form of the constraint of the
isotropic model (and hence may not extend to other models), it does
demonstrate the possibility that there are inequivalent ways of
constructing {\em physical Hilbert space and observables} starting from
the {\em same kinematical structures}.  Secondly, singularity resolution
{\em need not} be seen only as a classical/quantum {\em bounce}, a
regular extension is also a distinct possibility. 

More details should be seen in \cite{Madhavan}.

%\subsection{Fiducial cell independence issue in spatially flat models}
\subsection{Role of the Fiducial cell in spatially flat models}
\label{CellIndependence}
Recall that in the description of spatially homogeneous and isotropic
models one begins with a metric of the form (\ref{FRWMetric}). The
spatial metric is a metric with spatially constant (but possibly time
dependent) curvature. This is conveniently taken to be a time dependent
scaling of a fixed {\em co-moving metric} with corresponding {\em
co-moving coordinates}. Although not strictly necessary, let us assign
length dimension to the co-moving coordinates and take the scale factor
to be dimensionless. For non-flat models the co-moving metric can be
normalized to have the Ricci scalar to be $\pm 1$ in appropriate units
(Ricci scalar has dimensions of (length)$^{-2}$). Note that this is a
{\em local} condition, and by homogeneity, holds everywhere on the
spatial manifold. It is independent of the {\em size} of the spatial
manifold. For flat models, such a normalization of the co-moving metric
is not possible.  In this case, there is an arbitrariness in the {\em
definition} of the scale factor.  Clearly, by focusing only on those
quantities which are invariant under constant scaling of the scale
factor, eg $\dot{a}/a, \ddot{a}/a$, the energy density etc we can
obviate the need for choosing a co-moving metric/coordinates.  The
equations of motion - the Friedmann equation and the
Raychaudhuri/continuity equation - reflect this feature. 

However, spatial flatness, homogeneity and isotropy also implies
existence of (global) Cartesian coordinates with a metric $g^0_{ij} =
\delta_{ij}$ with the coordinate differences giving distances in the
chosen unit of length. This unit is arbitrary, but also determines the
unit of time by putting speed of light to be one. Change in this unit
results in an overall scaling of the {\em space-time metric} but does
not affect the scale factor. The scale factor is now unambiguously
identified and co-moving coordinates and metric are also fixed.

Construction of a quantum theory of the scale factor degree of freedom
(and matter homogeneous degrees of freedom) begins with a {\em four
dimensional action} principle restricted to homogeneous modes of the
fields. The action contains a spatial integration which is {\em
divergent} for spatially flat models, thanks to homogeneity. To have a
well defined phase space formulation, we need to regulate this
divergence. This is done by introducing an {\em arbitrarily chosen
fiducial cell}, specified by finite ranges of the co-moving coordinates
(thus having a finite co-moving volume $V_0$) and restricting the
integrations to this cell.  Note that this {\em need and the freedom} in
the choice of the fiducial cell arises strictly due to the need for an
action formulation for the full theory and the assumption of spatial
homogeneity\footnote{The action formulation (Lagrangian or Hamiltonian)
in turn is required for a quantum theory. The classical theory needs
only equations of motion which are independent of any cell.}. All
subsequent computations will carry a dependence on this cell, either
explicitly or implicitly. In the end, this dependence is to be removed
by taking a suitable limit $V_0 \to \infty$\footnote{The limit $V_0 \to
\infty$ can be viewed as a convenient way to pick-out $V_0-$independent
terms and/or could also be heuristically motivated by noting that the
definition of homogeneity identifies the spatial manifold with the group
manifold and this group manifold is $\mathbb{R}^3$ for the present
case.}.  Precisely at what stage and how should one take the limit?

In the canonical formulation of the full theory, the fiducial volume,
$V_0$, appears in the symplectic structure. This can however be absorbed
away by redefining the canonical coordinates ($\tilde{c}, \tilde{p} \to
c, p$). This makes the canonical coordinate $p$ to have dimensions of
(length)$^2$. Note that the physical volume of the cell is $a^3(t)\times
V_0$ is now directly given by $|p|^{3/2}$.  {\em Apparently}, there is
no reference to the fiducial cell any more in the model.  However this
is not so. The $|p|^{3/2}$ is the physical volume {\em of the fiducial
cell}. All subsequent computations, whether classical or quantum, done
using the ($c, p$) variables\footnote{During the process of {\em loop
quantization}, fiducial scales could appear again eg through the
holonomies along edges. However as explained in the footnote
\ref{SubCells}, the $V_0$ disappear.} have {\em no explicit reference}
to the cell. For example, the classical solution obtained in terms of
phase space trajectory, (eqn. \ref{ClassRelationalSoln}), does not
depend on $V_0$. 

As discussed above, the subsequent steps in the quantization, do not
introduce any further dependence on the fiducial cell.  It is no where
in sight even in the computation of the phase space trajectory
(expectation values of the Dirac observables).  These trajectories of
course differ from the corresponding classical trajectories. The problem
of Big Bang singularity is however phrased in the framework of
space-time geometry, specifically, in terms of backward evolution of the
{\em scale factor}. So we need to {\em transcribe} the phase space
trajectories (computed in terms of expectation values of Dirac
observables) into evolution of the space-time geometry i.e. the scale
factor. At this stage, a scale factor (and an explicit reference to the
fiducial cell) is re-introduced via the triad variable as, $a := \xi
\sqrt{p}$ where $\xi$ has dimensions of (length)$^{-1}$ and can be
identified with the fiducial volume: $\xi^{-1} = V_0^{1/3}$ (since
$p^{3/2}$ is the physical volume {\em of the fiducial cell}). The phase
space evolution then gives $a(t)$. The scale factor evolution so deduced
could have some dependence on $\xi$.  After taking the limit $\xi \to 0$
($V_0 \to \infty$), the evolution that survives, is the prediction of
the quantum theory, Whether or not this evolution is {\em singularity
free (i.e. all physical quantities remain bounded through out the
evolution)} is the central question of interest. Since the phase space
curves are inferred from the expectation values, the states in which
these are computed are also important to specify. These are expected to
be computed in physical states peaked on large volume $p \gg \lP^2$ and
small energy density (say), corresponding to a classical regime. The
singularity free evolution is required to hold for {\em all} such
states. 

The classical evolution given in eqn (\ref{ClassRelationalSoln})
provides an example of this transcription. We see that $\xi^{-2}$
cancels out from both $p$ and $p_*$ and the classical evolution of the
scale factor is {\em independent} of the fiducial cell as it should be.
The LQC computed solution for $p$, always shows a bounce, is not very
explicitly expressed and also contains an implicit dependence on the
fiducial cell.  It matches pretty closely with the classical solution in
the large $p$ regime and therefore could be expected to be $V_0$
independent in these regimes. This removes the cell dependence in the
initial condition and the question boils down to whether the bounce
feature and value of $p$ at the bounce, is independent of $V_0$. 

The APS investigations\cite{APSOne, APSTwo} found that, in the
$\mu_0-$scheme, a bounce can occur even for low values of energy density
something which is not exhibited by the observed isotropic universe.
Furthermore, the energy density at the bounce - which is a physical
observable - has a $V_0-$dependence. So the quantization scheme has some
problems. What exactly does this mean? 

Note that this does not necessarily mean that there is any mathematical
inconsistency in the process of quantization. However, the constructed
quantum theory should agree with GR for low energy densities (i.e. have
acceptable infra-red behaviour) and hopefully also {\em imply a
non-singular evolution}. It is possible that it may fail this
expectation. APS analysis concludes that the $\mu_0-$scheme with
symmetric ordering fails this test. 

{\em In retrospect}, this failure could have been inferred in the
following way.  The earlier methods of analysis were based on WKB
approximation and effective Hamiltonians\cite{EffHam} derived from it.
This allowed us to encode quantum modifications in terms of effective
density and effective pressure, defined by computing the left hand sides
of the Friedmann and the Raychaudhuri equations using the effective
Hamiltonian. In these papers, the scale factor was introduced by setting 
%
%In the first paper (eqns. 22 - 29), only the inverse triad corrections
%are incorporated while in the second paper (eqns 23 - 25) the holonomy
%corrections are also incorporated. In these papers the scale factor is
%
$p := a^2/4$ (dimensionful) and therefore still refers (implicitly) to
fiducial cell. To make the fiducial cell explicit, replace this scale
factor as $a \to a\xi^{-1}$. The expressions for the energy density
(say) can be transcribed in terms of the (dimensionless) scale factor
and $\xi$.  As explained above, prediction of the quantum theory for the
scale factor evolution is obtained by taking the limit $V_0 \to \infty
(\xi \to 0)$. Note that this is now done at the level of equations as
opposed to at the level of individual curves which need initial
conditions to be chosen (which we have argued to be independent of
$V_0$). 

Recall the effective density given in eqn.[\ref{EffDensity}],
\begin{eqnarray} \label{HolonomyCorrection}
\rho_{\mathrm{eff}} & = &
\left(\frac{H_{\mathrm{matter}}}{p^{3/2}}\right)\left[1 -
\frac{2\kappa\mu_0^2\gamma^2}{3}p
\left(\frac{H_{\mathrm{matter}}}{p^{3/2}}\right)\right] ~~~,
~~~\mathrm{where} \nonumber \\
H_{\mathrm{matter}} & = & \frac{1}{2} p_{\phi}^2 \left\{(2jp_0)^{-3/2}
\left(F_{\ell}(q)\right)^{\Case{3}{2(1 - \ell)}} \right\} ~~,~~ q :=
\frac{p}{2jp_0}
\end{eqnarray}
The square bracket contains the modification implied by {\em holonomy
corrections}\footnote{In the first paper of \cite{EffHam}, the effective
density contained only the leading terms of the holonomy corrections
which have been summed up in the second paper.}. The inverse volume
corrections are contained in the matter Hamiltonian (see
eqns.(\ref{InvTriad}, \ref{Ffunction})). 

Consider first the inverse volume corrections.  Observe that with $p \to
a^2\xi^{-2}/4$\ , \ $q \sim a^2\xi^{-2} \gg 1$ in the limit $\xi \to 0$.
The limiting form of $F_{\ell}$ then implies that $p^{-3/2}
H_{\mathrm{matter}} \sim p_{\phi}^2\xi^6a^{-6}$. For massless scalar
matter, the classical equation of state has $P/\rho = 1$ and hence the
classical density behaves as $\sim a^{-6}$. Thus $p^2_{\phi}\xi^6$ must
be a constant for any particular solution. For the leading term to give
the classical evolution, we have to take the limit $\xi \to 0$ {\em
along with} $p_{\phi} \to \infty$ keeping $p_{\phi}\xi^3$ a {\em
constant} specifying a particular initial condition.  This understood,
the $p^{-3/2}H_{\mathrm{matter}}$ factors go over to the {\em cell
independent} classical density plus corrections down by $q^{-2} \to 0$
in the limit. Thus, inverse volume {\em corrections}, simply vanish when
$V_0 \to \infty$ is imposed. 

Now consider the holonomy corrections. By the same logic as above, the
holonomy corrections, second term in the square bracket in the effective
density expression, goes as $\xi^{-2} \to \infty$!. This is clearly
unacceptable. Thus, in the $\mu_0$ scheme of quantization, the inverse
volume modifications do not survive the limit while the holonomy
modifications give an inconsistency and neither shed any light on the
singularity resolution issue. 

True as these features are, they are not immediately conclusive to look
for alternative quantizations because the fault may be with the WKB
approximation and the corresponding effective Hamiltonians. For instance
if one took the effective density from the first paper of \cite{EffHam},
the holonomy corrections would also be down by inverse powers of $q$ and
would vanish which is okay for the classical limit but the extrapolation
to the quantum regime is unreliable since WKB is unreliable at turning
points. Perhaps, physical level computations would clarify the issue.
This is indeed the case. With the physical level computations, APS
results show the unacceptability of the $\mu_0-$scheme while in
Madhavan's approach, with no holonomy corrections, the inverse volume
corrections would simply vanish by the argument given above.  Both APS
and Madhavan have also suggested ways out.

The APS analysis discussed above shows that the $\mu_0 \to \bar{\mu}(p) =
\sqrt{\Delta}\lP/\sqrt{p}$ substitution in the holonomies used in
replacing the $c$ variable, suffices to obtain a non-singular evolution
with good infra-red behaviour.  It implies that the deviations from
classical evolution (eg close to the bounce) occur when the energy
density reaches a universal, maximal value. This substitution also
renders the inverse triad correction from the matter sector highly
suppressed\footnote{From eqn. \ref{Mu1GenVals}, with $\lambda = 1$, one
sees that the corrections go as $v_1^{-2} \sim \lP^6 p^{-3} \sim \lP^6
(\xi)^6 a^{-6} \to 0$ as $\xi \to 0$.}.

Madhavan suggests that along with the APS suggested substitution, one
should also introduce a multiplicative parameter $\lambda$ as $\mu_0 \to
\lambda\bar{\mu}(p)$. Since only corrections that drive the quantum
modifications are the inverse volume correction and these go as
$(\lambda/v_1)^2 \sim (\lambda\xi^3\lP^3/a^3)^2$ (see eqn
\ref{Mu1GenVals}), $\xi$ independence is achieved by choosing $\lambda
\sim (\bar{\xi}/\xi)^{3}$ where the new dimensionful parameter
$\bar{\xi}$ is supposed to reflect a scale provided by an underlying LQG
state supporting the homogeneity approximation. For details, please see
\cite{Madhavan}.

To summarize: For the spatially flat, isotropic models, let us choose
the Cartesian coordinates with the standard Euclidean metric for the
spatial slice and choose proper time as the time coordinate so that the
space-time metric takes the form (\ref{FRWMetric}). The scale factor is
now specified unambiguously. For constructing a quantum theory, we need
to choose a regulator cell with volume $V_0$. While the cell dependence
can be hidden by choosing scaled variables, it manifests again because
quantum computed evolution must be transcribed in terms of the scale
factor evolution. This is necessary because the classical Big Bang
singularity is understood as a singular evolution of the scale factor so
its resolution lies in making the evolution non-singular. 

The scale factor evolution can be cast in the form of the Friedmann
equation with possible deviations from classical evolution, encoded in
the effective density. A prediction of the quantum theory is the
surviving correction terms after taking the limit $V_0 \to \infty$. A
quantum theory could be understood to have resolved the Big Bang
singularity if the surviving evolution is non-singular.  Cell
independence of quantum corrections automatically implies that
non-trivial limit exists. Not every quantization scheme passes this
test. 

There are two types of corrections - the holonomy corrections and the
inverse triad corrections. These have different properties in the limit.
The APS quantization with $\mu_0-$scheme implies that holonomy
corrections dominate and lead to unphysical implications. These are
cured by the $\bar{\mu}-$scheme. The Madhavan quantization scheme, even
with the $\bar{\mu}$ substitution in the inverse volume definition,
these corrections again vanish unless additional $\lambda$ parameter is
introduced.  In either case, extra ingredients (scales) have to be
`imported' to get non-trivial results. Both the schemes have ingredients
(role played by the area operator in the APS scheme\footnote{The logic
used to motivate the role of area operator, is also extended to other
Bianchi models such that when isotropy is imposed, the
$\bar{\mu}$-scheme of isotropic model is recovered back\cite{AWE}.} and
the specific form of constraint in the Madhavan scheme) which do not
have counterparts in the full theory as it is understood at present. 

So far the discussion has been within the context of full theory being
classically reduced directly to a homogeneous and isotropic model. In
the next subsection, we briefly discuss how homogeneity and isotropy can
be viewed from within a particular quantized inhomogeneous model. 
\subsection{A View from Inhomogeneity}
\label{LatticeView}
To keep the flow of the in focus, basic details of the inhomogeneous
model are given in the appendix 5.4.

The fundamental change in the way homogeneity is viewed, is that it is a
property exhibited by a {\em state} of an inherently (spatially)
inhomogeneous model. For definiteness, a lattice model with a lattice
spacing $\ell_0$ is taken. This allows for {\em states} of the model
which can be considered as {\em homogeneous on a certain scale}, eg
$\ell_0 N^{1/3}$. This also allows the fields to be restricted to be
periodic on this scale. Thus what is fundamental is the lattice spacing
$\ell_0$ below which it makes no sense to consider inhomogeneity (or
inhomogeneities are not probed) and a scale $N^{1/3} \ell_0$ provided by
a state of the model. Let us call the former as the {\em micro-}scale
and the latter as the {\em macro-}scale. The fundamentally isotropic
model refers to the macro-scale. The fiducial cell of the isotropic
model is determined by these two scales with fiducial volume given by,
$V_0 := N \ell_0^3$. Notice that in this view, $V_0$ (or $N$) is a
property of a quantum state and there is {\em no reason} to contemplate
a limit $V_0 \to \infty$.

Now all quantum effects due to inverse volume and holonomies, arise at
the micro-scale. To see how these translate or correspond to the quantum
effects seen in the fundamentally isotropic model, let us begin by
identifying variables.

In the lattice model, isotropic connection is defined by $\tilde{k}(x) =
\tilde{c}, \forall\ x, I$. Identifying this constant value with
$\tilde{c}_{\mathrm{iso}}$ and comparing the basic link holonomies with
the holonomies of the isotropic model implies, 
\begin{equation} \label{CIdentification}
c_{\mathrm{iso}} ~:=~ V_0^{1/3}\tilde{c}_{\mathrm{iso}} ~=~ V_0^{1/3}
\tilde{k}_I ~:=~ V_0^{1/3} \ell_0^{-1} c_{\mathrm{lat}} ~=~
N^{1/3}c_{\mathrm{lat}}
\end{equation}
The first equality is the definition from the isotropic model, the
second one identifies the isotropic connection with the lattice
connection $\tilde{k}$, the third one defines $c_{\mathrm{lat}}$ and
forth equality gives the final relation between the `$c$' variables of
the isotropic and the lattice models.

The state exhibiting isotropy on the macro-scale, may be characterised
by stipulating that the $\tilde{p}_I(\vec{v})$ values are all mutually
equal, and equal to $\tilde{p}_{\mathrm{lat}}$, for the vertices
comprising the fiducial cell and that this value is identified with the
isotropic variable, $\tilde{p}_{\mathrm{iso}}$. Recall that the
isotropic operator is obtained by an averaging of the lattice operators.
The stipulation says that the average value is realized at each of the
lattice vertices. This leads to,
\begin{equation} \label{PIdentification}
p_{\mathrm{iso}} ~:=~ V_0^{2/3}\tilde{p}_{\mathrm{iso}} ~=~
V_0^{2/3}\tilde{p}_{\mathrm{lat}} ~=~
V_0^{2/3}p_{\mathrm{lat}}\ell_0^{-2} ~=~ p_{\mathrm{lat}} N^{2/3}
\end{equation}
These identification give a relation between isotropic variables and a
state of lattice model with a scale parameter $N$. A dynamically
evolving isotropic universe may be thought of as a family of lattice
states with the scale $N$ being a function of the volume eg larger
number of elementary cells get `homogeneised/isotropised'. As an
example, if $N \propto p_{\mathrm{iso}}^{3/2}$, then $p_{\mathrm{lat}}$
will be a constant! The expressions seen in the context of isotropic
models with fiducial cell of size $V_0$ are now to be applied to the
elementary cell of size $\ell_0^3$.

In view of these identifications, let us consider the two specific
corrections seen in the isotropic model, namely the the inverse volume
corrections and the holonomy corrections. In the lattice model, these
corrections arise in the same manner as in the isotropic model, but from
the micro-cell. The inverse volume corrections are in powers of $p_*/p =
p_*/p_{\mathrm{lat}} = (p_*/p_{\mathrm{iso}})N^{2/3}$. For $N \propto
p_{\mathrm{iso}}^{3/2}$, these are independent of $p_{\mathrm{iso}}$.
(ii) With the holonomy corrections included, the effective density
(\ref{HolonomyCorrection}), is of the form $\rho(1 -
\rho/\rho_{\mathrm{crit}})$ with $\rho^{-1}_{\mathrm{crit}} \sim
\kappa\gamma^2p_{\mathrm{lat}} =
\kappa\gamma^2p_{\mathrm{iso}}N^{-2/3}$. Again for $N \propto
p_{\mathrm{iso}}^{3/2}$, $\rho_{\mathrm{crit}}$ is independent of
$p_{\mathrm{iso}}$. For a different dependence of $N$ on
$p_{\mathrm{iso}}$, the corrections will have non-trivial dependence on
$p_{\mathrm{iso}}$ and therefore on $V_0$, but $V_0$ is no longer a
purely mathematical artifact but is dictated by the underlying
inhomogeneous state.

In effect, a perspective from an underlying inhomogeneous model suggests
that the fiducial cell of a homogeneous model is selected by a state of
the inhomogeneous model and a dependence on the $V_0 := \ell_0^3 N$ is
not necessarily unphysical. For a more detailed discussion of
ramifications of these ideas, please see \cite{MartinLattice}.

\chapter{Appendix}
\section{Symmetric connections}
We assume that we have a manifold $M$ on which are defined connection
$A_{\mu}^a(z)$ with $a$ taking values in the Lie algebra $\underline{G}$
of a {\em gauge group} $G$. Assume further that there is an action of a
{\em symmetry group} $S$ on $M$ under which we want to have appropriate
notion of invariance. The infinitesimal action of the symmetry group is
generated by a set of vector fields $\xi_m^{\mu}\partial_{\mu}$ which
represent the Lie algebra of $S$: $[\xi_m, \xi_n] = f^p_{~mn}X_p$. The
Lie algebra of $G$ is generated by matrices $T_a$ satisfying: $[T_b,
T_c] = C^a_{~bc}T_c$.

When we have an ordinary tensor field, $T$, on a manifold, it is defined
to be invariant under (or symmetric w.r.t.) the action of an
infinitesimal diffeomorphism generated by a vector field $\xi$ if its
Lie derivative with respect to $\xi$ vanishes: $L_{\xi} T = 0$. When the
tensor fields also transform under the action of a gauge group, then the
invariance condition allows the Lie derivative to be an infinitesimal
gauge transform of the tensor field: $L_{\xi}T = \delta_{W(\xi)}T$,
where $W(\xi)$ is valued in the Lie algebra $\underline{G}$. Notice that
this associates a gauge transformation with a diffeomorphism.

This association has to satisfy two conditions: (a) If we took a gauge
transform of the tensor field and then applied the diffeomorphism, the
defining condition must be gauge covariant: $L_{\xi}(T^g) =
\delta_{W^g(\xi)} T$, where $W^g(\xi) = g^{-1}W(\xi)g +
g^{-1}L_{\xi}(g)$, and (b) The Lie derivatives represent the Lie algebra
of the symmetry group: $[L_{\xi_m}, L_{\xi_n}] T = L_{[\xi_m, \xi_n]} T
= f^p_{~mn}L_{\xi_p} T$ and the $W_m$ must obey the consequent
conditions. The task is to find those tensor fields which satisfy the
invariance conditions subject to the allowed gauge transformations.

This is aided by another consequence of the symmetry action. The action
of the the symmetry group $S$ on $M$ implies that $M$ can be expressed
as a collection of {\em orbits} of $S$. We will assume the simpler case
where {\em all} orbits are mutually diffeomorphic and are given by $S/F$
where $F$ is the stability subgroup of the $S-$action. Thus we obtain $M
\sim B + S/F$. Here $S/F$ is an orbit which is necessarily a coset space
while $B$ is a manifold whose points label the orbits. Note that a
non-trivial subgroup $F$ of $S$ means that a subset of vector fields
$\xi_m$ vanish at some point.  Corresponding to this structure of $M$,
its tangent and cotangent spaces are also decomposed. 

The solutions of the invariance conditions are constructed by using the
available structure on the group manifold $S$ and projecting these onto
$S/F$. Here are some details for the gauge connection\footnote{In the
context of Kaluza-Klein approach to unification, the analysis of
invariant quantities was carried out to construct suitable ansatz.
There, the space-time is taken to be of the product form $M_4 \times B$
with $B$ a compact manifold. The isometry groups of the compact manifold
played the role of symmetries and the forms of the fields on the
space-times were obtained.} \cite{ForgacsManton}.

Let $A := A^a_{\mu}T_a dx^{\mu}$ denote the $\underline{G}$ valued
connection 1-form (the gauge potential). Under a gauge transformation it
transforms as: $A^g := g^{-1} A g + g^{-1} d g$ and infinitesimally, $g
= 1 + \epsilon W$, $\delta_{\epsilon W}A = \epsilon D(W) := \epsilon(d W
+ [A, W]), ~ W := W^a T_a.$ Under an infinitesimal diffeomorphism, $x^{'
\mu} := x^{\mu} + \epsilon\xi^{\mu}$, it transforms as
$\delta_{\epsilon\xi}A := - \epsilon L_{\xi}A =  -
\epsilon(\partial_{\mu}\xi^{\nu}A_{\nu} + \xi^{\nu}\partial_{\nu}
A_{\mu})dx^{\mu} ~= - \left\{i_{\xi}dA + d(i_{\xi}A)\right\}$. 

The invariance conditions, when there are many symmetries, are: 
\begin{equation}\label{InvCondition}
L_{\xi_m} A = D(A)\ W_m := d W_m + [A, W_m] ~\Leftrightarrow~
\partial_{\mu}\xi^{\nu}A_{\nu} + \xi^{\nu}\partial_{\nu} A_{\mu} =
\partial_{\mu} W_m + [A_{\mu}, W_m]\ .
\end{equation}
Here, $W_m$ are some $\underline{G}$ valued scalars on the manifold $M$
associated with $\xi_m$. We induce a gauge transformation on the $W_m$
by demanding that the above condition be {\em gauge covariant}:
\begin{equation}\label{GaugeTransform}
L_{\xi_m}A^g := D(A^g)\ W_m^g ~~~\Leftrightarrow~~~ W_m^g := g^{-1}W_m g
+ g^{-1} L_{\xi_m}g~~,~~L_{\xi_m}\ g := \xi^{\mu}_m\ \partial_{\mu}\ g
\end{equation}
The Lie algebra of the vector fields, $\xi_m$, implies that the $W_m$'s
must satisfy: 
\begin{eqnarray}\label{AlgebraCondition}
[L_{\xi_m}, L_{\xi_n}]A & = & L_{[\xi_m, \xi_n]}A ~=~f^p_{~mn}L_{\xi_p}A
\hspace{3.0cm}\Rightarrow \nonumber\\
D(W_{mn}) & = & f^p_{~mn}D(W_p) ~~~,~~~ W_{mn} := L_{\xi_m}W_n -
L_{\xi_n} W_m + [W_m, W_n]~,~ ~~ or \nonumber \\
0 & = & D(W_{mn} - f^p_{~mn}W_p) 
\end{eqnarray}
where $[W_m, W_n]$ is the bracket in the Lie algebra, $\underline{G}$.

{\em Exercise:} For the field strength $F_{\mu\nu}^aT_a$, verify that
$L_{\xi_m} F = [F, W_m]$.

{\em Exercise:} Let $E$ be a vector field valued in $\underline{G}$
which transforms as $E^g = g^{-1} E g$. The condition for symmetric $E$
would be $L_{\xi_m} E = [E, W_m]$. Show that the $W_m$ transforms as
before and the symmetry Lie algebra implies $[E, W_{mn} - f^p_{~mn}W_p]
= 0$.

Suppose that $\chi_{mn} := W_{mn} - f^p_{~mn}W_p \neq 0$. Then the above
equations imply conditions on symmetric field strength (and indeed on
all symmetric quantities). For example, $D\chi_{mn} = 0$ implies that
$[F, \chi_{mn}] = 0$\footnote{Note that $D\chi = 0$ follows only for the
symmetric connections. If it were to hold for all connections, then $[F,
\chi] = 0$ would hold for all field strengths and this would correspond
to a reduction of the gauge group to the little group of $\chi_{mn}$.
This is analogous to the non-trivial Higgs vacua situation.}. This would
mean that the field strengths must commute with the $\chi_{mn}$'s. We
will assume that there does not exist any $\chi$ valued in
$\underline{G}$ such that $D \chi = 0$. This implies that $\chi_{mn} =
0$ i.e. 
\begin{equation}\label{Star}
L_{\xi_m}W_n - L_{\xi_n} W_m + [W_m, W_n] - f^p_{~mn}W_p  ~=~ 0 \ .
\end{equation}
Note that this is a condition involving only the $W_n$'s and the
symmetry generators $\xi_m$'s. Also observe that for vector fields
corresponding to the stability subgroup $F$, the (\ref{Star}) reduces to
$[W_{m}, W_{n}] = f^p_{~mn} W_p$ at the points where these vector fields
vanish.

The task is to characterise the symmetric connections satisfying eqn.
(\ref{InvCondition}) with $W_n$ satisfying eqn. (\ref{Star}) modulo
gauge transformations (\ref{GaugeTransform}) on a manifold $M \sim B
\times S/F$. The strategy is to show that the gauge freedom allows $W_n
$ to be taken in appropriate form and then determine the form of
symmetric connections in the same gauge.

Consider first the case where $F = \{e\}$ so that $S/F = S$ itself.
Introduce local coordinates $(x^i, y^{\alpha})$ on $B \times S$. Without
loss of generality, we can take the symmetry generators to be functions
of $y$ and with zero components along $B$. Noting that the vector fields
$\xi_m$ on $S$ are independent, it follows that the matrices
$\xi^{~\alpha}_m(y)$ are invertible and therefore we can define new
$\underline{G}$ valued 1-forms as: $W_m(x, y) := \xi^{~\alpha}_m
W_{\alpha}(x, y)$. It is easy to see that (i) $W_{\alpha}$ transform
exactly as a $G-$connection and (ii) the condition (\ref{Star}) is just
the statement that this connection is flat. Therefore, {\em locally} it
is always possible to choose $W_{\alpha}(x,y) = 0$ and hence $W_m(x,y) =
0$. Having chosen $W_m$'s to be zero, the gauge transformation freedom
is restricted to $L_{\xi_m}(g) = \xi_m^{~\alpha}\partial_{\alpha}g = 0$
i.e. the gauge functions must depend only on the $x$ coordinates.

In this gauge, the invariance conditions can be written separately for
$\mu = i$ and for $\mu = \alpha$ as: 
\[ \xi_m^{~\alpha}\partial_{\alpha}A_i = 0 ~~\mathrm{and}~~
(L_{\xi_m}A)_{\alpha} = 0
\]
The first implies that $A_i$ depends only on $x$ and so do the gauge
transformations. Hence $A_i$ is a $G-$connection on $B$. The solution
for $A_{\alpha}$ is obtained as follows.

On a group manifold, there are left and right actions of the group onto
itself which commute. Consequently, these generate left(right) invariant
vector fields and 1-forms (the Maurer-Cartan forms). Apply these to the
group manifold $S$. Assume that $\xi_m$ generate {\em left} action on
the group manifold and use the {\em left} invariant 1-forms eg the
unique, $\underline{S}-$valued Maurer-Cartan form
$\Theta_{MC}$\footnote{For classical matrix groups, these are given by
$g^{-1}dg$ and $dg g^{-1}$ and are left and right invariant
respectively.}. It is immediate that $L_{\xi_m} \Theta_{MC} = 0$. To
obtain a $G-$connection, we need a map $\Lambda: \underline{S} \to
\underline{G}$. Given such a map, we can define $A := \Lambda
(\Theta_{MC}) \leftrightarrow A_{\alpha}^a := \Phi^a_m
(\Theta_{MC})^m_{\alpha}$. Now $L_{\xi_m} A = 0 = L_{\xi_m}(\Phi)
\Theta_{MC} + 0$ implies that the ``Higgs'' fields, $\Phi_m$ are
constants on $S$ i.e. are functions only of $x^i$.

Thus, for the case where the $S-$action is free ($F$ is trivial), the
symmetric connections can be written as $A = {\cal A}_i(x)dx^i +
\Phi_m(x) \omega^m_{~\alpha}(y)dy^{\alpha}$ where ${\cal A}$ is an
arbitrary $G-$connection on $B$ and $\Phi_m(x)$ are ``Higgs'' scalars
valued in $\underline{G}$.

When the $S-$action is not free, the vector fields $\xi_m^{~\alpha}$ are
tangent to $S/F$ and the above steps do not go through immediately.
Nevertheless we can still find invariant, $\underline{S}-$valued 1-forms
on $S/F$ and a suitable map $\Lambda$ to construct invariant
connections. To see this, note that there is the natural projection map
$\pi: S \to S/F$. Choose an {\em embedding} $i:S/F \to S$. As discussed
above, on $S$ we have vector fields $\bar{\xi}_m$ generating left action
and the corresponding Maurer-Cartan form $\Theta_{MC}$. Using $\pi_*$ we
push-forward the vector field on to $S/F$ and using $i^*$ we pull-back
$\Theta_{MC}$ on to $S/F$. The projected vector fields match with the
$\xi_m$ (by definition of the symmetry action). Thus we get,
\[
\xi_n := \pi_*(\bar{\xi}_n) ~~,~~ \omega := i^*(\Theta_{MC}) ~~;~~
L_{\bar{\xi}_n}\Theta_{MC} = 0 ~\Rightarrow~ L_{\xi_n}\omega = 0 \ .
\]
As before, $\omega$ is valued in $\underline{S}$ since $\Theta_{MC}$ is.
Introduce $\Phi^a_n$ as before and define $A^a_{\alpha} := \Phi^a_n
\omega^n_{\alpha}$. Using these definitions let us rewrite the defining
equations as:
\begin{eqnarray}
\mbox{Invariance condition} \hspace{0.9cm} & :~~ & \left(\xi^{\alpha}_n
\omega^k_{\alpha}\right)\left( L_{\xi_m} \Phi_k - [\Phi_k, W_m]\right) ~
= ~ L_{\xi_n} W_m \ ; \label{Inv} \\
\mbox{Lie Algebra condition} \hspace{0.65cm} & :~~ & L_{\xi_m} W_n -
L_{\xi_n} W_m - [W_m, W_n] ~ = ~ f^p_{~mn} W_p \ ; \label{AlgebraCond}\\
\mbox{Gauge transformations} \hspace{0.5cm} & :~~ & W^g_m ~ = ~ g^{-1}
W_m g + g^{-1} L_{\xi_m} g \ ; \label{GaugeW}\\
& :~~ & \left(\xi^{\alpha}_n \omega^k_{\alpha}\right)\left( \Phi^g_k -
g^{-1} \Phi_k g \right) ~ = ~ g^{-1} L_{\xi_n} g \label{GaugePhi}
\end{eqnarray}
In writing the first equation we have used $L_{\xi_m}\omega^n = 0$ and
also multiplied by $\xi^{\alpha}_n$.

The last equation implies that $\Phi_n$'s transform as the adjoint
representation of $G$ {\em iff} the gauge transformations are constant
over $S/F$. We can take $\Phi_n$ to transform by the adjoint
representation of $G$, thereby restricting the gauge transformation to
be constant over $S/F$. In such a case, $W_n$'s also transform the same
way. For trivial $F$, we {\em can} transform away $W_n$ to {\em zero}
and recover the previous case. For non-trivial $F$, this is not the
case.

For non-trivial $F$, there exist a point, $y_0$ say, in $S/F$ at which
the vector fields $\xi_{\underline{m}}, \underline{m} = 1, \cdots,
\mathrm{dim}(F)$ vanish.  Then $L_{\xi_{\underline{m}}}$ terms drop out.
Consider the equation (\ref{AlgebraCond}) for $\underline{m},
\underline{n}$. Then, at $y_0$, we must have $[W_{\underline{m}},
W_{\underline{m}}] = f^{\underline{p}}_{~\underline{m}\underline{n}}
W_{\underline{p}}$.  Since $F$ is a subgroup, the sum on the right hand
side is restricted to $\underline{p}$. If there is a non-trivial
homomorphism $\lambda: F \to G$, it will induce a corresponding
homomorphism $\Lambda:\underline{F} \to \underline{G}$ on the Lie
algebras and we can choose the $W_{\underline{m}}(y^0)$ to represent it. 

Next, at $y_0$, consider the (\ref{Inv},\ref{AlgebraCond}) for
$\underline{m}$.  Eliminating $L_{\xi_n} W_{\underline{m}}$, and noting
that $\Phi_k$ alone depends on $x$, we must have (a)
$[W_{\underline{m}}, W_n] = f^p_{~\underline{m} n} W_p$ and (b)
$[\Phi_n, W_{\underline{m}}] = 0$. Note that this implies that the
residual gauge group is reduced to those elements of $G$ which commute
with $W_{\underline{m}}$ i.e. to the {\em centralizer of $\lambda(F)
\subset G$}. The gauge transformations are already restricted to be
functions of $x$ alone.

Considering the Jacobi identity for $\Phi_{\underline{k}}, \Phi_l,
W_{\underline{m}}$ it follows that $[\Phi_{\underline{k}}, \Phi_l] =
d^m_{~\underline{k} n}\Phi_m$ must hold for some $d$'s. This has the
same form as the condition (a) on the $W$'s. Hence
$d^m_{~\underline{k}l} = f^m_{~\underline{k}l}$ is obviously a solution.

In fact, it is a result that $S-$invariant connections, when they exist,
are in on-to-one correspondence with homomorphisms of the groups
$\lambda: S \to G$ and can be expressed as $A(x, y) = {\cal A}_i dx^i +
\Phi(x)_n i^*(\Theta_{MC})^n(y)$ where ${\cal A}$ is a connection on $B$
with the gauge group {\em reduced to the centralizer of $\lambda(F)$ in
G} (i.e.  group of all elements of $G$ which commute with the image of
$F$ in $G$ under the homomorphism $\lambda$). Furthermore the Higgs
fields have to satisfy the constraints: $[\Phi_{\underline{m}}, \Phi_n]
= f^p_{~\underline{m} n} \Phi_p$\footnote{The classification of
connections invariant under some group of automorphisms of appropriate
bundles is given by {\em generalised Wang theorem}. There are many
mathematical fine prints in the above discussion which should be seen in
the references in \cite{SymmetricConnections}.}.

What about other invariant fields, such as vector fields in the adjoint
of the gauge group (eg the triad fields)? Now the invariance condition
(\ref{Inv}) will change and also the corresponding gauge transformations
of the field. For $E^{\mu}_a\partial_{\mu}$, we will have $L_{\xi_m} E =
[E, W_m], E^g = g^{-1}E g$. Now use the projections $X_m$ of the {\em
left invariant vector fields} $\bar{X}_m$ on $S$ (these generate the
right action and are dual to the $\Theta_{MC}$) and write: $E^{\alpha}_a
:= \Psi^n_a X^{\alpha}_n$, $L_{\xi_m} X_n = 0$. The invariance condition
then becomes $(\omega^k_{\alpha}X^{\alpha}_n)(L_{\xi_m}\Psi^n - [\Psi^n,
W_m]) = 0$ and $(\omega^k_{\alpha}X^{\alpha}_n)((\Psi^n)^g - g^{-1}
\Psi^n g ) = 0$. The gauge transformations imply $\Psi^n$ transforms by
the adjoint representation and for $m = \underline{m}$, the invariance
condition implies: $[\Psi_n, W_{\underline{m}}] = 0$.  Exactly as
before, the $\Psi^n$ must satisfy constraints analogous to the
$\Phi_n$'s. Similar logic will hold for other tensor fields.

As a very simple illustration, consider the case of static magnetic
field in three dimensions invariant under translations along the z-axix.
We want to obtain the form of the vector potential $A_i$. In this case,
the gauge group $G = U(1)$ and the symmetry group $S = \mathbb{R}$ which
acts on $\mathbb{R}^3$ by translations. This action is free and
therefore $F = \{e\}$. We have $\mathbb{R}^3 \sim \mathbb{R}^2 \times
\mathbb{R}$. The Maurer-Cartan form on $\mathbb{R}$ is just $dz$. The
map from $\underline{\mathbb{R}} \to \underline{U(1)}$ is given by a
single `Higgs' scalar, $\Phi(x,y)$. The symmetric connection is then
given by $A_i(x,y,z)dx^i = {\cal A}_x(x,y)dx + {\cal{A}}_y(x,y)dy +
\phi(x,y) dz$. This just says that all the three components of the
vector potential depend {\em only} on $(x,y)$. Note that this is a
statement in the gauge where ``$W$'' has been set to zero. This implies
that the magnetic field is also independent of $z$. Note that since $A_z
:= \phi(x,y) \neq 0$, the magnetic field could be along any fixed
direction.

{\em Exercise:} Work out spherically symmetric Yang-Mills fields in three
dimensions. Now $G = SU(2), \ S = SO(3), \ F = U(1), \ \mathbb{R}^3 \sim
\mathbb{R}^+ \times S^2$.

Further examples may be seen in \cite{SymmetricConnections}.

\section{Schrodinger and Polymer Quantization}
We illustrate inequivalent quantization as well as the GNS procedure in
a simple example.

\subsection{The Weyl-Heisenberg C*-Algebra}
Consider the usual Schrodinger quantization of a single degree of
freedom. We have the usual Hilbert space $L^2(\mathbb{R}, dx)$, on which
are defined the self-adjoint operators $x, p$, satisfying the canonical
commutation relations: $[x, p] = i\hbar$. 

Define the corresponding unitary operators:
\begin{eqnarray}
U(\alpha) ~:=~ e^{i \alpha x} & , & V(\beta) ~:=~ e^{i\hbar^{-1}\beta
p}~~,~~\alpha, \beta \in \mathbb{R} \\
U^{\dagger}(\alpha) = e^{-i\alpha x} = U(-\alpha) = U(\alpha)^{-1} & , &
V^{\dagger}(\beta) = e^{-i\hbar^{-1}\beta p} = V(-\beta) = V(\beta)^{-1}
\nonumber
\end{eqnarray}

Using the BCH formula, 
\[
e^{A}\cdot e^{B} = e^{A + B + \Case{1}{2}[A, B]
+ \Case{1}{12}[A, [A, B]] - \Case{1}{12}[B, [A, B]] + \cdots}
\] 
it follows,
\begin{equation}
U(\alpha)U(\alpha') ~=~ U(\alpha + \alpha')~~,~~ V(\beta)V(\beta') ~=~
V(\beta + \beta')~~,~~ U(\alpha)V(\beta) ~=~
e^{-i\alpha\beta}V(\beta)U)(\alpha)
\end{equation}

Define, for $z := (\alpha + i \beta)/\sqrt{2} \in \mathbb{C}$, the
unitary operator, 
\begin{eqnarray}
W(z) & := & e^{i\frac{\alpha\beta}{2}} U(\alpha)V(\beta) \hspace{2.9cm}
= ~ e^{i(\alpha x + \hbar^{-1}\beta p)} \nonumber \\
& = & \mathrm{exp}\left[{i\left\{\frac{z + \bar{z}}{\sqrt{2}} x -i
\frac{z - \bar{z}}{\sqrt{2}} \frac{p}{\hbar}\right\}}\right] ~=~
\mathrm{exp}\left[{i\left\{z\frac{x - ip/\hbar}{\sqrt{2}} + \bar{z}
\frac{x + ip/\hbar}{\sqrt{2}}\right\}}\right] \nonumber \\
& = & e^{i\left(z a^{\dagger} + \bar{z} a\right)} \hspace{3.85cm} = ~
e^{-\frac{|z|^2}{2}}\ e^{iza^{\dagger}}\ e^{i\bar{z}a}~~~~
\mathrm{where,} \\
a & := & \frac{1}{\sqrt{2}}\left(x + i \frac{p}{\hbar}\right)
\hspace{3.05cm} \Rightarrow ~ [a, a^{\dagger} ] = 1 \ .
\end{eqnarray}
It follows, 
\begin{eqnarray}
W(z_1) W(z_2) & = & e^{-\frac{i}{2}(\alpha_1\beta_2 -
\alpha_2\beta_1)}W(z_1 + z_2) \nonumber \\
& = & e^{\frac{1}{2}(z_1\bar{z}_2 - z_2\bar{z}_1)} W(z_1 + z_2)
\nonumber \\
& = & e^{\frac{1}{2}\mathrm{Im}(z_1\bar{z}_2)} W(z_1 + z_2)
\hspace{3.0cm} \mathrm{and,} \label{Algebra}\\
W(z)^{\dagger} & = & W(-z) ~=~W(z)^{-1} \ . \label{StarRelation}
\end{eqnarray}

Taking finite linear combinations of products of the unitary operators
$W(z)$, we get an algebra called the {\em Weyl-Heisenberg algebra},
${\cal W}$. This is *-algebra due to the Hermitian dagger defined for
the operators. The unitary operators $W(z)$ are bounded and so are
polynomials in them. With respect to the operator norm (which satisfies
$||A^{\dagger}|| = ||A||, ||A^{\dagger}A|| = ||A||^2$), The
Weyl-Heisenberg algebra is a C*-algebra. Notice that the ${\cal W}$
C*-algebra is {\em non-commutative} and has two {\em commutative}
sub-algebras, namely those generated by the elements, $W(\Case{\alpha +
i0}{\sqrt{2}})$ and $W(\Case{0 + i\beta}{\sqrt{2}})$ respectively.

Thus at this stage we have constructed a C*-algebra of bounded operators
on the specific Hilbert space. We will now define a {\em positive linear
functional} on the C* algebra, ${\cal W}$, construct a unitary
representation of the algebra and show its equivalence to that provided
by the Schrodinger quantization. The same procedure will then be used to
construct another representation, the {\em Polymer Representation}, of
the same algebra.
\subsection{Re-construction of the Schrodinger Representation}
In the Hilbert space, consider the wavefunction, $\langle x|0\rangle :=
\psi_0(x) := \pi^{-1/4} e^{-x^2/2}$ so that $\langle 0|0 \rangle := \int
dx |\psi_0(x)|^2 = 1$. Following results hold:
\begin{eqnarray}
a |0\rangle & = & \frac{1}{\sqrt{2}}\left(x -
\frac{d}{dx}\right)\psi_0(x) ~~=~~0 \\
\left[W(z)\psi_0\right](x) & = & e^{\Case{i}{2}\alpha\beta}
\left[U(\alpha) V(\beta) \psi_0\right](x) ~~=~~
e^{\Case{i}{2}\alpha\beta}e^{i\alpha}\left[V(\beta)\psi_0\right](x)\nonumber
\\
& = & e^{\Case{i}{2}\alpha\beta} e^{i\alpha x}\psi_0(x + \beta) \\
& = & \langle x|W(z)|0\rangle ~~=~~ e^{-\frac{|z|^2}{2}}\langle x|e^{i z
a^{\dagger}} |0\rangle ~~:=~~ \langle x|z\rangle \\
\therefore \int dx \psi_0^*(x) \left[W(z)\psi_0\right](x) & = &
\pi^{-1/2} \int dx \ e^{-\Case{x^2}{2}} e^{i \alpha x - \Case{(x +
\beta)^2}{2}} \nonumber \\
& = & e^{-\Case{|z|^2}{2}} ~~:=~~ \langle 0|W(z)|0\rangle ~~=~~ \langle
0|z\rangle 
\end{eqnarray}

Define a linear functional $\Omega_{\mathrm{Sch}}$, on the
Weyl-Heisenberg algebra
\[
\Omega_{\mathrm{Sch}}\left(\sum_i c_i W(z_i)\right) ~:=~ \sum_i c_i
\Omega_{\mathrm{Sch}}(W(z_i)) ~:=~ \sum_i c_i \langle 0|W(z_i)|0\rangle
~:=~ \sum_i c_i e^{-|z_i|^2/2} \ .
\]
The first definition ensures linearity and the third one completes the
definition. The second definition (notational) makes it obvious that
$\Omega_{\mathrm{Sch}}$ is a positive linear functional since
$\Omega_{\mathrm{Sch}}(A^*A) = \langle 0|A^{\dagger}A|0\rangle =
||A|0\rangle||^2 \ge 0, \forall ~ A := c_i W(z_i) \in $ the C*-algebra.
The equality holds only if $A|0\rangle = c_i|z_i\rangle = 0$. Since
$|z_i\rangle$ states are linearly independent, there are no non-trivial
states in the Hilbert space which satisfy $A|0\rangle = 0$. The positive
linear functional then defines an {\em inner product} on the algebra by,
\[
\langle W(z), W(z')\rangle ~:=~ \Omega_{\mathrm{Sch}}(W(z)^{\dagger}
W(z')) ~~=~~ \langle 0|W(-z)W(z')|0\rangle
~~=~~e^{-\frac{\mathrm{Im}(z\bar{z}')}{2}}\langle 0|z'
- z\rangle \ .
\] 
and extended by linearity to the algebra. This turns the algebra into a
Hilbert space (distinct from the original Hilbert space).

Next, define operators, $\hat{W}(z)$ acting on the algebra, by,
\[
\hat{W}(z) [W(z')] ~:=~ W(z)W(z') ~=~
e^{\frac{\mathrm{Im}(z\bar{z}')}{2}} W(z + z')
\] 
and extended to the algebra by linearity. Similarly, one defines an
operator for each element of the algebra, in an obvious manner. 

{\em Exercise:} Show that $(\widehat{W(z)})^{\dagger} = \hat{W}(-z)$.

This implies that $\hat{W}(z)$ are unitary operators.

That $W(z) \to \hat{W}(z)$ provides a homomorphism of the ${\cal W}$
algebra is obvious from the action of the operators. Thus the algebra
with the inner product defined, carries a representation of itself in
which the $W(z)$ are represented unitarily.

Consider general matrix elements of the operators $\hat{W}(z)$: 
\begin{eqnarray}
\left\langle W(z_1), \hat{W}(z) \left[W(z_2)\right]\right\rangle & = &
\left\langle 0 |W(-z_1)W(z)W(z_2)|0\right\rangle \nonumber \\
& = & e^{\frac{1}{2}\left(z\bar{z}_2 - z_1\bar{z} - z_1\bar{z}_2\right)}
\left\langle 0|W(z + z_2 - z_1|0\right\rangle\nonumber \\
& = & e^{\frac{\left(z\bar{z}_2 - z_1\bar{z} - z_1\bar{z}_2\right)}{2}}
e^{- \frac{| z + z_2 - z_1 |^2}{2}} 
\end{eqnarray}

Observe that for $z = \alpha$ or $z = i\beta$, the above matrix elements
are continuous in $\alpha, \beta$ respectively. General matrix elements
are obtained from finite combinations of these and hence {\em the
$\hat{W}(\alpha)$ and $\hat{W}(i\beta)$} are both weakly continuous
families of unitary operators. Actually, these are also {\em strongly}
continuous families i.e. w.r.t. vector space norm. The strong continuity
can be checked by evaluating the norm $||(\hat{W}(z) -
\hat{W}(0))[W(z')]||$ and checking the limits for $z = \alpha/\sqrt{2}$
and $z = i \beta/\sqrt{2}$.

This allows us to define two self-adjoint operators (on the algebra) as,
\begin{eqnarray}
\hat{X} & := & \lim_{\alpha \to 0} \frac{\hat{W}(\alpha/\sqrt{2}) -
\mathbb{I}}{i \alpha} ~~,~~ \hat{P} ~ := ~ \hbar ~ \lim_{\beta \to 0}
\frac{\hat{W}(i\beta/\sqrt{2}) - \mathbb{I}}{i \beta} \\
& \Longrightarrow & \left[ \hat{X} , \hat{P} \right] ~=~ i\hbar
\mathbb{I}
\end{eqnarray}
The commutator can be evaluated directly using the definitions for
$\alpha, \beta \ne 0$ and using the existence of the limits guaranteed
by strong continuity. 
\subsection {Another positive linear functional and the Polymer
representation:}
Now view the Weyl-Heisenberg algebra defined above as an abstract
structure i.e. an algebra generated by elements $W(z), z \in
\mathbb{C}$, obeying the relations (\ref{Algebra},\ref{StarRelation})
with a norm defined by $||W(z)|| = 1 ~ \forall ~ z \in \mathbb{C}$ and
extended by linearity. Define a linear functional by,
\begin{equation} \label{PolyPositive}
\Omega_{\mathrm{Poly}}(W(z)) ~:=~ \left\{ \begin{array}{ll}1 &
\mathrm{if ~ Im}(z) = 0 \\ 0 & \mathrm{otherwise} \end{array} \right.
\end{equation}
This is positive because 
\begin{eqnarray}
\Omega_{\mathrm{Ploy}}\left(\sum_i \left\{C_iW(z_i)\right\}^{\dagger}
\sum_j\left\{C_j W(z_j)\right\}\right) & = & \sum_{ij} C^*_iC_j
e^{-\frac{1}{2}\mathrm{Im}(z_i\bar{z}_j)}\Omega_{\mathrm{Ploy}}(W(z_j -
z_i)) \nonumber \\
& = & \sum_i |C_i|^2 + \sum_{i\ne j} C^*_iC_j
e^{\frac{i}{2}\beta_i(\alpha_j - \alpha_i)}\delta_{\beta_i, \beta_j}
\hspace{1.0cm}
\end{eqnarray}
In the sum, only those pairs $(i,j)$ which have the same $\beta$,
contribute. Group together all the terms whose $\beta_i$ are equal (eg
$\{z_1, z_2, \ldots, z_m\}, \{z_{m+1}, \ldots z_{m+n}\} \ldots $), and
consider one such group at a time. In each such group, the phases in the
second term can be absorbed in the $C_i$'s (since the $\beta$ is common)
and then combining with the first term gives
$|\sum_i\{C_ie^{i\beta\alpha_i/2}\}|^2$ which is non-negative. This
completes the proof. Note that positive linear functional must evaluate
to 1 on the identity element of the algebra (namely, $\mathbb{I} :=
W(0)$) and therefore we {\em cannot} interchange $\beta \leftrightarrow
\alpha$ in the defining condition in (\ref{PolyPositive}).

The $\Omega_{\mathrm{Poly}}$ defines a {\em degenerate} inner product on
the algebra,
\[
\langle W(z'), W(z) \rangle ~:=~
\Omega_{\mathrm{Poly}}(W(z')^{\dagger}W(z)) ~=~
e^{-\frac{\mathrm{Im(z'\bar{z})}}{2}} \Omega_{\mathrm{Poly}}(W(z - z'))
\]
and extended by linearity. The elements whose norm w.r.t. this
degenerate inner product is zero, forms a closed subspace ${\cal N}$, of
the algebra and consists of elements of the form $\chi := \sum_i C_i
W(\Case{\alpha_i + i\beta}{\sqrt{2}}), \beta \in \mathbb{R} \,
\mathrm{such~that} \sum_i C_i e^{-i\beta\alpha_i/2} = 0$. Elements of
${\cal N}$ also satisfy the property: $\Omega_{\mathrm{Poly}}(A \chi) =
0 \ \forall \ A \ \in {\cal W}$, which is useful in the exercise below.
The quotient space, ${\cal W}/{\cal N}$, is an inner product space and
its Cauchy completion defines a Hilbert space of the {\em Polymer
Representation}.

{\em Exercise:} Let $A$ denote a general element of the algebra and
$\chi$ an element of ${\cal N}$. Define $[A] := \{B \in {\cal W}/B = A +
\chi\}$. Define $\langle [A], [B]\rangle := \langle A, B\rangle$ and
$\hat{W}(z)\{ [A] \} := [ \hat{W}\{A\} ]$. Show that these definition
are well defined and conclude that ${\cal W}/{\cal N}$ provides a {\em
unitary} representation of the quotient algebra. From now on, we refer
to the quotient representation without being explicit about it.

Observe that $\Omega_{\mathrm{Poly}}$ is continuous in $\alpha$ and {\em
discontinuous} in $\beta$. This directly implies that
$\hat{W}(i\beta/\sqrt{2})$ {\em cannot be weakly continuous} and
therefore we cannot define the analogue of $\hat{P}$. This follows by
noting that $\langle W(z'), \hat{W}(i\beta/\sqrt{2})\{W(z')\}\rangle
\sim \Omega_{\mathrm{Poly}}(W(i\beta/\sqrt{2} + z' - z'))$. The weak
continuity (actually also {\em strong continuity}) in $\alpha$ however
allows the definition of $\hat{X}$ self-adjoint operator. It remains to
make the representation explicit.

In both the cases above, with the Schrodinger and the polymer
functionals, we constructed a representation of the ${\cal W}$ in which
the $W(z)$ are represented by unitary operators.  This is the
Gelfand-Naimark-Segal (GNS) construction. In the Schrodinger case, we
obtained the $\hat{X}, \hat{P}$ operators satisfying the canonical
commutation relations. In the Polymer case we obtained only $\hat{X}$.
In both cases we get the following relations,  directly by applying the
definitions:
\begin{eqnarray}
\hat{W}^{\dagger}\left(i\beta/\sqrt{2}\right)~\hat{X}~\hat{W}\left(i\beta/\sqrt{2}\right)
& = & \hat{X} - \beta\mathbb{I} ~~ \Rightarrow \label{Translation}\\
\hat{X}\hat{W}\left(i\beta/\sqrt{2}\right)  -
\hat{W}\left(i\beta/\sqrt{2}\right) \hat{X} & = & -
\beta\hat{W}\left(i\beta/\sqrt{2}\right)
\end{eqnarray}

From these relations follow an important result. Note that $\hat{X}$ is
a self adjoint operator and therefore its spectrum is real. What can one
say about its eigenvectors? The above relation implies that {\em if
$|x\rangle$ is an eigenvector of $\hat{X}$ with eigenvalues $x$, then
$\hat{W}(i\beta/\sqrt{2})|x\rangle$ is also an eigenvector with
eigenvalues $(x + \beta)$. Hence, either every real number is an
eigenvalue or none is.}

{\em Suppose every $x$ is an eigenvalue}. Then we have the orthogonality
relation  $\langle x|x'\rangle = \delta_{x,x'}$ - the Kronecker
$\delta$. Taking expectation value of the second equation above, it
follows that $\beta f(x, \beta) = 0, \forall \, x, \beta \in
\mathbb{R}$, where $f(x, \beta) := \langle
x|\hat{W}(i\beta/\sqrt{2})|x\rangle$. This implies that $f(x,\beta)$ is
zero for $\beta \ne 0$ and $f(x, 0) = 1$ directly from the definition.
Thus $f(x, \beta)$ {\em cannot} be continuous at $\beta = 0$, for any
$x$. This means $\hat{W}(i\beta/\sqrt{2})$ cannot be weakly continuous
at $\beta = 0$. 

This also means that {\em if} $\hat{W}(i\beta/\sqrt{2})$ {\em is} weakly
continuous (as for the Schrodinger representation), then $\hat{X}$
cannot have {\em any eigenvector}. Each $x \in \mathbb{R}$ is a {\em
generalised eigenvalue} and hence, in the formal notation, $\langle
x|x'\rangle = \delta(x - x')$ - the Dirac $\delta$-function.

Thus, in the Schrodinger representation, $\hat{X}$ necessarily has only
generalised eigenvalues, while in the polymer representation, it {\em
could} have proper eigenvalues, but generalized eigenvalues is not ruled
out.

However, in the polymer representation, we note:
\begin{eqnarray}
\left\langle W(i\beta/\sqrt{2}), \, W(i\beta'/\sqrt{2}) \right\rangle &
= &
\Omega_{\mathrm{Poly}}\left(W(-i\beta/\sqrt{2})W(i\beta'/\sqrt{2})\right)
~ = ~ \delta_{\beta, \beta'} \ ; \\
\left\langle W(i\beta/\sqrt{2}), \,
\hat{X}\left\{W(i\beta'/\sqrt{2})\right\} \right\rangle & = &
- \beta\delta_{\beta, \beta'} \\
\left\|(\hat{X} - \lambda\mathbb{I})[W(i\beta/\sqrt{2})\right\| & = &
\lim_{\alpha \to 0} \left\| \frac{1}{i\alpha}\left[W(\Case{\alpha +
i\beta}{\sqrt{2}}) - (1 +
i\alpha\lambda)\right]W(i\beta/\sqrt{2})\right\| \nonumber \\ & = & 0
\hspace{2.0cm} \mathrm{for} ~ \lambda = -\beta \ .
\end{eqnarray}
which show explicitly that $W(-i\beta/\sqrt{2})$ is a normalized
eigenvector of $\hat{X}$ with eigenvalue $\beta$, for every $\beta \in
\mathbb{R}$. This means that the Polymer Hilbert space is {\em
non-separable}. 

This concludes the illustration of the GNS construction of
representations of C* algebras. In the next sub-section we see the
analogue of the spin network construction.
\subsection{Polymer representation via `spin networks'}
We begin by introducing `graphs' in a `0-dimensional manifold', define
`holonomies' and `spin network functions', define an inner product and
densely defined operators. More details may be seen in \cite{Shadows}.
\begin{enumerate}
\item {\em Graphs, holonomies, cylindrical functions:} Any {\em
countable} set of real numbers, $\{x_i\}$ represents a {\em graph} and
is denoted by $\gamma$\footnote{For precise technical conditions, please
refer to \cite{Shadows}}. Note that the `points' $x_j$ on the real line
correspond to {\em edges}. Associated to each edge, $x_j$, we define a
{\em point holonomy}, $e^{ikx_j}$, $k \in \mathbb{R}$ plays the role of
a {\em connection}. For each graph $\gamma$, define complex valued
functions, $f_{\gamma}(k) := \sum_{j}f_j e^{ikx_j}$. Let Cyl$_{\gamma}$
denote the vector space of $f_{\gamma}(k)$. Elements of this vector
space are said to be {\em functions cylindrical with respect to the
graph $\gamma$}. Let Cyl := $\sum_{\oplus}$ Cyl$_{\gamma}$, where the
sum is over all possible graphs. 

Thus a general element of Cyl is a function of $k$ expressible as a
countable linear combination of the {\em elementary functions}
$f_{x_j}(k) := e^{ikx_j}$'s.
\item {\em Inner Product on Cyl:} Define 
\[
\langle f, g \rangle_{\mathrm{Poly}} ~:=~ \lim_{L \to \infty}
\frac{1}{2L} \int_{-L}^{L} dk f^*(k)g(k)
\]
which for elementary functions $f_{x_j}(k), f_{x_l}(k)$ gives
$\delta_{x_j, x_l}$. Introducing the notation, $|x_j\rangle
\leftrightarrow e^{ikx_j}$ this is expressed as $\langle x_j, x_l\rangle
:= \delta_{x_j, x_l}, \ \forall \ x_j, x_l \in \mathbb{R}$.  Cauchy
completion of Cyl with respect to this inner product defines a Hilbert
space, ${\cal H}_{\mathrm{Poly}}$.
\item {\em Action of ${\cal W}$:} Define operators $\hat{W}(z)$ on Cyl
by,
\[
\hat{W}\left(\Case{\alpha + i \beta}{\sqrt{2}}\right) f(k) ~:=~
e^{-i\alpha\beta/2}e^{i\beta k}f(k - \alpha) ~~,~~\forall ~ f \in
\mathrm{Cyl}, ~ \forall ~ \alpha, \beta \in \mathbb{R}
\]
These are densely defined and can be extended to bounded unitary
operators on ${\cal H}_{\mathrm{Poly}}$. It is easily verified that this
provides a representation of the abstract algebra defined in
(\ref{Algebra}).

{\em Exercise:} Show that the 1-parameter families of unitary operators,
$\hat{W}(\alpha/\sqrt{2})$ and $\hat{W}(i\beta/\sqrt{2})$ are weakly
continuous at $\alpha = 0$ and weakly discontinuous at $\beta = 0$
respectively. This implies that while $\hat{X}$ can be defined from the
$\hat{W}(\alpha/\sqrt{2})$ family, there is no corresponding operator
from the second family. Thus the holonomies - $h_{x_j}(k)$ are well
defined but not the connection - $k$ itself.
\end{enumerate}
\subsection{Harmonic Oscillator in the polymer representation}
So we see two distinct representations of the same abstract algebra with
the polymer representation being similar to the LQG representation. Are
there observable quantities which would reveal which representation
occurs in nature?

An obvious candidate is to consider the dynamics of the Harmonic
oscillator, with the classical Hamiltonian, $H(x, p) := p^2/(2m) +
m\omega^2 x^2/2, \{x, p\} = 1$. In the quantum theory, the $x, p$ are
expected to be replaced by the corresponding operators. However, in the
polymer representation, there is no $\hat{p}$! So in proposing the
quantum Hamiltonian we need to introduce a {\em scale, $\mu_0$} and {\em
define} $\widehat{p^2} := \{ 2 - \hat{W}(i\mu_0/\sqrt{2}) +
\hat{W}(-i\mu_0/\sqrt{2}) \}/\mu_0^2$. We could of course define
$\hat{p}$ first and then take its square. This is a quantization
ambiguity not too important for our purposes here \cite{Shadows}. In the
Schrodinger representation, we could use exactly the same definition,
work out quantities of interest eg spectrum and then take the limit
$\mu_0 \to 0$. In the polymer representation, we {\em cannot} take this
limit.

Consider the eigenvalue equation for the Hamiltonian:
$\hat{H}|\psi\rangle  = E|\psi\rangle$. Writing $|\psi\rangle := \sum_{x
\ \in \ \mathrm{countable~set}} \psi(x)|x\rangle$, and noting that the
unitary operators in $\widehat{p^2}$ shift the $|x\rangle \to |x \pm
\mu_0\rangle$, the eigenvalue equation becomes a {\em difference}
equation, involving $\psi(x), \psi(x \pm \mu_0)$. This means that
$\psi(x)$ with $x$ in a lattice $L_{x_0} := \{ x = x_0 + \mu_0 N, N \in
\mathbb{Z}\}$ constitute a solution while those belonging to different
lattices are unrelated. Span of the vectors in any lattice form a {\em
separable} subspace of the polymer Hilbert space. The spectrum can be
determined for each lattice independentally.  This is analyzed in detail
in \cite{Shadows}. Suffice it to say that the spectrum differs from that
in the Schrodinger representation by terms down by powers of $\mu_0/d$.
Here, $d := \sqrt{\hbar/m\omega}$ is the length scale defined by the
system while $\mu_0$ is the length scale introduced by the approximation
for the momentum operator. For the physical systems modelled well by an
oscillator (eg for vibrational spectra of molecules), $\mu_0/d$ is
extremely small and so Schrodinger vs polymer representation cannot be
resolved by observations. 

Additional comments may be seen in \cite{Shadows}.

\section{Inverse Triad Operator(s)}
As noted before, the discrete nature of the spectrum of the triad
operator implies that its inverse is not densely defined. Consequently
the counterpart of the classical function $p^{-1}$ needs to be defined
indirectly, by a suitable prescription. Being a prescription, it
introduces quantization ambiguities. We will consider a prescription
which is sufficiently general.

We aim to define an operator $\widehat{\mathrm{sgn}(p) |p|^{-1}}$.
Introduce the following notation: $n^i$ is a unit, 3-dimensional vector
and $\tau_i$ are anti-hermitian generators of $SU(2)$ in the
$J^{\mathrm{th}}$ representation, satisfying 
\[
[\tau_i, \tau_j] = \epsilon_{ijk}\tau_k ~~,~~
\mathrm{Tr}_J(\tau_i\tau_j) = - \frac{1}{3}J(J + 1)(2J + 1)\delta_{ij}
~~~ := ~~~ - {\cal N}_J \delta_{ij}
\]
For $f_{\alpha}(p) := \mu_{\alpha}|p|^{-\alpha/2}$, and define
$g_{\alpha}(p) := \int^p f_{\alpha}^{-1}(x)dx = \mu_{\alpha}^{-1}
\mathrm{sgn}(p)|p|^{1 + \alpha/2}(1 + \alpha/2)^{-1}$. For $\alpha = 0$
we have $f_0 = \mu_0$ while for $\alpha = 1$ we have $f_1 = \mu_1
|p|^{-1/2}$ and we choose $\mu_1 := \lP\sqrt{\gamma\sqrt{3}/4}$. These
recover the $\mu_0$ and the $\bar{\mu}$ schemes.  We will suppress the
$\alpha$ label.  Define $h_f := e^{\lambda n^i\tau_i f(\alpha, p) c}$.
This is matrix of order $(2J + 1)$.  Classically, the following is true.
\begin{eqnarray}
h_f\{h_f^{-1}, |g|^l(p)\} & = & - \frac{\kappa\gamma}{3} \left(\lambda
n^i\tau_i\right) l|g|^{l -1}\mathrm{sgn}(p)
~\hspace{0.0cm}\mathrm{where~we~used~~~}f\frac{d|g|}{dp} =
\mathrm{sgn}(p). \\
|p|^{(l -1)(1 + \alpha/2)} & = & \mathrm{sgn}(p)\left[
\frac{3}{\kappa\gamma l\lambda}\left(\mu_{\alpha}( 1 +
\alpha/2)\right)^{l - 1}{\cal N}_J^{-1} \right] \nonumber \\ & &
\hspace{4.0cm} \times \left[\mathrm{Tr}_J\left( (n^i\tau_i)
h_f\{h_f^{-1}, |g|^l\}\right)\right] \\ \label{Defn1}
& = & \mathrm{sgn}(p)\left[ \frac{\kappa\gamma
l\lambda}{3}\left(\mu_{\alpha}( 1 + \alpha/2)\right){\cal N}_J
\right]^{-1}\nonumber \\ 
& & \hspace{4.0cm}\times \left[\mathrm{Tr}_J\left( (n^i\tau_i)
h_f\{h_f^{-1}, |p|^{l(1 + \alpha/2)} \}\right)\right] \label{Defn2}
\end{eqnarray}
Thus we have a classical expression for $|p|^{(l-1)(1 + \alpha/2)}$
which has {\em four} ambiguity parameters: $\alpha, J, l, \lambda$. $J$
is a positive half integer, $0 < l < 1, \alpha > -2$. The special cases
would be: (a) $\mu_0-$scheme: $\alpha = 0, \lambda = 1$; (b) improved
scheme: $\alpha = 1, \lambda = 1, j = 1/2$ and some special values of
$l$ explored; (c) Madhavan scheme: similar to the improved scheme except
$\lambda$ is correlated with the fiducial volume $V_0$ (more on this
later). We could {\em choose} $l -1 = (1 + \alpha/2)^{-1}$ to define
inverse triad, but we will postpone such choices.

The corresponding quantum operator is obtained by replacing the Poisson
bracket by $-i\hbar^{-1}$ times the commutator. The $-i$ is combined
with $n\cdot\tau$ to make the generators Hermitian and the $\hbar^{-1}$
combines with $\kappa$ to replace $\kappa$ by $\lP^2$. The commutator is
expanded as: $\hat{h}_f [ \hat{h}^{-1}_f, \widehat{|g|}^l ] =
\mathrm{I}.  \widehat{|g|}^l - \hat{h}_f \widehat{|g|}^l
\hat{h}^{-1}_f$. 

Observe that $n\cdot(-i\tau)$ can be diagonalised with diagonal elements
being $-J, -J + 1, \ldots J -1, J$. So the $h_f$ becomes the diagonal
matrix $e^{(i\lambda f_{\alpha}c)(J, J -1, ~, , ,~ -J + 1, -J)}$. So the
commutator terms are diagonal matrices.

The computations simplify if we label the basis states by \footnote{The
functions $f_{\alpha}(p)$ are taken to be dimensionless. This makes the
$\mu_{\alpha}$ to have dimensions of $\lP^{\alpha}$ and $g_{\alpha}(p)$
to have dimensions of $\lP^2$. The $v_{\alpha}$ is defined to be
dimensionless.}
\begin{equation}
v_{\alpha} ~ := ~
\left(\frac{1}{6}\gamma\lP^2\right)^{\alpha/2}\left(\frac{1}{\mu_{\alpha}(1
+ \alpha/2)}\right) \mathrm{sgn}(\mu) |\mu|^{1 + \alpha/2} ~~,~~
\hat{g}_{\alpha}|v_{\alpha}\rangle =
\frac{\gamma\lP^2}{6}v_{\alpha}|v_{\alpha}\rangle
\end{equation}
So that the $h_f$ shifts the $v_{\alpha}$ labels simply as,
\[
\widehat{e^{i \lambda k f_{\alpha}c}}|v_{\alpha}\rangle ~ = ~
|v_{\alpha} + 2 k \lambda\rangle  ~~~~ \because ~~ e^{ifc/2}|v\rangle =
|v + 1\rangle \ .
\]

With these, acting on a basis state $|v_{\alpha}\rangle$, the Tr$_J$
evaluates to,
\begin{eqnarray}
\left[\mathrm{Tr}_J\{ \cdots \}\right] & = &
\left(\frac{\gamma\lP^2}{6}\right)^{l}\sum_{k = -J}^J k\left\{
|v_{\alpha}|^l
- |v_{\alpha} - 2 k\lambda|^l \right\} \nonumber \\
& = & \left(\frac{\gamma\lP^2}{6}\right)^{l}\sum_{k = -J}^J k
|v_{\alpha} + 2 k\lambda|^l  \mathrm{~~~and~defining~} v_{\alpha} := 2
J\lambda q_{\alpha} \ , \nonumber \\
& = & \left(\frac{\gamma\lP^2\ \lambda}{6}\right)^{l} 2^l \sum_{k =
-J}^J k |J q_{\alpha}  + k|^l
~~:=~~\left(\frac{\gamma\lP^2\lambda}{3}\right)^{l}
\mathrm{sgn}(q_{\alpha})G_{J,l}(q_{\alpha}) \\
G_{J,l}(q_{\alpha}) & := & \mathrm{sgn}(q_{\alpha}) \sum_{k = -J}^J k |J
q_{\alpha}  + k|^l \label{GjlDefn}
\end{eqnarray}
The eigenvalues of $\widehat{|p|^{(l -1)(1 + \alpha/2)}}$ are then given
by
\begin{eqnarray}
\widehat{|p|^{(l -1)(1 + \alpha/2)}}|v_{\alpha}\rangle & := &
\mathrm{sgn}(v_{\alpha})\Lambda_{J, l,
\alpha}(v_{\alpha})|v_{\alpha}\rangle \\
\Lambda_{J, l, \alpha}(v_{\alpha}) & = & \left(\frac{\mu_{\alpha}
\gamma\lP^2\lambda}{3}\right)^{l -1}\frac{(1 + \alpha/2)^{l -
1}}{l}{\cal N}_J^{-1}\left[ G_{J,l}(q_{\alpha})\right] \nonumber 
\end{eqnarray}
The first bracket takes care of the dimensions and the remaining factors
are dimensionless. It remains to calculate the last square bracket which
is a $\lambda, \alpha$ independent, universal function of its argument
and depends only on $J, l$.

From its definition, it is easy to see that $G_{J,l}(0) = 0$ and
$G_{J,l}(-q_{\alpha}) = G_{J,l}(q_{\alpha})$, the
$\mathrm{sgn}(q_{\alpha})$ factor is crucial for this. Thus it suffices
to consider only $q_{\alpha} > 0$. 
\begin{eqnarray}
G_{j,l}(q_{\alpha} > 0) & = & \sum_{k = -J}^{J} k |k + Jq_{\alpha}|^l
\nonumber \\
& = & \sum_{k = -J}^{J} \left\{ (k + Jq_{\alpha})|k + Jq_{\alpha}|^l -
Jq_{\alpha}|k + Jq_{\alpha}|^l \right\} \nonumber \\
& = & \sum_{k = J(q_{\alpha} -1)}^{J(q_{\alpha} +1)}
\left\{\mathrm{sgn}(k)|k|^{l + 1} - Jq_{\alpha}|k|^l\right\} \\
G_{j,l}(q_{\alpha} \ge 1) & = & \sum_{k = J(q_{\alpha}
-1)}^{J(q_{\alpha} +1)} \left\{|k|^{l + 1} - Jq_{\alpha}|k|^l\right\}
\label{QGreater1}\\
G_{j,l}(0 < q_{\alpha} < 1) & = & - \sum_{k = J(q_{\alpha} -1)}^{0_-}
\left\{|k|^{l + 1} + Jq_{\alpha}|k|^l\right\} + \sum_{0_+}^{k =
J(q_{\alpha} +1)} \left\{|k|^{l + 1} - Jq_{\alpha}|k|^l\right\}
\label{QLess1}
\end{eqnarray}
In the second step, we have shifted $k \to k - Jq_{\alpha}$. $k$ is no
longer integral but still changes in steps of 1. Clearly, for
$q_{\alpha} \ge 1$, $k$ is positive and the sgn as well as the absolute
value are redundant (the $k = 0$ term for $q_{\alpha} = 1$ gives zero
and hence the sum is confined to positive $k$ only), as in
(\ref{QGreater1}). For $q_{\alpha} < 1$, the sum splits in two groups as
in (\ref{QLess1}), and the $0_{\pm}$ denote the respective limits on the
values of $k$ which must match with the other limits and shift in steps
of 1.

We will not simplify/approximate this further but consider the special
cases (i) $\alpha = 0, \lambda = 1$ and (ii) $\alpha = 1, J = 1/2, l =
2/3$.

(i) \underline{$\alpha = 0, \lambda = 1$}: (eigenvalues of $|p|^{l -1}$)
\begin{equation}
f_0(p) = \mu_0~~ , ~~ g_0(p) = \mu_0^{-1}\mathrm{sgn}(p)|p|~~ , ~~ V_0 =
\mu_0^{-1} \mathrm{sgn}(\mu)|\mu|
\end{equation}
\begin{eqnarray}
\Lambda_{J,l,0}(V_0) & = & \left(\frac{\mu_0\gamma\lP^2}{3}\right)^{l -
1} (l {\cal N}_J)^{-1} G_{J,l}(|\mu|/(2J\mu_0)) \\
\Lambda_{\Case{1}{2},\Case{1}{2},0}(V_0) & = &
\left(\frac{\gamma\lP^2}{6}\right)^{-1/2} \frac{1}{\sqrt{\mu_0}} \left(
\left|\frac{\mu}{\mu_0} + 1 \right|^{1/2} 
- \left|\frac{\mu}{\mu_0} - 1 \right|^{1/2}\right) \label{Mu0Vals}
\end{eqnarray}

(ii) \underline{$\alpha = 1, J = 1/2 $}: (eigenvalues of
$|p|^{\Case{3}{2}(l - 1)}$)
\begin{equation}
f_1(p) = \mu_1 |\mu|^{-1/2} ~~ , ~~ \mu_1 :=
\sqrt{\frac{\gamma\lP^2}{6}\frac{3\sqrt{3}}{2}} ~~,~~
g_0(p) = \mu_1^{-1}\mathrm{sgn}(p)|p|^{3/2} (3/2)^{-1} ~~ , 
\end{equation}
\begin{equation}
v_1 = \left(\frac{\gamma\lP^2}{6}\right)^{1/2} \mu_1^{-1} (3/2)^{-1}
\mathrm{sgn}(\mu)|\mu|^{3/2} ~~ = ~~ K \mathrm{sgn}(\mu)|\mu|^{3/2}
\end{equation}
\begin{equation}
\Lambda_{\Case{1}{2},l,1}(v_1) =
\left(\frac{\mu_1\gamma\lP^2}{3}\frac{3}{2}\lambda\right)^{l - 1} (l
{\cal N}_{\Case{1}{2}})^{-1} G_{1/2,l}(|v_1|/(2\Case{1}{2}\lambda))
\end{equation}
The $G$ can be computed directly from the (\ref{GjlDefn}), for $v_1 >
0$, as
\begin{equation} \label{Mu1GenVals}
G_{\Case{1}{2}, l}(v_1/\lambda) = \left(\frac{1}{2}\right)^{l + 1}
\left( \left| \frac{v_1}{\lambda} + 1 \right|^{l} - \left|
\frac{v_1}{\lambda} - 1 \right|^{l} \right)
\end{equation} 
For $l = 2/3$ the operator becomes $|p|^{-1/2}$ and the eigenvalue
becomes,
\begin{equation} \label{Mu1Vals}
\Lambda_{\Case{1}{2},\Case{2}{3},1}(v_1) ~=~
\left(\frac{\gamma\lP^2}{6}\right)^{-1/2}
\frac{3}{4}\left(\frac{K}{\lambda}\right)^{\Case{1}{3}}\left[
\left|\frac{v_1}{\lambda} + 1\right|^\frac{2}{3} -
\left|\frac{v_1}{\lambda} - 1\right|^\frac{2}{3} \right]
\end{equation}

For large $v_1$ and $\lambda = 1$, this matches with the eigenvalue
given by \cite{APSThree}\footnote{Actually, for large volume, the
leading term is {\em independent} of $\lambda$. The sub-leading
(correction) terms, do depend on $\lambda$.}. The difference arises
because the APS prescription takes the $\alpha = 0$ expression and
replaces $\mu_0$ by $\bar{\mu}$ in equation (\ref{Defn2}). 

For large $J$ the sum can be approximated using,
\[
\int_0^1 dx x^{r} ~=~ \frac{1}{r + 1} \approx \sum_{i = 1}^N
\left(\frac{i}{N}\right)^r \frac{1}{N} ~~~~ \Rightarrow
\sum_{i = 1}^N i^r \approx \frac{N^{r+1}}{(r + 1)}
\]
and applying it to the sums in the definition of the $G_{J,l}$.
\section{Inhomogeneous Lattice Models} 
In the main body, we focused on quantization of {\em symmetry reduced
models} which are based on a homogeneous and isotropic background. From
a perspective of the full theory, this background presumably corresponds
to a state of full theory. Generic states of the full theory would be
inhomogeneous. One way in which symmetric states of the full theory have
been understood in the LQC context is that the symmetric states are
those distributions in Cyl$^*$ of the full theory which have support on
the invariant connections \cite{SymmetricConnections}.  In the same
spirit, we may stipulate certain kinds of inhomogeneous states as those
distributions which have support on certain form of `inhomogeneous
connections'. Specific models can be then constructed using similar
strategies as used in LQC constructions. Such models can shed some light
on how homogeneous and isotropic models could be viewed from an
inhomogeneous perspective.  These so-called lattices models are briefly
summarised below. Details should be seen in \cite{MartinLattice}.

For definiteness, let us continue to work with homogeneous (not
necessarily isotropic), spatially flat, diagonalised model with a
fiducial cell of co-moving volume $V_0$ as before. The spatial isometries
provide directions (of the Killing vectors) and the fiducial metric
provides coordinates as background structures which are to be kept
fixed. Using these background structures, Bojowald constructs another
model as follows.

Choose a cubical lattice (say) aligned with the isometry directions and
with a spacing $\ell_0 := (V_0/N)^{1/3}$.  Let the vertices of the
lattices be denoted by $\vec{v}$ and the three oriented links be denoted
by $\vec{e}_{I,\vec{v}}(t) := \vec{v} + t \hat{e}_I, ~ t \in [0,
\ell_0]$ and $\hat{e}_I$ is the unit vector in the I$^{\mathrm{th}}$
direction.  For future reference, let $S_{I, \vec{v}}$ denote the
elementary surface perpendicular to the elementary link $\vec{e}_{I,
\vec{v}}$ and passing through its mid-point.

Restrict the connections and triad variables to be of the form,
\begin{equation}\label{LatticeFields}
A^i_a(x) := \tilde{k}_I(x)\delta_{(I)}^{i}\delta^I_a ~~,~~ E_i^a(x) :=
\tilde{p}^I(x)\delta^{(I)}_{i}\delta_I^a \ .
\end{equation}
These are the local versions of the diagonalised homogeneous models. The
diagonal form of the connections implies that the holonomies - path
ordered exponentials - become ordinary exponentials of line integrals.

The $\tilde{k}_I, \tilde{p}^I$ are further taken to be spatially
periodic with period $V_0^{1/3}$,
\begin{equation}\label{Periodicity}
\tilde{k}_I(x) ~=~ \sum_{\vec{m}} \tilde{k}_I(\vec{m})e^{i
\vec{m}\cdot\vec{x}} ~~,~~
\tilde{p}^I(x) ~=~ \sum_{\vec{m}} \tilde{p}^I(\vec{m})e^{i
\vec{m}\cdot\vec{x}} ~~,~~\vec{m} ~=~ 2\pi V_0^{-1/3}\vec{n}, ~~ \vec{n}
\in \mathbb{Z}^3 \ .
\end{equation}
The Poisson brackets between the connection and the triad lead to,
\begin{equation}
\{\tilde{k}_I(x), \tilde{p}^J(y)\} ~=~ \kappa\gamma\delta^J_I\delta^3(x,
y) \ .
~~\Rightarrow~~ \{\tilde{k}_I(\vec{m}), \tilde{p}^J(\vec{m'})\} ~=~
\kappa\gamma V_0^{-1}\delta^J_I\delta^3(\vec{m},-\vec{m'}) \ .
\end{equation}

In loop quantization, basic variables of the model will be
holonomies of the lattice connection along the three elementary links at
each vertex and the three fluxes of the lattice triad variables along
the elementary surfaces through the mid-points of the elementary links
and perpendicular to the link. As noted above, these holonomies will be
ordinary exponentials thanks to the diagonal form of the connection.

The line integrals of the connection along elementary links of the
lattice are given by,
\begin{eqnarray}
{\cal I}_{I,\vec{v}} & := & \int_{\vec{e}_{I, \vec{v}}} dt \
\tilde{k}_I(\vec{e}_I(t)) ~=~ \int_{\vec{e}_{I, \vec{v}}} dt
\sum_{\vec{m}}\tilde{k}_I(\vec{m})e^{i\vec{m}\cdot\vec{e}_{I,\vec{v}}(t)}
~=~ \sum_{\vec{m}}
\tilde{k}_I(\vec{m})e^{i\vec{m}\cdot\vec{v}}\int_0^{\ell_0} dt e^{i t
\vec{m}\cdot\hat{e}_I}
\nonumber \\
& = & \sum_{\vec{m}} \tilde{k}_I(\vec{m})\left\{2\
e^{i\vec{m}\cdot\vec{v}}e^{im_I\ell_0/2}
\left(\frac{\mathrm{sin}(m_I\ell_0/2)}{m_I}\right)\right\} ~~~
\mathrm{where,}~ m_I := \vec{m}\cdot\hat{e}_I \ .
\label{HolonomyIntegrand} \\
& \approx & \tilde{k}_I(\vec{v})\ell_0 ~~\because ~~~ m_I\ell_0 \ll 1
\mathrm{~dominates\ the\ sum.} \nonumber \\
h_{I, \vec{v}} & := & e^{\Case{i}{2} {\cal I}_{I,\vec{v}}} ~~ \approx ~~
e^{\Case{i}{2}\tilde{k}_I(\vec{v})\ell_0} ~~ := ~~
e^{\Case{i}{2}{k}_I(\vec{v})} \hspace{2.0cm}
(\mathrm{elementary~holonomies}) \label{LatticeHolonomy}
\end{eqnarray}
Likewise, the fluxes of the lattice triad along elementary surfaces are
given by,
\begin{eqnarray}
{\cal F}^J_{\vec{v}} & := & \int_{S_{J,\vec{v}}}
\sum_{\vec{m}}\tilde{p}^J(\vec{m})e^{\vec{m}\cdot\vec{y}}
~=~\sum_{\vec{m}}\tilde{p}^J(\vec{m})e^{\vec{m}\cdot\vec{v}}e^{im_J\ell_0/2}\int_{-\ell_0/2}^{\ell_0/2}
e^{i t m_K}dt \int_{-\ell_0/2}^{\ell_0/2} e^{i t m_L}dt \nonumber \\
& = & \sum_{\vec{m}}\tilde{p}^J(\vec{m})\left\{4
e^{i\vec{m}\cdot\vec{v}}e^{im_J\ell_0/2}\left(\frac{\mathrm{sin}(m_K\ell_0/2)\mathrm{sin}(m_L\ell_0/2)}{m_K
m_L}\right)\right\} \label{Fluxes} \\
& \approx & \tilde{p}_I(\vec{v})\ell_0^2 \hspace{6.7cm}
(\mathrm{elementary \ fluxes}) \label{ElementaryFlux}
\end{eqnarray}
The $J,K,L$ indices are chosen such that $\epsilon_{JKL} = 1$. This
takes care of the orientations. There are no smearing functions above
because the $\tilde{p}^I$ variables are (U(1)) gauge invariant thanks to
diagonalised form.

These variables satisfy the Poisson brackets,
\begin{eqnarray}
\{{\cal I}_{I,\vec{v}}, {\cal F}^J_{\vec{v}'}\} & = & \kappa\gamma
\delta^J_I \left[8 V_0^{-1}\sum_{\vec{m}}e^{i\vec{m}\cdot(\vec{v} -
\vec{v}')}\frac{\mathrm{sin}(m_I\ell_0/2)\mathrm{sin}(m_K\ell_0/2)\mathrm{sin}(m_L\ell_0/2)}{m_I
m_K m_L}\right] \nonumber \\
& = & \kappa\gamma\delta^J_I \left[\chi_{\ell_0}(\vec{v} -
\vec{v}')\right] ~~=~~ \kappa\gamma\delta^J_I\delta_{\vec{v}, \vec{v}'}
\end{eqnarray}
The square brackets above is the characteristic function of width
$\ell_0$ and centered at $(\vec{v} - \vec{v}')$ which is just the
Kronecker delta.

The kinematical Hilbert space is then described in terms of the flux
representation as:
\begin{eqnarray}
\hat{{\cal F}}^I_{\vec{v}} |\ldots, \ \mu_{I, \vec{v}}\ , \
\ldots\rangle & = & \left(\frac{\gamma\lP^2}{2} \mu_{I, \vec{v}}\right)
|\ldots, \ \mu_{I, \vec{v}}\ , \ \ldots\rangle ~~~,~~~\mu_{I,\vec{v}} \
\in \ \mathbb{Z} \label{FluxBasis} \\
\hat{h}_{I,\vec{v}}|\ldots, \ \mu_{I, \vec{v}}\ , \ \ldots\rangle & = &
|\ldots, \ \mu_{I, \vec{v}} ~ + ~ 1\ , \ \ldots\rangle
\label{HolonomyAction}
\end{eqnarray}
The flux eigenvalues are in {\em integer} steps of $\gamma\lP^2/2$
because the elementary holonomies suffice to separate the {\em lattice
connections} (periodic) and thus only their integer powers appear.

Subsequent steps are similar to what is done in the homogeneous models.
In particular, the volume corresponding to the cell with $N^3$ lattice
sites, can be expressed as 
\begin{eqnarray}
V & = & \int d^3x \sqrt{|\tilde{p}^1\tilde{p}^2\tilde{p}^3|} \approx
\sum_{\vec{v}}\ell_0^3\sqrt{|\tilde{p}^1(\vec{v})\tilde{p}^2(\vec{v})\tilde{p}^3(\vec{v})|}
= \sum_{\vec{v}}\sqrt{|{p}^1(\vec{v}){p}^2(\vec{v}){p}^3(\vec{v})|}
\nonumber \\ & \approx & \sum_{\vec{v}}\sqrt{|{\cal F}^1_{\vec{v}}{\cal
F}^2_{\vec{v}}{\cal F}^3_{\vec{v}}|} 
\end{eqnarray}
leading to the corresponding operator expression. 

There is no diffeomorphism constraint since the background coordinates
are fixed in defining the lattice, the SU(2) gauge invariance is first
reduced to the U(1)$^3$ due to restriction to diagonal connection and
triad and by the form of these variables, the $\tilde{k}_I, \tilde{p}^I$
are gauge invariant variables. Hamiltonian constraint remains as in the
case of homogeneous models. In essence, we have $N^3$ `homogeneous
models' (labelled by the Fourier label $\vec{m}$) at the level of basic
variables and the kinematical Hilbert space.  The role that $V_0$ played
in the homogeneous model is now played by $\ell_0$.  The {\em
inhomogeneity} is reflected by basis states having {\em different
values} of $\mu_{I,\vec{v}}$ variables. 

How do we relate this set-up to the isotropic one discussed before?

Observe that a generic basis state in the lattice model will be,
\begin{equation}
\psi_{\{\mu_{I,\vec{v}}\}}[h_{I,\vec{v}}] = \prod_{I,\vec{v}}
(h_{I,\vec{v}})^{\mu_{I,\vec{v}}} ~:=~ \langle k_J(x)|\ldots, \ \mu_{I,
\vec{v}}\ , \ \ldots\rangle ~~~,~~~k_J(x) := \tilde{k}_J(x)\ell_0 \ .
\end{equation}
If we choose $\tilde{k}(x) := \tilde{c} := V_0^{-1/3}c ~ \forall~ x, I$,
then the basis function becomes a function of a single variable $c$
(which is independent of $x$), and is of the form: 
\begin{equation}
\psi_{\mu}(c) ~ = ~ e^{i\mu c/2} ~~,~~ \mu := N^{-1/3}\sum_{I, \vec{v}}
\mu_{I, \vec{v}} ~~ \in \mathbb{Q} \ ;
\end{equation}
which can be viewed as a basis element of Cyl$_{\mathrm{isotropic}}$.
Thus we can define a map $\pi:$ Cyl$_{\mathrm{lattice}} \to $
Cyl$_{\mathrm{isotropic}}$, 
\begin{eqnarray}
\pi: |\ldots,\ \mu_{I,\vec{v}}\ ,\ \ldots\rangle ~ \to ~ |\mu\rangle &
\Leftrightarrow & 
\langle c|\mu\rangle ~ := ~ \langle k_J(x)|\ldots,\ \mu_{I,\vec{v}}\ ,\
\ldots\rangle\left.  \right|_{\tilde{k}(x) = \tilde{c}} \nonumber \\
\mathrm{with}~~ \mu & := & N^{-1/3}\sum_{I, \vec{v}} \mu_{I, \vec{v}}
\end{eqnarray}
Note that the image of $\pi$-map is a separable subspace of
Cyl$_{\mathrm{isotropic}}$, spanned by $|\mu\rangle, \mu \in
\mathbb{Q}$.

Clearly we cannot {\em uniquely} identify a cylindrical state of the
lattice model, given a cylindrical state of the isotropic model.
However, we can define a map $\sigma:$ Cyl$_{\mathrm{isotropic}}$ $\to$
Cyl$^*_{\mathrm{lattice}}~$, $~\sigma: |\mu\rangle \to (\mu|, \ \mu \in
\mathbb{R}$, such that,
\begin{equation}
(\mu|\ldots, \ \nu_{I, \vec{v}}\ , \ \ldots\rangle ~ = ~
\langle\mu|\pi\left(|\ldots, \ \nu_{I, \vec{v}}\ , \
\ldots\rangle\right)  ~~ = ~~\delta_{\mu,\nu}~~,~~ \nu :=
N^{-1/3}\sum_{I, \vec{v}} \nu_{I, \vec{v}} \ .
\end{equation}
In the second equality, we have used the inner product of the isotropic
model. This map embeds cylindrical states of the isotropic model into
the distributional states of the lattice model\footnote{Notice that
$\mu, \nu$ defined above are {\em rationals with a common denominator
$N^{1/3}$}. Therefore the distributions $(\mu|$ are non-trivial only for
$\mu \in \mathbb{Q}$ with the same denominator.
%\{$\psi_{\mu}(c), \ \mu \in \mathbb{Q}$\} does {\em not} form a
%separating set of functions for the {\em isotropic} variable $c$.
}.

Now, we have Operators $A^*$ acting on Cyl$^*_{\mathrm{lattice}}$
corresponding to operators $A$ acting on the Cyl$_{\mathrm{lattice}}$,
defined in the usual manner.  Those of these operators which act {\em
invariantly} on the image of $\sigma$ in Cyl$^*_{\mathrm{lattice}}$, can
be identified with operators of the isotropic model. For these
operators, we can define $A_{\mathrm{isotropic}}$ via the equation:
$\sigma( A_{\mathrm{isotropic}}|\mu\rangle ) := (\sigma
|\mu\rangle)A^*_{\mathrm{lattice}}$. Since we have embedded isotropic
states in the distributions of the lattice model and also have
correspondence between operators, matrix elements computed in the
isotropic model can be understood as actions of lattice distributions on
lattice cylindrical states.

Consider an operator $A_{\mathrm{lattice}}$ on Cyl$_{\mathrm{lattice}}$.
This defines an operator $A^*_{\mathrm{lattice}}$ on
Cyl$^*_{\mathrm{lattice}}$: $(A^*_{\mathrm{lattice}} \phi|\ldots,\
\nu_{I,\vec{v}}\ ,\ \ldots\rangle := (\phi |\left\{A_{\mathrm{lattice}}
|\ldots,\ \nu_{I,\vec{v}}\ ,\ \ldots\rangle\right\}$. {\em If}, for
every distribution $(\phi| = (\mu| =: \sigma(|\mu\rangle)$, the operator
$A^*_{\mathrm{lattice}}$ gives another distribution $(\mu'| =:
\sigma|\mu'\rangle$, {\em then} we get an operator on
Cyl$_{\mathrm{isotropic}}\ $: $A_{\mathrm{isotropic}}|\mu\rangle :=
|\mu'\rangle$.

%\begin{eqnarray} %
%\langle\nu|A_{\mathrm{isotropic}}|\mu\rangle_{\mathrm{isotropic}} & :=
%& \left\{(\nu|A^*_{\mathrm{lattice}}\right\}|\mu\rangle ~ = ~
%(\nu|\left\{A_{\mathrm{lattice}}|\ldots, \ \mu_{I, \vec{v}}\ , \
%\ldots\rangle\right\} ~:=~ \nonumber \\
%(\mu|\left\{A_{\mathrm{lattice}}|\ldots, \ \nu_{I, \vec{v}}\ , \
%\ldots\rangle\right\} &:=& % \end{eqnarray}
%
It is easy to see that the multiplicative operators on
Cyl$_{\mathrm{lattice}}$, give multiplicative operators on
Cyl$_{\mathrm{isotropic}}$. For example, taking $A_{\mathrm{lattice}} =
h_{J,\vec{v}'}$, the lattice state $|\ldots,\ \nu_{I,\vec{v}}\ ,\
\ldots\rangle$ will have the  $\nu_{J,\vec{v}'}$ incremented by 1. The
action of the $(\mu|$ will give $\delta_{\mu, \nu + 1}$. This can be
understood as the action of $(\mu - 1|$ on the original lattice state.
Thus $(A^*_{\mathrm{lattice}}\mu| = (\mu - 1|$ which implies the a
multiplicative action $A_{\mathrm{isotropic}}|\mu\rangle := |\mu -
1\rangle$.

For elementary flux operators, little more work is needed. For example,
action of $\hat{{\cal F}}^I_{\vec{v}}$ on a basis state, $|\ldots,\
\nu_{J,\vec{v}'}\ ,\ \ldots\rangle$ is {\em zero} unless
$\nu_{J,\vec{v}'} \neq 0$ for some $J = I$ and at some $\vec{v}' =
\vec{v}$ i.e. 
\begin{eqnarray}
\hat{{\cal F}}^I_{\vec{v}}|\nu_{J,\vec{v}'}\rangle & = &
\Case{1}{2}\gamma\lP^2\nu_{J,\vec{v}'}\delta^I_J\delta_{\vec{v},
\vec{v}'}|\nu_{J,\vec{v}'}\rangle ~~~~~~~or \nonumber \\
(\mu|\hat{{\cal F}}^I_{\vec{v}}|\nu_{J,\vec{v}'}\rangle & = &
\left[\frac{1}{2}\gamma\lP^2\nu_{J,\vec{v}'}\right]\delta^I_J\delta_{\vec{v},
\vec{v}'}\delta_{\mu,\nu}~~~,~~~\nu := N^{-1/3}\nu_{J,\vec{v}'}
~~~However, \\
(\mu'|\nu_{I,\vec{v}'}\rangle & = & (\mu'|\nu_{I,\vec{v}''}\rangle
~~~~~\forall ~~~~~ (\mu'|,\ \vec{v}', \ \vec{v}'' \ .
\end{eqnarray}
Thus $(\hat{{\cal F}}^I_{\vec{v}})^*$ {\em cannot} act invariantly on
the image of $\sigma$ in Cyl$^*_{\mathrm{lattice}}$. It is clear though
that if we {\em sum} the elementary flux operators (with the directional
index $I$) over {\em all} the lattice sites (this is a finite sum due to
the cell), then the sum will act invariantly. By averaging over the
directions as well, we can construct an operator corresponding to a
`flux' operator on Cyl$_{\mathrm{isotropic}}$. In equations,
\begin{eqnarray}
\hat{p}^I_{\mathrm{lattice}} & := & N^{-1/3} \sum_{\vec{v}} \hat{{\cal
F}}^I_{\vec{v}} ~~,~~ \hat{p}_{\mathrm{lattice}} ~ := ~
\frac{1}{3}\sum_I \hat{p}^I_{\mathrm{lattice}} ~~ \Rightarrow \\
(\mu| \left\{\hat{p}_{\mathrm{lattice}}|\ldots,\ \nu_{J,\vec{v}'}\ ,\
\ldots\rangle\right\} & = & \frac{1}{6}\gamma\lP^2 N^{-1/3}\sum_{J,
\vec{v}} \nu_{J, \vec{v}}(\mu|\ldots,\ \nu_{J,\vec{v}'}\ ,\
\ldots\rangle \nonumber \\
& = & \frac{1}{6}\gamma\lP^2\ \nu\ \delta_{\mu,\nu} ~~,~~ \nu :=
N^{-1/3} \sum_{I, \vec{v}}\nu_{J, \vec{v}}
\end{eqnarray}
The last expression is exactly the matrix element of the $\hat{p}$
operator defined in the isotropic model. As noted in the footnote, the
identification of the matrix elements is restricted to $\mu, \nu \in
\mathbb{Q}$.

This completes our summary of the lattice model and how isotropic model
is `embedded' in the lattice model. This `embedding' refers to embedding
of the particular separable subspace of Cyl$_{\mathrm{isotropic}}$.

\end{document}